\documentclass[12pt]{article}

\usepackage{jheppub} % for details on the use of the package, please                    % see the JHEP-author-manual
\pdfoutput=1 % if your are submitting a pdflatex (i.e. if you have
\usepackage{amsmath}
\allowdisplaybreaks[4]        
\usepackage{amssymb}
\usepackage{euscript}     
\usepackage{color}         
\usepackage{tensor}        
\usepackage{amsthm}
\usepackage{bbm}
\usepackage{float}
\usepackage{pgfplots}
\usepackage{graphicx,subcaption}
\usepackage{mathtools}	

\usepackage{hyperref}
\hypersetup{colorlinks=true,allcolors=blue}
\usepackage[most]{tcolorbox}
\tcbuselibrary{skins, breakable, theorems}
\usepackage[header,title,page,titletoc]{appendix}  
\usepackage[T1]{fontenc} % if needed
\usepackage[numbers]{natbib}  
\usepackage{romannum}

\newcommand{\arccosh}{\text{arccosh}}

\newcommand{\eg}{{\it e.g.,}\ }
\newcommand{\ie}{{\it i.e.,}\ }

\newcommand{\mt}[1]{\textrm{\tiny #1}}
\renewcommand{\(}{\left(}
\renewcommand{\)}{\right)}

\newcommand{\be}{\begin{equation}}
\newcommand{\ba}{\begin{eqnarray}}
\newcommand{\ea}{\end{eqnarray}}
\newcommand{\ee}{\end{equation}}

\newcommand{\bea}{\begin{eqnarray}}
\newcommand{\eea}{\end{eqnarray}}
\newcommand{\bes}{\begin{equation*}}
\newcommand{\beas}{\begin{eqnarray*}}
\newcommand{\eeas}{\end{eqnarray*}}
\newcommand{\bas}{\begin{array*}}
\newcommand{\eas}{\end{array*}}
\newcommand{\ees}{\end{equation*}}
\newcommand{\nn}{\nonumber}

\newcommand{\ep}{\epsilon}

\newcommand{\GN}{G_{\mt{N}}}

\newcommand{\mD}{\mathcal{D}}
\newcommand{\mP}{\mathcal{P}}

\begin{document}

\subheader{\today}
\title{Toward a Consistent Definition of Holographic Entanglement Entropy in de Sitter Space}

\author[a,b]{Shan-Ming Ruan,}
\author[c]{Yu-ki Suzuki}

\affiliation[a]{School of Physics, Peking University,\\
Beijing 100871, China}
\affiliation[b]{Center for High Energy Physics, Peking University,\\
Beijing 100871, China}

\affiliation[c]{Center for Gravitational Physics and Quantum Information, \\ Yukawa Institute for Theoretical Physics, Kyoto University,\\
Kitashirakawa Oiwakecho, Sakyo-ku, Kyoto 606-8502, Japan}

% e-mail addresses: one for each author, in the same order as the authors

\emailAdd{ruanshanming@pku.edu.cn}
\emailAdd{yu-ki.suzuki@yukawa.kyoto-u.ac.jp}

\abstract{We investigate a new definition of holographic entanglement entropy in the framework of static patch holography for de Sitter space. Using the replica trick and twist operator formalism, we derive an entropy functional in dS$_3$ expressed through de Sitter Green’s functions. Since the naive Ryu–Takayanagi prescription in de Sitter spacetime fails to satisfy strong subadditivity, we demand that our proposed formula be consistent with the fundamental entropic inequalities. The resulting conditions place nontrivial constraints on the undetermined parameter space of static patch holography. Our results demonstrate that entanglement inequalities provide a sharp diagnostic for candidate definitions of entanglement entropy in de Sitter holography and offer quantitative evidence in support of static patch holography as a consistent framework.}

%%%%%%%%%%%%%%%%%%%%%%%%%%%%%%%%%%%%%%%%%%%%%%%%%%%%
%%%%%%%%%%%%%%%%%%%%%%%%%%%%%%%%%%%%%%%%%%%%%%%%%%%%
%%%%%%%%%%%%%%%%%%%%%%%%%%%%%%%%%%%%%%%%%%%%%%%%%%%%
%%%%%%%%%%%%%%%%%%%%%%%%%%%%%%%%%%%%%%%%%%%%%%%%%%%%
\begin{flushright}
YITP-25-127
\\
\end{flushright}
%%%%%%%%%%%%%%%%%%%%%%%%%%%%%%%%%%%%%%%%%%%%%%%%%%%%
	\maketitle
	\flushbottom

%%%%%%%%%%%%%%%%%%%%%%%%%%%%%%%%%%%%%%%%%%%%%%%%%%%%%%%%%%%%%%%%%%%%%%%%%%%%%%%%%%%%%%%%%%%%
\section{Introduction to de Sitter Holography}
Entanglement entropy has emerged as a fundamental quantity in quantum field theory (QFT) and quantum gravity, serving as a powerful probe of quantum correlations between spatially separated regions \cite{Callan:1994py,Calabrese:2004eu}. In the AdS/CFT correspondence \cite{Maldacena:1997re}, the celebrated Ryu-Takayanagi (RT) prescription \cite{Ryu:2006bv,Ryu:2006ef} provides a geometric dual of the entanglement entropy in the boundary QFT, \ie the area of a minimal surface in the AdS bulk. A proof of the RT formula was later obtained using gravitational path integrals \cite{Lewkowycz:2013nqa}. Importantly, holographic entanglement entropy defined in this way is consistent with basic information-theoretic principles: it satisfies standard entropy inequalities such as subadditivity, the Araki–Lieb inequality, and strong subadditivity (SSA) \cite{Lieb:1973zz,Lieb:1973cp}. In particular, the holographic proof of SSA was established in \cite{Headrick:2007km}. There have also been attempts to generalize the RT formula to a finite radial cutoff, for example, through the $T\bar T$ deformation \cite{Zamolodchikov:2004ce,McGough:2016lol,Guica:2019nzm}. In this setting, it has been observed that boosted versions of strong subadditivity, associated with time-dependent subsystems, may be violated \cite{Sanches:2016sxy,Lewkowycz:2019xse,Grado-White:2020wlb}, indicating that holographic entanglement entropy (HEE) is not well defined in the AdS/CFT correspondence with a finite cut-off surface.

The situation is even more subtle in de Sitter (dS) space. In dS$_3$, for instance, there are no spacelike geodesics connecting two points on future infinity, which obstructs a straightforward application of the RT prescription to the dS/CFT correspondence of \cite{Strominger:2001pn}. Analytic continuation to Lorentzian dS spacetime involves complex geodesics whose lengths yield an imaginary contribution to the entropy, corresponding to half the de Sitter entropy \cite{Doi:2022iyj,Doi:2023zaf}. The difference with the AdS/CFT case originates from the fact that the dual CFT living on the future infinity of dS is non-unitary.

More recently, alternative approaches to de Sitter holography have been proposed by introducing a timelike boundary, in analogy with AdS/CFT \cite{Susskind:2021omt,Kawamoto:2023nki,Anninos:2024wpy,Silverstein:2022dfj,Batra:2024kjl}. A straightforward generalization to explore the holographic entanglement entropy in de Sitter space is to apply a dS analogue of the RT formula (for further developments in this direction, refer to  \cite{Arias:2019pzy,Arias:2019zug,Narayan:2022afv,Narayan:2023zen,Nomura:2017fyh,Murdia:2022giv,Franken:2023pni,Franken:2023jas}). However, it has been shown that the HEE defined in this way fails to satisfy (strong) subadditivity \cite{Kawamoto:2023nki,Chang:2024voo}, a basic property of entanglement entropy associated with any density matrix. In this work, we focus on static patch holography with the holographic screen placed at the stretched horizon, and aim to resolve this difficulty by proposing a novel candidate for holographic entanglement entropy in de Sitter space. Our construction is based on the replica trick derivation of entanglement entropy in quantum field theories, originally developed by Calabrese and Cardy \cite{Calabrese:2004eu,Calabrese:2009qy}. For simplicity, we focus on the case of dS$_3$. Applying the replica trick to the putative dual field theory on the stretched horizon, we are led to define the holographic entanglement entropy in the static patch holography as
\begin{equation}
S{\mt{EE}} = -C\cdot \log G_{\rm dS}(x,x')
= - C \cdot\log\left(\frac{\sin \mD_\ep}{\sin\mD} \frac{\sinh\mu(\pi-\mD)+ N \sinh (\mu\mD)}{\sinh\mu(\pi-\mD_\ep)+ N \sinh (\mu\mD_\ep)} \right) \,.
\end{equation}
Here, $G_{\rm dS}(x,x')$ denotes the generic two-point function of scalar operators in the Euclidean vacuum of de Sitter space; $\mD$ is the dimensionless geodesic distance between the two endpoints $x$ and $x'$ of the subsystem; and $\mu=\sqrt{m^2L^2-1}$ characterizes the scaling dimension of the scalar operator in dS$_3$.\footnote{We assume $m \geq 1/L$ here, though later we also consider the case $m \leq 1/L$.} We then investigate the constraints imposed on the free parameter space by demanding consistency with basic entropy inequalities.
 
This paper is organized as follows. In the next subsection, we review various proposals for de Sitter holography, with particular emphasis on the static patch construction. Section \ref{sec:reviewsection} introduces the twist operator formalism and discusses the relevant Green’s functions in de Sitter space. In section \ref{eq:defineEEindS}, we present our proposed definition of holographic entanglement entropy in the static patch and analyze the conditions under which the fundamental entropy inequalities are satisfied. Finally, section \ref{sec:dis} discusses several subtle aspects of the proposal and outlines directions for future investigation.

In the following, we review several proposals for de Sitter holography that are particularly relevant to the present study. For other approaches not covered here, as well as discussions on different aspects of entanglement entropy in de Sitter space, see \cite{Nomura:2016ikr,Cotler:2019nbi,Arias:2019pzy,Arenas-Henriquez:2022pyh,Hikida:2021ese,Silverstein:2022dfj,Narovlansky:2023lfz,Anninos:2024wpy,Batra:2024kjl}.

\subsection{dS/CFT correspondence}
The dS/CFT correspondence was originally conjectured by Strominger in \cite{Strominger:2001pn}. This proposal posits that quantum gravity in $(d+1)$-dimensional de Sitter space is dual to a $d$-dimensional CFT defined at future infinity, \ie the asymptotic boundary of dS spacetime. Analogous to the holographic dictionary in the AdS/CFT correspondence \cite{Gubser:1998bc,Witten:1998qj}, correlation functions on both sides can be derived from
the equivalence between the bulk wavefunction of the universe in de Sitter space and the partition function of the boundary CFT. In \cite{Maldacena:2002vr}, it is suggested that this correspondence may be understood via an analytic continuation of the curvature radius from AdS to dS space. In particular, for the  dS$_3$/CFT$_2$ case, the central charge $c$ of the boundary CFT is proposed to be
 \begin{equation}
    c= i \frac{3L }{2\GN} \,,
 \end{equation}
where $L$ denotes the dS radius. This expression implies that the central charge is purely imaginary at tree level, suggesting that the boundary CFT is non-unitary. This non-unitarity is partially attributed to the absence of a well-defined time direction on the future boundary where the CFT resides. One realization of the dS/CFT correspondence was proposed in \cite{Anninos:2011ui}, wherein the analytic continuation ($N\rightarrow-N$) of the $O(N)$ vector model leads to a CFT with  Sp$(N)$ symmetry in CFT. Another realization derived in \cite{Hikida:2021ese,Hikida:2022ltr} involves performing an analytic continuation of the level $k$ in the Wess-Zumino-Witten model.

\subsection{Static Patch Holography}
Driven by the AdS/CFT correspondence, it is natural to look for a holographic framework in which the boundary theory has a Lorentzian time dimension. To this end, distinct models involving timelike boundaries have been proposed. In this subsection, we review the static patch holography proposals developed in references \cite{Susskind:2021omt,Susskind:2021dfc,Shaghoulian:2021cef,Shaghoulian:2022fop} and explain why the cosmological horizon is considered the most suitable location for the holographic screen.

\subsubsection{Bousso's covariant entropy bound}
As preparation, we first recall Bousso’s covariant entropy bound \cite{Bousso:1999xy}, later proven in \cite{Bousso:2014sda,Bousso:2014uxa} (see also \cite{Bousso:2002ju} for a detailed review). Broadly speaking, this conjecture can be regarded as a covariant generalization of the holographic principle originally proposed in \cite{tHooft:1993dmi,Susskind:1994vu}. This bound provides the conceptual foundation for associating the holographic degrees of freedom with the cosmological horizon in de Sitter space.

\begin{figure}[t]
 \centering
 \includegraphics[width=4.5in]{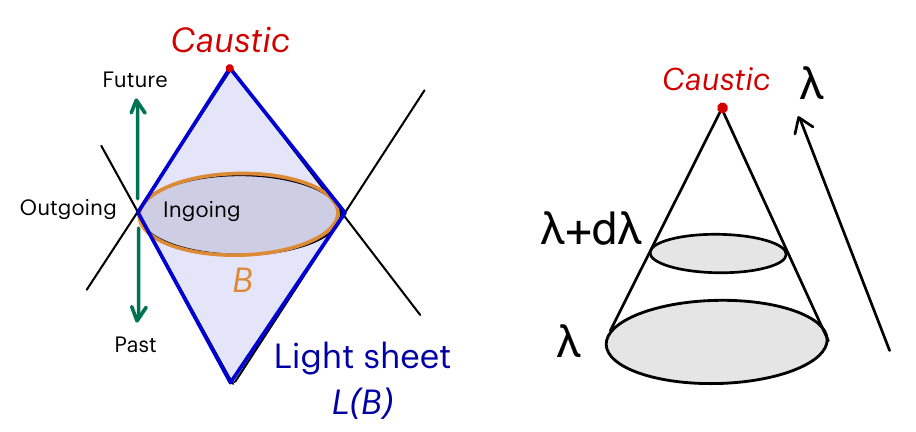}
 \caption{Left: the light sheet $L(\mathcal{B})$ (purple) associated with a spacelike codimension-two surface $\mathcal{B}$. The orthogonal null geodesics are classified into future- (past-) directed ingoing or outgoing congruences. The light sheet is defined by the ingoing ones. Right: the evolution of a null congruence under an infinitesimal change in the affine parameter $\lambda$.}\label{fig:Bound}
\end{figure}

Consider a $d$-dimensional Lorentzian manifold $\mathcal{M}$, and let $\mathcal{B}$ be a codimension-two spacelike surface. We denote its area by $A(\mathcal{B})$. From $\mathcal{B}$, as illustrated in figure \ref{fig:Bound}, there are four families of orthogonal null geodesics (or lightrays): future-directed ingoing, future-directed outgoing, past-directed ingoing, and past-directed outgoing. To distinguish between ``ingoing'' and ``outgoing,'' consider a null congruence parametrized by an affine parameter $\lambda$. After an infinitesimal displacement $d\lambda$ along the congruence, we evaluate the area of the deformed surface $\mathcal{B}(\lambda+d\lambda)$. If the area decreases, $A(\mathcal{B}(\lambda+d\lambda)) \leq A(\mathcal{B}(\lambda))$, we classify the direction as ingoing; otherwise, it is outgoing. Equivalently, defining the expansion as
\begin{equation}
\theta(\lambda) = \frac{1}{A(\mathcal{B})} \frac{dA(\mathcal{B})}{d\lambda} \,,
\end{equation}
a null geodesic is said to be ingoing if $\theta \leq 0$. At least one of the future-directed null congruences must satisfy this condition.

According to the Raychaudhuri equation, the expansion $\theta$ decreases monotonically along a null congruence if the initial condition $\theta \leq 0$ is satisfied and appropriate energy conditions hold. For example, assuming the null energy condition (NEC), the Raychaudhuri equation implies the following inequality:
\begin{equation}
\dot{\theta} \leq -\frac{\theta^2}{2} \,.
\end{equation}
This ensures that a caustic inevitably forms within a finite affine parameter, at which point the null congruences focus and intersect. We thus define a codimension-one null hypersurface with $\theta \leq 0$ as a light sheet of $\mathcal{B}$, denoted $L(\mathcal{B})$. The covariant entropy bound then states that
\begin{equation}
S[L(\mathcal{B})] \leq \frac{A(\mathcal{B})}{4G} \,,
\end{equation}
where $S[L(\mathcal{B})]$ is the entropy of matter fields propagating on the light sheet. In cosmological spacetimes, this quantity can be computed by integrating the local entropy density along $L(\mathcal{B})$.

The covariant entropy bound has been verified in general Friedmann–Robertson–Walker (FRW) universes with equation of state parameter in the range $-1 \leq w \leq 1$. A general proof was later obtained in \cite{Bousso:2014sda,Bousso:2014uxa}, using techniques involving the modular Hamiltonian.

\subsubsection{Static patch holography: Susskind's suggestion}

\begin{figure}[t]
\centering
\includegraphics[width=5in]{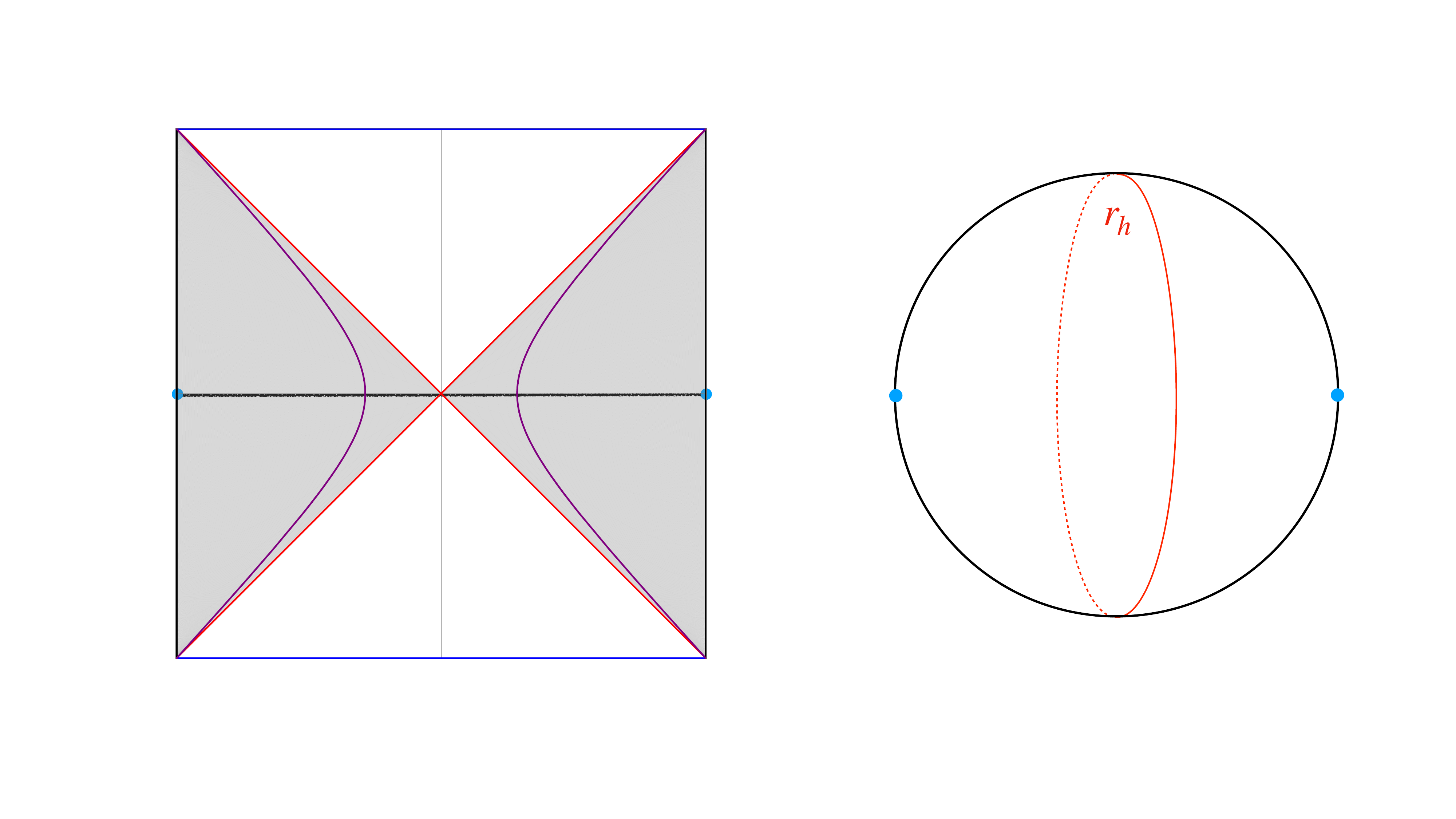}\caption{Left: Penrose diagram of de Sitter space. The gray shaded region represents the left/right static patch parametrized by the metric eq.~\eqref{eq:dSstaticpatch}. Right: The spatial geometry of a constant time slice is a sphere where the stretched horizon is located at $r=r_\ast$.}
\label{fig:dSPenrose}
\end{figure}
Motivated by the holographic principle and Bousso’s covariant entropy bound, Susskind proposed a new framework of static patch holography \cite{Susskind:2021omt,Susskind:2021dfc,Shaghoulian:2022fop}.The static patch of de Sitter space is described by the metric
\begin{equation}\label{eq:dSstaticpatch}
ds^2 = - \left(1 - \frac{r^2}{L^2}\right) dt^2 + \frac{dr^2}{1 - \frac{r^2}{L^2}} + r^2 d\Omega^2_{d-2},,
\end{equation}
where $L$ is the de Sitter curvature radius. These coordinates cover only one quarter of the global de Sitter spacetime, as illustrated in figure \ref{fig:dSPenrose}. The point $r=0$ corresponds to the north pole, while the region $0<r<L$ describes the static patch accessible to an observer at the pole. The surface at $r=L$ is the cosmological horizon, beyond which ($r>L$) an observer at either pole cannot receive or send signals. Unlike the eternal AdS black hole, de Sitter space is compact and does not possess a spatially asymptotic boundary. It is known that the (connected) isometry group of global de Sitter $\mathrm{dS}_{d+1}$ space is $\mathrm{SO}(1,d+1)$. 
However, the presence of the cosmological horizon reduces the symmetry preserved by the static patch to the subgroup $\mathrm{SO}(d) \times \mathbb{R}$, generated by spatial rotations and the time translations associated with the Killing vector $\xi^\mu = \partial_t$.\footnote{Introducing a spatial boundary at a stretched horizon $r=r_\ast$ within the static patch does not alter this reduced symmetry. The preserved isometry remains $\mathrm{SO}(d) \times \mathbb{R}$.}

Bousso’s entropy bound suggests that the holographic screen should be chosen so that its area is sufficient to encode all information contained within the spacetime region of interest. To illustrate this, consider first the AdS case. If the holographic screen were placed near a black hole horizon, the boundary degrees of freedom would capture only the physics behind the horizon, while missing the information associated with the asymptotic region. This motivates the standard AdS/CFT where the holographic screen near the asymptotic boundary and the full set of bulk degrees of freedom can be encoded. Turning to the de Sitter case, if the holographic screen is placed near the north or south pole, it would capture only the degrees of freedom in the small region between the pole and the screen. In contrast, choosing the screen at the cosmological horizon ensures that all degrees of freedom within the entire static patch are accessible. Thus, the cosmological horizon naturally emerges as the appropriate location for the holographic screen in de Sitter space. Based on these considerations, Susskind proposed the framework of static patch holography, which may be summarized as follows:
\begin{quote}
\emph{Quantum gravity in the static patch is dual to a quantum mechanical system defined on the cosmological horizon.}
\end{quote}
To regulate the physics near the horizon which is a null surface, it is natural to define the boundary theory on the stretched horizon located at $r = L - \epsilon < L$.

\subsection{Half de Sitter holography}

As an alternative to introducing a timelike boundary at the cosmological horizon, the framework of half de Sitter holography was proposed in \cite{Kawamoto:2023nki}. In this approach, a timelike boundary is introduced directly in global de Sitter space, where Dirichlet boundary conditions are imposed in order to obtain a non-gravitating boundary theory, in close analogy with the AdS/CFT correspondence.

The central idea is the introduction of a finite cutoff, echoing earlier developments in the AdS/CFT context through the $T\bar T$ deformation \cite{McGough:2016lol}. In the de Sitter setting, this cutoff effectively restricts the spacetime to a portion of global de Sitter, bounded by the chosen timelike hypersurface. The resulting construction provides a tractable holographic setup, while highlighting the similarities and crucial differences between finite-cutoff AdS/CFT and potential holographic descriptions of de Sitter space.

\subsection{RT formula in de Sitter space}

It is natural to ask whether the Ryu–Takayanagi (RT) prescription extends to de Sitter space, since the original proof of the prescription does not explicitly rely on the choice of background geometry \cite{Lewkowycz:2013nqa}. However, serious obstacles arise in attempting such a generalization. For instance, in the dS$_3$/CFT$_2$ correspondence, there are no bulk spacelike geodesics connecting the endpoints of a subsystem located at future infinity. One possible solution is to prepare a Hartle–Hawking state \cite{Hartle:1983ai} and consider complex geodesics that extend into the Euclidean continuation of the geometry \cite{Hikida:2022ltr}. In general, however, this procedure yields complex-valued holographic entanglement entropies.

A further difficulty is that the holographic entanglement entropy computed via the RT prescription in de Sitter space violates (strong) subadditivity. Within the RT framework, this issue raises the question of which extremization procedure should be applied. In particular, it has been argued that the appropriate prescription is the min–max procedure, rather than the max–min one, as discussed in \cite{Shaghoulian:2021cef,Shaghoulian:2022fop,Franken:2023pni,Franken:2023jas}. The min–max procedure entails maximizing along spatial directions and minimizing along the time direction. In the AdS/CFT context, the max–min prescription \cite{Wall:2012uf} is known to be equivalent to the Hubeny–Rangamani–Takayanagi (HRT) prescription \cite{Hubeny:2007xt}. In de Sitter space, however, spatial slices are compact, and minimization along spatial directions trivially yields zero, effectively selecting the degenerate contribution from the poles of the sphere. To obtain a non-trivial result, it is natural to expect that entanglement entropy must instead receive contributions from the cosmological horizon (see, \eg \cite{Balasubramanian:2002zh,Cotler:2023xku}). Taken together, these observations suggest that the RT prescription, at least in its straightforward form, is not well suited for de Sitter holography.

%%%%%%%%%%%%%%%%%%%%%%%%%%%%%%%%%%%%%%%%%%
%%%%%%%%%%%%%% new section %%%%%%%%%%%%%%%
%%%%%%%%%%%%%%%%%%%%%%%%%%%%%%%%%%%%%%%%%%
\section{Review on the replica trick and Green’s function}\label{sec:reviewsection}

In this section, we review the derivation of entanglement entropy via the replica method in QFTs and introduce the relevant Green’s functions in de Sitter space. These ingredients will be essential for constructing a holographic derivation of entanglement entropy in the static patch holography discussed later.

\subsection{Entanglement entropy in field theory and the replica trick}\label{sec:replica}
Throughout this work, we adopt the standard definition of the entanglement entropy of a quantum field theory as well as its holographic dual as the von Neumann entropy of a reduced density matrix. Before turning to the de Sitter case, it is useful to recall the replica trick, a powerful method for computing entanglement entropy in QFTs, originally developed by Calabrese and Cardy in the context of two-dimensional CFTs \cite{Calabrese:2004eu,Calabrese:2009qy}.

For a mixed quantum state described by a normalized density matrix $\hat{\rho}$, the von Neumann entropy is defined as
\begin{equation}
   S_{\mathrm{vN}} (\hat{\rho})= -\text{Tr} \, \left(  \hat{\rho}  \log \hat{\rho}  \right)\,. 
\end{equation}
which automatically satisfies a number of fundamental inequalities such as subadditivity, the Araki–Lieb inequality, and strong subadditivity.

In a QFT, consider a pure state $|\Psi\rangle$ defined on a time slice, and divide the system into two subsystems $A$ and $B$, such that $|\Psi_{AB}\rangle = |\Psi\rangle$. The reduced density matrix of subsystem $A$ is obtained by tracing out the degrees of freedom in $B$:
\begin{equation}
\hat{\rho}_A = \text{Tr}_B \, | \Psi_{AB} \rangle | \langle \Psi_{AB} | \,.
\end{equation}
The entanglement entropy of $A$ is then given by the von Neumann entropy of $\hat{\rho}A$:
\begin{equation}
    S_{\mt{EE}} (\hat{\rho}_A) := -\text{Tr}_A \, \left(\hat{\rho}_A \log \hat{\rho}_A \right) \,.
\end{equation}
Direct calculation of the entanglement entropy for arbitrary subsystems in a QFT can be challenging due to the presence of infinite degrees of freedom. An elegant approach to circumvent this challenge is the \textit{replica trick}, introduced by Calabrese and Cardy \cite{Calabrese:2004eu, Calabrese:2009qy} for 1+1 dimensional CFTs. This method could also be applied to general QFTs in any dimension, though the analytic calculations can be challenging. It involves constructing an $n$-sheeted Riemann surface by replicating the system $n$ times and joining adjacent sheets along the entangling region. For integer $n$, this construction enables us to compute  the so-called \textit{R\'enyi entropies}, \ie
\begin{equation}
    S_A^{(n)} = \frac{1}{1 - n} \log \text{Tr} \left(\hat{\rho}_A^n \right).
\end{equation}
In the limit $n \to 1$, the R\'enyi entropy converges to the entanglement entropy\footnote{For a non-normalized density matrix $\tilde{\rho}_A$, it is also convenient to define the modular entropy $\tilde{S}_A^{(n)}$ and take the following limit
\begin{equation}
  \lim_{n \to 1} \tilde{S}_A^{(n)}  =\lim _{n \rightarrow 1}  \left( 1 - n \partial_n  \right) \log \mathrm{Tr} \,\tilde{\rho}_A^n    = S_{\mt{EE}} (\hat{\rho}_A)   
\end{equation}
to get the entanglement entropy.}:
\begin{equation*}
     \lim_{n \to 1} S_A^{(n)}  =-\lim _{n \rightarrow 1} \frac{\partial}{\partial n} \mathrm{Tr} \,\rho_A^n    = S_{\mt{EE}} (\hat{\rho}_A)   \,.
\end{equation*}

To perform the analytic continuation in $n$, the path integral representation of the density matrix provides a natural construction. Calabrese and Cardy's replica trick provides an elegant solution by transforming the non-analytic logarithmic computation into a differentiable analytical procedure. The key insight lies in constructing the $n$ factors of $\hat{\rho}_A$ by the path integral on the $n$-sheeted Riemann surface
(replica manifold $\mathcal{M}_n$) with a cut and the analytic continuation of $n$ to non-integer values. 
For a normalized reduced density matrix satisfying $ \mathrm{Tr} \hat{\rho}_A =1$, the trace can then be expressed as the ratio of partition functions:
\begin{equation}
   \text{Tr} \,  \hat{\rho}_A^n  = \frac{Z\left[\mathcal{M}_{n,A}\right]}{(Z[\mathcal{M}_A])^n} \equiv \frac{Z_n}{(Z_1)^n}  \,,
\end{equation}
where $\mathcal{M}_{n,A}$ denotes the $n$-sheeted Riemann surface with a cut located at the subsystem $A$. 

In practice, computing the partition function on a $n$-sheeted Riemann surface $\mathcal{M}_n$ directly is challenging.
Different from the worldsheet perspective, the path integral for the partition function $Z\left[\mathcal{M}_{n,A}\right]$ can be equivalently interpreted as the path integral for the Lagrangian density of the multi-copy model over the orbifold manifold in the presence of the twist fields, which is the target space perspective. As a result, the calculation of R\'enyi entropies can be reformulated in the context of two-dimensional field theories by employing \textit{twist operators}, which are local operators inserted at the endpoints of the entangling region on each sheet. These twist operators, denoted $\sigma_n$ and $\bar{\sigma}_n$, encode the branch cut structure of the replicated sheets and effectively introduce a cyclic boundary condition around the entangling region. In terms of twist operators, the R\'enyi entropy reduces to a two-point function:
\begin{equation}\label{eq:Renyi}
    S_A^{(n)} = \frac{1}{1 - n} \ln \langle \sigma_n (x_1) \bar{\sigma}_n (x_2) \rangle \,,
\end{equation}
where $x_1$ and $x_2$ are the endpoints of the entangling interval $A$. Although first developed in two-dimensional CFTs, this operator-based formulation of the replica trick can be extended to more general QFTs, and will form the basis for our construction of holographic entanglement entropy in de Sitter space.

\subsubsection{CFT$_2$}
As a concrete example, let us consider a two-dimensional CFT. In this case, conformal symmetry completely determines the two-point function of twist operators up to an overall normalization. On the complex plane, the universal form is
\begin{equation}
    \langle \sigma_n (x_1) \bar{\sigma}_n (x_2) \rangle = \frac{c_n}{|x_1 - x_2|^{2\Delta_n}} \,, 
\end{equation}
where $\Delta_n$ denotes the conformal dimension of the twist operator $\sigma_n$, given by
\begin{equation}
    \Delta_n = \frac{c}{12} \left( n - \frac{1}{n} \right),
\end{equation}
with $c$ the central charge of the CFT. Because conformal symmetry in two dimensions strongly constrains the form of correlation functions, the R\'enyi  entropies \eqref{eq:Renyi} and hence the entanglement entropy can be computed explicitly. Substituting the above correlator into eq.~\eqref{eq:Renyi} and taking the replica limit yields the entanglement entropy,
\begin{equation}
   S_{\mt{EE}} (\hat{\rho}_A) \equiv    \lim_{n \to 1}  S_A^{(n)} = \frac{c}{3} \ln \frac{|x_1 - x_2|}{\epsilon} \,, 
\end{equation}
where $\epsilon$ is a UV cutoff that regularizes the short-distance divergence.

The replica trick methods developed by Calabrese and Cardy thus provide a powerful toolkit for computing entanglement entropy in two-dimensional CFTs and have shed light on the entanglement structure of quantum field theories more generally. More interestingly, these techniques have been extended to holographic settings, where entanglement entropy is captured by minimal surfaces in AdS spacetimes through the Ryu-Takayanagi formula \cite{Ryu:2006bv,Ryu:2006ef}, establishing a profound connection between quantum entanglement and spacetime geometry.

\subsubsection{Generic field theories}
In more general quantum field theories that do not possess conformal symmetry, the computation of R\'enyi entropies requires evaluating the partition function on an $n$-sheeted Riemann surface. This task is typically formidable and often analytically intractable. Formally, it has been suggested that the twist operator formalism can still be applied to general QFTs provided the theory admits a global replica symmetry \cite{Calabrese:2009qy}. However, in such cases the twist operator approach \eqref{eq:Renyi} becomes much more difficult to implement. Without the strong constraints of conformal symmetry, the functional form of twist operator correlation functions is no longer fixed, and explicit calculations generally require further assumptions or approximations.

Holographic duality provides an alternative route for addressing this challenge, as the holographic dictionary offers another way to compute the relevant correlation functions from bulk gravity side. In this paper, we adopt this perspective and focus on calculating entanglement entropy via the replica trick expressed in terms of twist operators, 
\begin{equation}\label{eq:twist}
 S_{\mt{EE}} (\hat{\rho}_A) = \lim_{n \to 1}   \left( \frac{1}{1 - n} \ln \langle \sigma_n (x_1) \bar{\sigma}_n (x_2) \rangle \right) \,. 
\end{equation}
In the following, we will examine the two-point function
\begin{equation}
\langle \mathcal{O}(x_1)\mathcal{O}(x_2) \rangle_{\rm{bdy}}  \,,
\end{equation}
of boundary operators, which we assume to be dual to two-point functions in the de Sitter bulk spacetime. This will serve as the cornerstone for constructing a holographic definition of entanglement entropy in the static patch.

\subsection{Holographic dictionary of correlation functions}\label{sec:dictionary} 
Before addressing the challenges of de Sitter holography, it is instructive to review the well-established AdS/CFT correspondence, which provides the prototypical framework for holographic duality. The lessons drawn from AdS/CFT serve as a useful guide in formulating a expected holographic dictionary for de Sitter space.

The AdS/CFT correspondence \cite{Maldacena:1997re,Gubser:1998bc,Witten:1998qj} posits a duality between a $d$-dimensional CFT and a $(d+1)$-dimensional gravitational theory in AdS. A central formulation of this duality is the GKPW prescription \cite{Gubser:1998bc,Witten:1998qj}, which states the equivalence between the AdS bulk partition function and the generating functional of CFT correlators:
\begin{equation}\label{eq:gkpw}
Z_{\mt{AdS}}[\phi_0] = \left\langle e^{\int_{\partial \mt{AdS}} \phi_0(x) \, \mathcal{O}(x)} \right\rangle_{\mt{CFT}} \,,
\end{equation}
where $\phi_0(x)$ is the boundary value of a bulk field $\phi(x,z)$, and $\mathcal{O}(x)$ is the dual boundary operator. Functional differentiation with respect to $\phi_0(x)$ yields the connected correlation functions of $\mathcal{O}(x)$:
\begin{equation}
	\langle \mathcal{O}(x_1) \mathcal{O}(x_2) \cdots \mathcal{O}(x_n) \rangle_{\mt{CFT},c} = \frac{\delta^n \ln Z_{\text{AdS}}[\phi_0]}{\delta \phi_0(x_1) \delta \phi_0(x_2) \cdots \delta \phi_0(x_n)} \bigg|_{\phi_0=0} \,.
\end{equation}

An alternative formulation of the AdS/CFT dictionary, known as the extrapolate dictionary \cite{Banks:1998dd, Balasubramanian:1998sn, Balasubramanian:1998de}, identifies AdS bulk correlators near the boundary with the corresponding CFT correlators. To wit, 
\begin{equation}
\lim _{r \rightarrow \infty} r^{n \Delta}\left\langle\phi\left(r, x_1\right) \phi\left(r, x_2\right) \ldots . . \phi\left(r, x_n\right)\right\rangle_{\mt{AdS }} = \langle \mathcal{O}(x_1) \mathcal{O}(x_2) \cdots \mathcal{O}(x_n) \rangle_{\mt{CFT}} \,,
\end{equation}
where $r$ is the AdS radial coordinate in the global AdS with a metric $ds^2=-(1+r^2)dt^2+\frac{dr^2}{1+r^2}+ r^2d\Omega^2$. The equivalence between the extrapolate dictionary and the GKPW prescription was established in \cite{Harlow:2011ke}. For example, in the case of the two-point function, the CFT correlator is reproduced by the boundary-to-boundary propagator of a bulk scalar field. This follows from the near-boundary expansion
\begin{equation}
\phi(x, z \sim 0)=\phi_0(x) z^{d-\Delta}+\tilde{\phi}(x) z^{\Delta} \,,
\end{equation}
with $z$ represents the radial direction in Poincar\'e coordinates. In this sense, the regulated bulk propagator reduces to the two-point function of the dual boundary operator, \ie 
\begin{equation}
   G_{\partial \partial}(x_1, x_2)=\left\langle\phi\left(x_1, r_1 \rightarrow \infty \right) \phi\left(x_2, r_2 \rightarrow \infty \right)\right\rangle_{\mt{AdS}} \sim\left\langle \mathcal{O}_{\Delta}\left(x_1\right) \mathcal{O}_{\Delta}\left(x_2\right)\right\rangle_{\mt{CFT}} \,, 
\end{equation}
up to a normalization factor.

In contrast to the AdS/CFT correspondence, which benefits from a well-established and extensively tested holographic dictionary, a universally accepted holographic dictionary for static patch holography in de Sitter space remains lacking. Nonetheless, it is natural to expect that the guiding principles of AdS/CFT should continue to play a role. One plausible conjecture is the existence of a GKPW-like relation, namely 
\begin{equation}
Z_{\mathcal{O}}\left[\Phi_0\right]_{\mathrm{bdy}}=Z_\Phi\left[\Phi_0\right]_{\rm{dS} } \,.
\end{equation} 
where $Z_{\mathcal{O}}$ denotes the boundary partition function with source $\Phi_0$, and $Z_\Phi$ the corresponding bulk partition function in de Sitter spacetime. Here, the boundary theory is assumed to reside on the stretched horizon. While such an equivalence would provide a straightforward dictionary between bulk and boundary theories, we will not assume this holographic dictionary in the present work due to the absence of a rigorous derivation.

Instead, we adopt a more practical approach, inspired by the extrapolate dictionary: we posit that boundary correlators in static patch holography can be extracted directly from bulk correlators in de Sitter space. 
This is particularly suited to our primary goal of investigating holographic entanglement entropy in the context of static patch holography.
Specifically, we propose the following working assumption: 
\begin{equation}\label{eq:dSdictionary}
\boxed{\text{Conjectured Holographic Dictionary:} \, \langle \mathcal{O}(x_1)\mathcal{O}(x_2) \rangle_{\rm{bdy}} \sim  \lim_{r \to r_\ast} \langle \Phi(X_1)\Phi(X_2) \rangle_{\rm{dS}} \,. }
\end{equation}
where $r_\ast$ denotes the location of the stretched horizon in the static patch, and the symbol $\sim$ indicates equality up to a normalization constant. The relative coefficient between the boundary and bulk correlator would be fixed by imposing a normalization constant. In the later section, we fix the normalization by requiring that the entanglement entropy vanishes at the cut-off scale. This bulk-to-boundary relation provides a minimal framework for static patch holography and furnishes the necessary input for our construction of holographic entanglement entropy.

%%%%%%%%%%%%%%%%%%%%%%%%%%%%%%
%%%%%%%%%%%%%%%%%%%%%%%%%%%%%
\subsection{Green's functions in de Sitter space}\label{sec:green}
%%%%%%%%%%%%%%%%%%%%%%%%%%%%%%
%%%%%%%%%%%%%%%%%%%%%%%%%%%%%
We now review the construction of Green’s functions in de Sitter space, following \cite{Strominger:2001pn,Bousso:2001mw,Spradlin:2001pw}.Since our primary interest lies in twist operators on the boundary, we focus on a free scalar field $\Phi$ of mass $m$ propagating in dS$_{d+1}$, with action
\begin{equation}
S=-\frac{1}{2}\int d^{d+1}x\sqrt{-g}\left(  (\nabla\Phi)^2+m^2\Phi^2\right) \,, 
\end{equation}
where $g$ is the determinant of the dS background metric. We are interested in the Wightman function (or Green's function) of the free massive scalar\footnote{Of course, all physical information in this case has been encoded in the two-point function since it is only a free field theory.}, which is defined by
\begin{equation}
G_{\Omega}\left(x, x^{\prime}\right)=\left\langle\Omega\left|\Phi(x) \Phi\left(x^{\prime}\right)\right| \Omega\right\rangle=\sum_n \Phi_n(x) \Phi_n^*\left(x^{\prime}\right) \,,
\end{equation}
where $|\Omega\rangle$ denotes the vacuum state, and the eigenmodes $\Phi_n(x)$ solve the Klein–Gordon equation in dS.  If one demands invariance under the de Sitter isometry group $\mathrm{SO}(d+1,1)$, the vacuum state must be dS-invariant. However, such vacua are not unique: there exists a one-parameter family, the so-called $\alpha$-vacua $|\alpha\rangle$. Correspondingly, the Green's functions in dS space characterize the various dS invariant vacua.

Due to the high degree of symmetry, the Green’s function depends only on the invariant geodesic length between two points $x,x'$. To make this explicit, we embed dS$_{d+1}$ into $(d+2)$-dimensional Minkowski spacetime with metric
\begin{equation}
ds^2=-dX_0^2+dX_1^2+\cdots+dX_{d+1}^2 \,.
\end{equation}
The dS$_{d+1}$ space is represented as a hyperboloid constrained by 
\begin{equation}
-X_0^2 + X_1^2 + \cdots + X_{d+1}^2 = L^2 \,,
\end{equation} 
where $L$ denotes the de Sitter radius. Using this embedding formalism, we can parametrize the geodesic distance by a de Sitter invariant length, \ie
\begin{equation}\label{eq:geodesic_distance} 
 \mathcal{P}(x,x') \equiv  \frac{\eta_{AB} X^A X'^B}{L^2}= \frac{1}{L^2}(-X_0\cdot X_0'+X_1\cdot X'_1+\cdots +X_{d+1}\cdot X'_{d+1}) \,.
\end{equation} 
The corresponding (dimensionless) geodesic length $\mD(x,x')$ is given by\footnote{If $\cos\frac{D}{L}>1$, it corresponds to a timelike geodesic, admitting an imaginary geodesic distance. For $\cos\frac{D}{L}<1$ and $\cos\frac{D}{L}=1$, it is a spacelike and null geodesic, respectively.}  
\begin{equation}
\mP(x, x') = \cos \left( \mD(x,x') \right)  \,.
\end{equation}
The Green’s function $G_{\rm dS}(x,x')$ satisfies the Klein–Gordon equation
\begin{equation}\label{eq:laplace} 
(\nabla^2-m^2)G_{\rm{dS}}(x,x')= 0 \,. 
\end{equation}
If one assumes that the vacuum state and its correlation function are de Sitter invariant ($\mathrm{SO}(1,d+1)$ de Sitter symmetry group), one can substitute the general ansatz $G_{\rm{dS}}(\mP(x,x'))$ into the Laplace equation \eqref{eq:laplace} and get a much simpler ordinary differential equation, \ie  
\begin{equation}\label{eq:Green}
(1-\mP^2)\frac{d^2G_{\rm{dS}}}{d\mP^2}-(d+1)\mP\frac{dG_{\rm{dS}}}{d\mP}-m^2L^2 G_{\rm{dS}}=0 \,.
\end{equation}
Details of this derivation are given in Appendix \ref{app:Laplacian}. A caveat must be emphasized: unlike in the dS/CFT correspondence, the dual theory on the stretched horizon in static patch holography \emph{need not} preserve the full $\mathrm{SO}(d+1,1)$ symmetry. Such a  symmetry breaking is in fact a generic feature of any de Sitter holography with a timelike boundary.  From the bulk perspective, the most general Green’s function may depend on covariant distances modified by boundary effects. In the following, we continue to assume the ansatz $G_{\rm{dS}}(\mP(x,x'))$,\footnote{If the contribution of the stretched horizon is included, the geodesic distance is generically modified into a covariant function $f(x,x';\alpha)$, and the Green’s function is then obtained by replacing the geodesic distance with $f(x,x';\alpha)$, where $x,x'$ are two points and the $\alpha$ is the position of the stretched horizon.} Although this assumption only selects a subclass of all possible physical solutions, it captures the most interesting cases that admit a geometric interpretation, namely the geodesics in the de Sitter bulk in analogy with the Ryu–Takayanagi formula.

A particularly important state is the Euclidean vacuum $|\mathrm{E}\rangle$, also known as the Bunch–Davies or Hartle–Hawking vacuum \footnote{If we trace the Euclidean vacuum over the Hilbert space of the northern modes, we can obtain a thermal density matrix in the southern diamond \cite{Bousso:2001mw}.}, obtained by analytic continuation from the Euclidean sphere. Its Green’s function is the solution of eq.~\eqref{eq:Green} and given by
\begin{equation}\label{eq:GBD}
G_{\mt{G}}(x,x')=c_{m,d}\,\cdot {}_2F_1\left[h_+,h_-,\frac{d+1}{2},\frac{1+\mP}{2}\right] \,,
\end{equation}
with 
\begin{equation}
h_{\pm}=\frac{1}{2}\left(d\pm\sqrt{d^2-4m^2L^2}\right) \,. 
\end{equation}
This hypergeometric function contains a singularity located at $\mD=0$ and also a branch cut for $1<\cos \mD <\infty$. This singularity corresponds to the coincident limit of the two points in the Green's function. The standard way to fix the coefficient is to require that its short distance behavior matches with that of flat space. Note that the time-ordered Green's function (Feynman propagator) for a massive scalar field in 
$(d+1)$-dimensional Minkowski spacetime is given by
\begin{equation}
G_{\rm Flat}^{(d+1)}(X)=\frac{m^{\frac{d-1}{2}}}{(2 \pi)^{\frac{d+1}{2}}} \frac{K_{\frac{d-1}{2}}\left(m \sqrt{-X^2+i \epsilon}\right)}{\left(\sqrt{-X^2+i \epsilon}\right)^{\frac{d-1}{2}}} \,.
\end{equation}
The short distance divergence is thus described by 
\begin{equation}
\lim_{X \to 0}  G_{\rm Flat}^{(d+1)}(X)
\sim  \frac{1}{4 \pi }\left(\frac{1}{\pi  \mD^2}\right)^{\frac{d-1}{2}} \Gamma \left(\frac{d-1}{2}\right)\,. \qquad   \mD \sim |X|
\end{equation}
In order to match the short-distance divergence:
\begin{equation}
\lim_{X \to 0}  G_{\rm Flat}^{(d+1)}(X) = \lim_{\mD \to 0}  G_{\mt{G}}^{(d+1)}(\mD) \,.
\end{equation}
one can fix the coefficient as \cite{Strominger:2001pn}
\begin{equation}
c_{m,d}=\frac{\Gamma(h_+)\Gamma(h_-)}{(4\pi)^{\frac{d+1}{2}}\Gamma\left(\frac{d+1}{2}\right)} \,.
\end{equation}
by using the following expansion of hypergeometric function: 
\begin{equation}
 \lim_{\mD \sim 0} \, {}_2F_1\left[h_+,h_-,\frac{d+1}{2},\frac{1+P}{2}\right]  \approx \left(\frac{\mD^2}{4}\right)^{\frac{1-d}{2}}
 \frac{\Gamma\left(\frac{d+1}{2}\right) \Gamma\left(\frac{d-1}{2}\right)}{\Gamma\left(h_{+}\right) \Gamma\left(h_{-}\right)} \,. 
\end{equation}
At short distances, the Green's function $G_{\mt{BD}}$ does not capture the curvature of de Sitter space and approaches the propagator in the Minkowski space. 

The Laplace equation for the Green's function \eqref{eq:Green}) admits a second independent solution, obtained by $\mP\to -\mP$:  
\begin{equation}
G_{\mt{A}}(x,x')=c_{m,d}\,\cdot {}_2F_1\left[h_+,h_-,\frac{d+1}{2},\frac{1-\mP}{2}\right]  = G_{\mt{G}}(x_{\mt{A}}, x')\,. 
\end{equation}
where $x_A$ is the antipodal point of $x$. This solution is regular at coincident points with $\mD(x,x')=0$ (\ie $\mP(x,x')=1$). However, we denote this as the antipodal branch since there is a divergence when two points are placed at antipodal points. 

Different from the Bunch-Davies vacuum, one can also perform the Bogoliubov transform to obtain the so-called $\alpha$-vacuum which is labeled by a complex parameter $\alpha$. These infinite states are also invariant under full $\mathrm{SO}(1,d+1)$ de Sitter symmetry group. The corresponding correlation functions are given by 
\begin{equation}
G_\alpha(x,x')=\frac{1}{1-e^{\alpha+\alpha*}}(G_{\mt{G}}(x,x')+e^{\alpha} G_{\mt{G}}(x_A,x')+e^{\alpha*} G_{\mt{G}}(x,x'_A)+e^{\alpha+\alpha*}G_{\mt{G}}(x_A,x'_A)) \,. 
\end{equation}
For real $\alpha$ with $\alpha\leq 0$, assuming CPT invariance, one finds
\begin{equation}
G_\alpha(x,x')=\frac{1}{1-e^{2\alpha}}\left( \left(1+e^{2\alpha}\right)G_{\mt{G}}(x,x')+2e^{\alpha} G_{\mt{A}}(x,x') \right)  \,,
\end{equation}
with the Euclidean vacuum recovered in the limit $\alpha\to -\infty$.

The most general de Sitter Green’s function is a linear combination of the two branches, namely 
\begin{equation}\label{eq:linearGA}
\boxed{G_{\rm dS}(x,x')= N_{\mt{G}}  \cdot G_{\mt{G}}(x,x')+ N_{\mt{A}} \cdot G_{\mt A}(x,x')\,,}
\end{equation}
with arbitrary constants $N_{\mt{G}}, N_{\mt{A}}$. The Bunch–Davies and antipodal vacua correspond to the limits $N_{\mt{A}}\to 0$ and $N_{\mt{G}}\to 0$, respectively. In static patch holography, such combinations naturally arise from imposing boundary conditions (Dirichlet, Neumann, or Robin) for bulk fields at the stretched horizon.

%%%%%%%%%%%%%%%%%%%%%%%%%%%%%%
%%%%%%%%%%%%%%%%%%%%%%%%%%%%%%
\section{Constraints of holographic entanglement entropy in dS space}\label{eq:defineEEindS}
%%%%%%%%%%%%%%%%%%%%
In the AdS/CFT correspondence, the Ryu-Takayanagi prescription provides a geometric realization of entanglement entropy in holographic theories. From the field theory side, the same results can also be derived from the twist operator formalism for CFTs. However, as emphasized earlier, the naive generalization of the Ryu-Takayanagi formula to de Sitter space fails to yield a consistent definition of entanglement entropy. In particular, it does not satisfy basic entropic inequalities. To address this problem, and in the absence of a well-established dS holographic dictionary, we instead turn to the replica trick and twist operator formalism for defining the holographic entanglement entropy in de Sitter space. Expressed in terms of bulk Green’s functions, this derivation provides a candidate for holographic entanglement entropy in dS spacetime, which is different from the Ryu-Takayanagi formula.

\subsection{Holographic entanglement entropy in the static patch}

Our working assumption is that  holographic dual field theory defined on the timelike boundary of de Sitter space has a non-negative norm between inner products and density matrix.  For a spatial subsystem $A$ on the holographic screen, one may insert two twist operators at the endpoints of $A$ and apply the replica trick, following the discussion in the previous section. This leads us to propose that the entanglement entropy of $A$ is captured by the logarithm of the two-point function of twist operators, namely 
\begin{equation}\label{eq:defineentropy}
\boxed{ S_{\mt{EE}} =-C\cdot \log G_{\rm dS}(x,x') \,. }
\end{equation}
where $x$ and $x'$ denote the endpoints of $A$, and $C$ is a dimensionless constant depending on the replica limit $n \to 1$. In the absence of explicit knowledge of twist operators in static patch holography, we treat $C$ as an undetermined positive prefactor. The minus sign reflects the standard replica limit from $n > 1$ to $n =1$. It is natural to assume that the twist operator corresponds to a scalar operator. As a result, its two-point function is given by the Green’s function of a scalar field of mass $m$ in dS. Importantly, we do not attempt to identify this mass with a definite operator dimension in the putative boundary theory, nor do we fix the normalization constant $C$. Instead, we study the most general form of eq.~\eqref{eq:defineentropy} and ask whether it can satisfy the consistency conditions of entanglement entropy. Since the boundary QFT in this setup is defined on a finite cutoff in de Sitter space, we assume that the twist operator formalism continues to apply despite potential non-locality \cite{Kawamoto:2023nki}. The overall normalization does not affect entropic inequalities, so we are free to rescale $S_{\mt{EE}}$ by a constant. Moreover, because the entropy depends only on the geodesic distance in dS, it is thus covariant by construction\footnote{Extensions to higher dimensions will be non trivial since we need surface-like information rather than curve. This is the same as the RT case.}. 

\begin{figure}[t]
\centering
\includegraphics[width=4.5in]{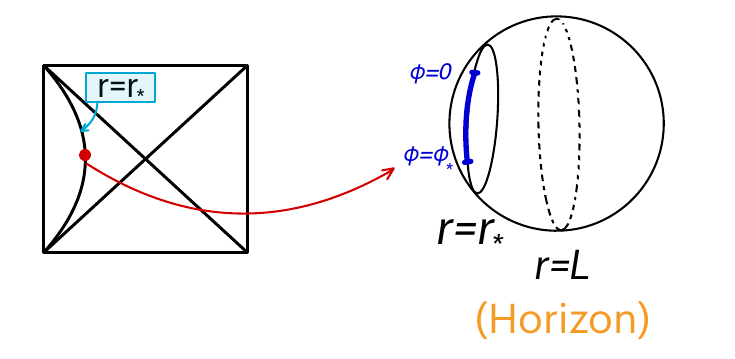}
\caption{A subsystem $A$ on the stretched horizon (located at $r=r_\ast$), represented by the blue curve on the right.}
\label{fig:timeslice}
\end{figure}

For concreteness, we focus on dS$_3$, whose static patch is described by
\begin{equation}
ds^2=-\left(1-\frac{r^2}{L^2}\right)dt^2+\frac{dr^2}{1-\frac{r^2}{L^2}}+r^2d\phi^2 \,, 
\end{equation}
with coordinates
\begin{equation}
0\leq r\leq L,\quad  -\infty<t<\infty,\quad 0\leq\phi\leq 2\pi \,. 
\end{equation}
The explicit embedding into the Minkowski space is given by
\begin{align}
    X_0&=\sqrt{L^2-r^2}\sinh\left(\frac{t}{L}\right),\quad X_2=r\cos\phi,\nn\\
     X_1&=\sqrt{L^2-r^2}\cosh\left(\frac{t}{L}\right),\quad
     X_3=r\sin\phi,
\end{align}
with
\begin{equation}
-X_0^2+X_1^2+X_2^2+X_3^2=L^2,\quad ds^2=-dX_0^2+dX_1^2+dX_2^2+dX_3^2 \,.
\end{equation}
It is thus straightforward to derive the geodesic distance $\mD$ of two points in the same static patch, \ie 
\begin{equation}
\cos \left( \mD(x,x')\right)=\sqrt{\left(1-\frac{r^2}{L^2}\right)\left(1-\frac{r'^2}{L^2}\right)}\cosh\left(\frac{t-t'}{L}\right)+\frac{rr'}{L^2}\cos(\phi-\phi') \,. 
\end{equation}
In dS$_3$, the two independent branches of the scalar Green’s function are
\begin{equation}\label{eq:defineGdS}
G_{\mt{G}}(x,x')=  \frac{\sinh\mu\left(\pi- \mD \right)}{4\pi \sinh (\mu\pi)\sin \mD}  \,, \qquad G_{\mt{A}}(x,x') = \frac{\sinh(\mu \mD)}{4\pi \sinh (\mu\pi) \sin \mD}   \,, 
\end{equation}
where the divergence appears at two identical points with $\mD=0$ and two antipodal points with $\mD=\pi$, respectively. As a result, the generic form of the Green's function of scalar operator in dS$_3$ is given by
\begin{equation}
\boxed{G_{\rm dS}(x,x')= N_{\mt{G}}  \cdot \frac{\sinh\mu\left(\pi- \mD \right)}{4\pi \sinh\mu\pi\sin \mD }+ N_{\mt{A}} \cdot \frac{\sinh(\mu \mD)}{4\pi \sinh\mu\pi\sin \mD}  \,,}
\end{equation}
We further consider the scalar field with a large mass such that $m L \ge 1 $ and $\mu=\sqrt{m^2L^2-1}\in \mathbf{R}^+$. For the light operator with a mass $m L \leq 1$, we can perform the analytic continuation by taking $\mu=i\nu$ and will discuss it later.

While the cosmological horizon at $r=L$ is the most natural choice for the holographic screen, we regularize this null surface by considering the stretched horizon which is located inside the static patch, \eg
\begin{equation}
 r = r_\ast \in (0, L) \,.  
\end{equation}
For later use, we introduce a dimensionless parameter $\alpha$ by 
\begin{equation}
  \sin \alpha \equiv \frac{r_\ast}{L} \,, \qquad \text{with} \qquad \alpha \in \left(0, \frac{\pi}{2}\right) \,,
\end{equation}
in order to simplify many analytical expressions in the following. It is straightforward to find that the induced metric of the holographic screen is nothing but the Minkowski spacetime, \ie 
\begin{equation}\label{eq:flatmetric}
\begin{split}
ds^2 \big|_{\rm bdy} &=   - \left(  1- \frac{r_\ast^2}{L^2} \right) dt^2 + r_\ast^2  d\phi^2  = - dT^2 + dX^2 \,, 
\end{split}
\end{equation}
where the flat coordinates are defined by 
\begin{equation}
T= t \, \sqrt{1- \frac{r_\ast^2}{L^2}}    \,, \qquad X = r_\ast \phi \,. 
\end{equation}

\begin{figure}[t]
	\centering
	\includegraphics[width=4.5in]{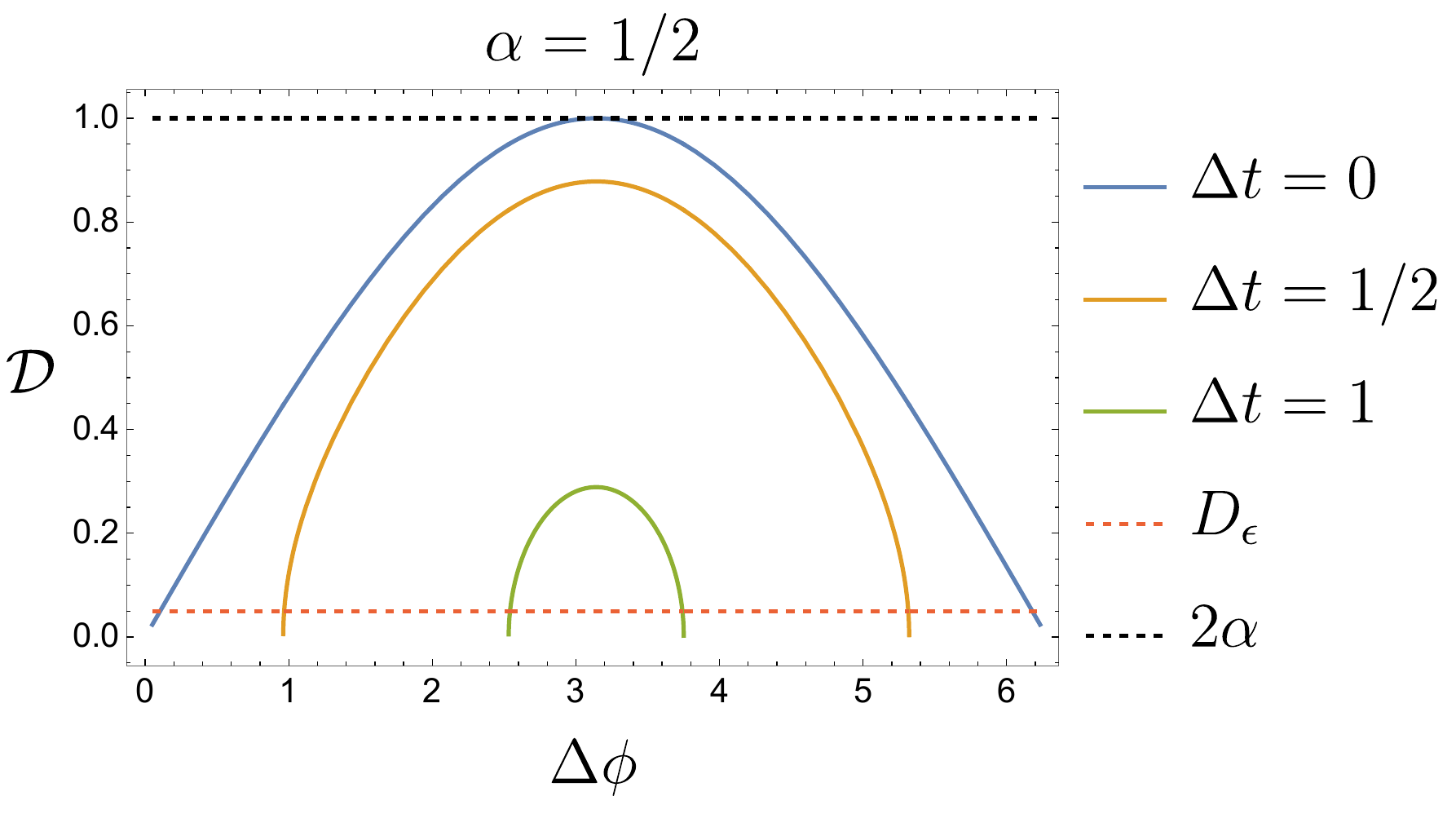}
	\caption{The dimensionless length \eqref{eq:defineDdS3} of spacelike geodesics in the static patch of dS$_3$. We fix the location of the stretched horizon at $\alpha= \arcsin (r_\ast/L)=1/2$ for this plot.}
	\label{fig:geodesiclength}
\end{figure}

Considering an arbitrary interval on the stretched, we denote the two endpoints as $x= (t_1, r_\ast, \phi_1)$ and  $x'= (t_2, r_\ast, \phi_2)$. Correspondingly, the geodesic length connecting two points on the stretched horizon is reduced to
\begin{equation}\label{eq:defineDdS3}
 \boxed{ \mD = \arccos\left(\sin^2\alpha \cos (\Delta \phi )+\cos^2\alpha \cosh (\Delta \tilde{t}) \right) } \,,
\end{equation}
with $\Delta \phi= \phi_1 - \phi_2$ and $\Delta \tilde{t}= \frac{t_1 - t_2}{L}$. See figure \ref{fig:geodesiclength} for a illustration of this function. It is interesting to notice that the bulk geodesic distance generally does not respect the boost invariance of the boundary flat metric \eqref{eq:flatmetric}, whose invariant interval reads
\begin{equation}
\Delta L= \sqrt{\Delta X^2 - \Delta T ^2} = L \sqrt{ \sin^2 \alpha \, (\Delta \phi )^2  - \cos^2\alpha \, (\Delta \tilde{t})^2} \,. 
\end{equation}
The discrepancy originates from the periodicity of the boundary circle, which breaks global Lorentz boosts. Only in the short-distance limit $\Delta\phi,\Delta t\to 0$ do the bulk and boundary lengths coincide:
\begin{equation}
D \overset{\Delta t, \Delta\phi \to0}{\approx} L \sqrt{ \sin^2 \alpha \, (\Delta \phi )^2  - \cos^2\alpha \, (\Delta \tilde{t})^2} + \mathcal{O}(\Delta\phi^3, \Delta t^3)  \approx \Delta L \,. 
\end{equation}
We also note that there are two geodesics connecting the same endpoints. We consider the one with a smaller length, \ie $\mD \ge \pi$. As a consequence, this type of spacelike geodesic whose two endpoints both lie in the static patch is free of conjugate points and therefore globally minimal among all spacelike curves anchored on the same endpoints. It is also straightforward to check that the value of the dimensionless geodesic length is constrained by
\begin{equation}
 0 \le  \mD \le 2 \alpha  \,.
\end{equation}
The lower bound with $\mD=0$ corresponds to the bulk null geodesic with  
\begin{equation}\label{eq:bulknulltime}
\Delta \tilde{t} \le  \Delta \tilde{t}\big|_{\rm bulk-null} =  \arccosh \left(  \frac{ 1- \sin^2 \alpha \cos (\Delta \phi)}{\cos^2\alpha }  \right) \,.
\end{equation}
For the interval with fixed size $\Delta \phi$, the spacelike geodesic connecting the two endpoints only exists before this critical time. On the other hand, the upper bound ($\mD=2\alpha$) is achieved at
\begin{equation}
\Delta \phi = \pi ,  \quad \Delta t=0 \,.  
\end{equation}

Combining all the results shown before and substituting this geodesic length into the bulk Green's functions \eqref{eq:defineGdS} and the generic form \eqref{eq:defineentropy}, it is straightforward to calculate the entanglement entropy of a given subsystem $A$ once the coefficient $C$ and the mass parameter $\mu$ are specified.

\subsubsection{large mass limit and RT formula}
We begin by clarifying why we focus on using the generic two-point function \eqref{eq:defineentropy} as a candidate for holographic entanglement entropy in de Sitter space, rather than adopting the standard Ryu-Takayanagi (RT) prescription. Indeed, it is straightforward to verify that eq.~\eqref{eq:defineentropy} reproduces the RT formula in the large-dimension limit $\mu\to\infty$, similar to the AdS/CFT case.\footnote{For a discussion of the geodesic approximation in de Sitter space, see \cite{Chapman:2022mqd}.} To make this connection precise, we need to restrict to the Bunch–Davies vacuum with taking
\begin{equation}
N_{\mt{A}} =  0 \,, \qquad \text{for Bunch-Davies vacuum} \,.
\end{equation}
Taking the large dimension limit of the twist operator, \ie $\mu \to \infty$,\footnote{As a reminder, the similar limit is taken in the AdS/CFT correspondence because the dimension of the twist operator $\Delta_n  \sim c \sim \frac{1}{\GN} \to \infty$ in the classical limit or large central charge limit.} the Green's function is dominated by 
\begin{equation}
 G_{\mt{dS}}(x,x') \big|_{N_{\mt{A}} \to 0, \mu \to \infty} \sim e^{-\mu \mD} \,,
\end{equation}
which is parallel to the geodesic approximation of the holographic correlation functions in the AdS/CFT correspondence \cite{Balasubramanian:1999zv,Louko:2000tp,Aparicio:2011zy,Balasubramanian:2012tu}. Assuming that the replica limit $n\to 1$ is performed with
\begin{equation}
  \lim_{n \to 1} \left( C \times  \mu  \right) = \frac{L}{4\GN}, 
\end{equation}
the definition \eqref{eq:defineentropy} yields the RT formula in de Sitter space, namely 
\begin{equation}\label{eq:dSRT}
 \boxed{ S_{A} =-C\cdot \log G_{\rm dS}(x,x')  \sim \frac{D_A}{4\GN} + \cdots }\,.
\end{equation}
Here, $D_A$ denotes the length of the bulk geodesic anchored at the endpoints of the interval $A$, playing the role of an extremal surface in dS$_3$.

However, each step in this derivation is problematic. First, the assumption that the boundary theory resides on the stretched horizon implies that the dual state cannot be identified with the global Euclidean vacuum. It is natural to expect that such a holographic screen introduces finite cutoff effects, as we will demonstrate below. Second, the shorter geodesics (with $\mD \le \pi$) in Euclidean dS$_3$ are the codimension-two minimal surface, which is similar to those in the Euclidean AdS$_3$. In Lorentzian dS they are extremal curves with one positive and one negative second-variation eigenvalue: they minimize length under spatial variations but maximize it under timelike infinitesimal variations.

To see this explicitly in Lorentzian dS$_3$, let $\gamma(s)$ be a spacelike geodesic parametrized by proper length. The crucial part is given by the second variation formula of geodesics. Considering a variation vector field $V$ that vanishes at the endpoints, the variation of the length functional at the second order (\ie index form) is derived as 
\begin{equation}
\text{Length Variation}=\delta^2 \ell(V) = \int_0^{\ell}\left(g\left(D_s V, D_s V\right)- \mathcal{R}(u, V, u, V)\right) d s \,, 
\end{equation}
where $D_s$ denotes the pull-back of the Levi-Civita connection and vector $u$ is the unit tangent vector of the original geodesic. Let us consider two types of variation generated by a spacelike normal $n_s$ as well as a timelike normal $n_t$. Since dS space is the maximally symmetric space, one get always get $\mathcal{R}(u, n, u, n) = \frac{1}{L^2} |n|^2$ for the second term with opposite sign for spacelike and timelike normal directions. To show that the sign is fixed, let us consider a purely timelike variation as an example, \eg 
\begin{equation}
V(s)= f(s)\, n_t(s) \,, \qquad \text{with} \qquad f(0)=f(\ell)=0\,.
\end{equation}
The index form reduces to 
\begin{equation}
\delta^2 \ell = \int_0^{\ell}\left[-f^{\prime}(s)^2+\frac{1}{L^2} f(s)^2\right] d s  \,. 
\end{equation}
Due to the Dirichlet boundary conditions, we can apply the Poincar\'e inequality for any $C^1$ function $f(s)$ and get 
\begin{equation}
\int_0^{\ell} f^{\prime 2} d s \geq\left(\frac{\pi}{\ell}\right)^2 \int_0^{\ell} f^2 d s \,. 
\end{equation}
Because we are interested in the shorter geodesic with $\ell \le \pi L$, the second variation of the geodesic length with respect to any timelike variation satisfies 
\begin{equation}
\delta^2 \ell  (n_t) \le \int_0^{\ell}\left[-f^{\prime}(s)^2+ \left(\frac{\pi}{\ell} \right)^2  f(s)^2\right] d s  \le 0  \,. 
\end{equation}
It indicates that the shorter spacelike geodesic in dS$_3$ is a locally maximal with respect to the timelike variation. The opposite conclusion $\delta^2 \ell  (n_s)  \ge 0$ associated with the spacelike variation simply follows. This is contrasted with the AdS$_3$ case since the maximization along the timelike variation is not trivial. It is known that the RT or HRT surface in AdS spacetime is the maximin surface. It first minimizes the area of a candidate surface on each bulk Cauchy slice, then maximizes that minimum over all slices. The special maximin property is tightly tied to the focusing theorem with the null energy condition, \ie a feature that de Sitter space lacks.

More seriously, the RT formula in dS fails to satisfy fundamental entropic inequalities. In particular, geodesic length in dS space breaks the strong subadditivity of entanglement entropy (SSA), which states
\begin{equation}
S(AB) + S(BC) - S(B) - S(ABC) \ge 0 \,. 
\end{equation}
as originally established in \cite{Lieb:1973zz,Lieb:1973cp}. In particular, subadditivity and strong subadditivity of von Neumann entropy hold for any quantum state (described by a positive semidefinite density operator) on a Hilbert space with a positive-definite inner product.
It indicates that the naive holographic entanglement entropy in eq.~\eqref{eq:dSRT} cannot serve as the holographic dual of entanglement entropy of a density operator. More generally, the violation occurs for the global Green’s function $G_{\mt{G}}(x,x')$ \eqref{eq:defineGdS} with any finite $\mu$, corresponding to the Bunch-Davies vacuum. It is tempting to argue that the violation of the SSA in the dS RT formula can be traced back to the fact that the potential dual boundary state, which lives on a finite stretched horizon, is not described by the global Euclidean vacuum. 

Taken together, these observations indicate that the RT prescription cannot serve as the consistent definition of holographic entanglement entropy in de Sitter space. Instead, they point to the necessity of including the antipodal contribution to the Green’s function. Motivated by this, we now proceed to investigate the entropy definition \eqref{eq:defineentropy} with the most general Green’s function as eq.~\eqref{eq:defineGdS}, leaving the coefficients $N_{\mt{G}},N_{\mt{A}}$ undetermined. By imposing fundamental entropic inequalities such as non-negativity, subadditivity, and strong subadditivity, we will derive constraints on the allowed parameter space for consistent definitions of holographic entanglement entropy in static patch holography.

%%%%%%%%%%%%%%%%%%%%%%%%%%%%%%%%%%%%%%%%
\subsection{Non-negativity of entanglement entropy}
%%%%%%%%%%%%%%%%%%%%%%%%%%%%%%%%%%%%%%%%
A fundamental property of entanglement entropy in quantum systems is its non-negativity. Given a quantum state described by a density matrix, the entanglement entropy of a subsystem $A$ is defined as the von Neumann entropy of its reduced density matrix $\rho_A$:
\begin{equation}
    S_{\mt{EE}} = -\text{Tr} \left( \rho_A \log \rho_A \right) \,.
\end{equation}
This expression captures the amount of quantum entanglement between $A$ and its complement. Since the density matrix $\rho_A$ is a positive semidefinite operator with unit trace, its eigenvalues represent probabilities. Consequently, the von Neumann entropy satisfies
\begin{equation}
      S_{\mt{EE}} \geq 0 \,,
\end{equation}
and vanishes if and only if the subsystem is in a pure state. This non-negativity is a universal feature of quantum entropy, independent of the specific dynamics or dimension of the system.

In order to be a consistent holographic dictionary, it is therefore necessary to demand the same property for any holographic definition of entanglement entropy. In static patch holography, we proposed in eq.~\eqref{eq:defineentropy} that holographic entanglement entropy in dS is related to bulk two-point functions as $S_{\mt{EE}} = - C \, \log G_{\rm dS}(x, x')$. For the entropy function to remain non-negative, the correlator $G_{\rm dS}(x, x')$ must satisfy
\begin{equation}
    G_{\rm dS}(x, x') \le  1.
\end{equation}
To examine this condition, we consider the explicit form of the Green’s function
\begin{equation}\label{eq:GA01}
    G_{\rm dS} (x, x') = N_{\mt{G}} \cdot  \frac{\sinh \mu \left(\pi - \mD \right)}{4\pi \sinh ( \mu \pi) \sin \mD } +  N_{\mt{A}} \cdot \frac{\sinh (\mu \mD)}{4\pi \sinh (\mu \pi) \sin \mD } \,.
\end{equation}
However, it is obvious that the de Sitter Green's function $ G_{\rm dS} (x, x')$ is unbounded: it diverges at coincident points $\mD\to 0$ as well as at antipodal points $\mD \to \pi$. This is because dS Green's function presents the short distance singularity at $\mD =0$, which is same as that the correlation function in flat space. More explicitly, we can find that the dS Green's function contains the short distance singularity as well as the antipodal counterpart, \ie 
\begin{equation}
\begin{split}
\lim_{\mD \to 0}   G_{\rm dS}(x, x') 
\sim &\frac{N_{\mt{G}}}{4 \pi  \mD}  \longrightarrow \infty \,,\\
\lim_{\mD \to \pi }   G_{\rm dS}(x, x') 
\sim &\frac{N_{\mt{A}}}{4 \pi (\pi -  \mD)}  \longrightarrow \infty
\end{split}
\end{equation}
In other words, there always exist small subsystems or larger subsystems with $G_{\rm dS}^\mu(x, x') > 1$, which lead to negative entropy.

This behavior closely parallels AdS, where bulk-to-bulk propagators also diverge in the short-distance limit. For $(d+1)$-dimensional AdS space, the bulk-to-bulk propagator for the bulk scalar field is known as  
\begin{equation}
G_{\mathrm{AdS}}^{\Delta}\left(x ; x^{\prime}\right)=G_{\mathrm{AdS}}^{\Delta}(\xi)=\frac{C_{\Delta}}{2^{\Delta}(2 \Delta-d)} \xi^{\Delta} \cdot{ }_2 F_1\left(\frac{\Delta}{2}, \frac{\Delta+1}{2} ; \Delta-\frac{d}{2}+1 ; \xi^2\right)\,,
\end{equation}
where we have introduced the so-called chordal distance $\xi$ that is associated with the (dimensionless) geodesic distance $\mD(x,x') $ as the following
\begin{equation}
\mD\left(x ; x^{\prime}\right)=\ln \left(\frac{1+\sqrt{1-\xi^2}}{\xi}\right), \quad \xi=\frac{1}{\cosh \left(\mD\left(x ; x^{\prime}\right)\right)} \,. 
\end{equation}
We can find that it is also divergent for short distance $\mD \to 0$ ($\xi \to 1$), \ie   
\begin{equation}
\lim_{\mD \to 0}  G_{\mathrm{AdS}}^{\Delta}\left(x ; x^{\prime}\right)  \approx    \frac{1}{4 \pi }\left(\frac{1}{\pi  \mD^2}\right)^{\frac{d-1}{2}} \Gamma \left(\frac{d-1}{2}\right) \,, 
\end{equation}
which is same as that of the flat/dS space. However, this potential problem is naturally resolved by introducing a UV cutoff in the AdS/CFT correspondence and normalizing the correlation function properly. As a result, the infinitesimal small subsystem at the cut-off scale corresponds to a vanishing entanglement entropy \footnote{Let us consider the (holographic) entanglement entropy of a CFT$_2$ living on the circle as an explicit example. For a subsystem with a size $\Delta L$, its entanglement entropy reads 
\begin{equation}
  S_{\mt{EE}} = \frac{c}{3} \log \left(\frac{L_{\mathrm{bdy}}}{\pi \epsilon} \sin \left(\frac{\pi \Delta L}{L_{\mathrm{bdy}}}\right)\right) \,,
\end{equation}
where $L_{\mathrm{bdy}}$ denotes the circumference of the circle. The cut-off scale $\epsilon$ at the leading-order can be fixed by taking 
$  S_{\mt{EE}} (\Delta L = \epsilon) = 0$. 
}.    
 
\begin{figure}[h]
	\centering
    \includegraphics[width=2.65in]{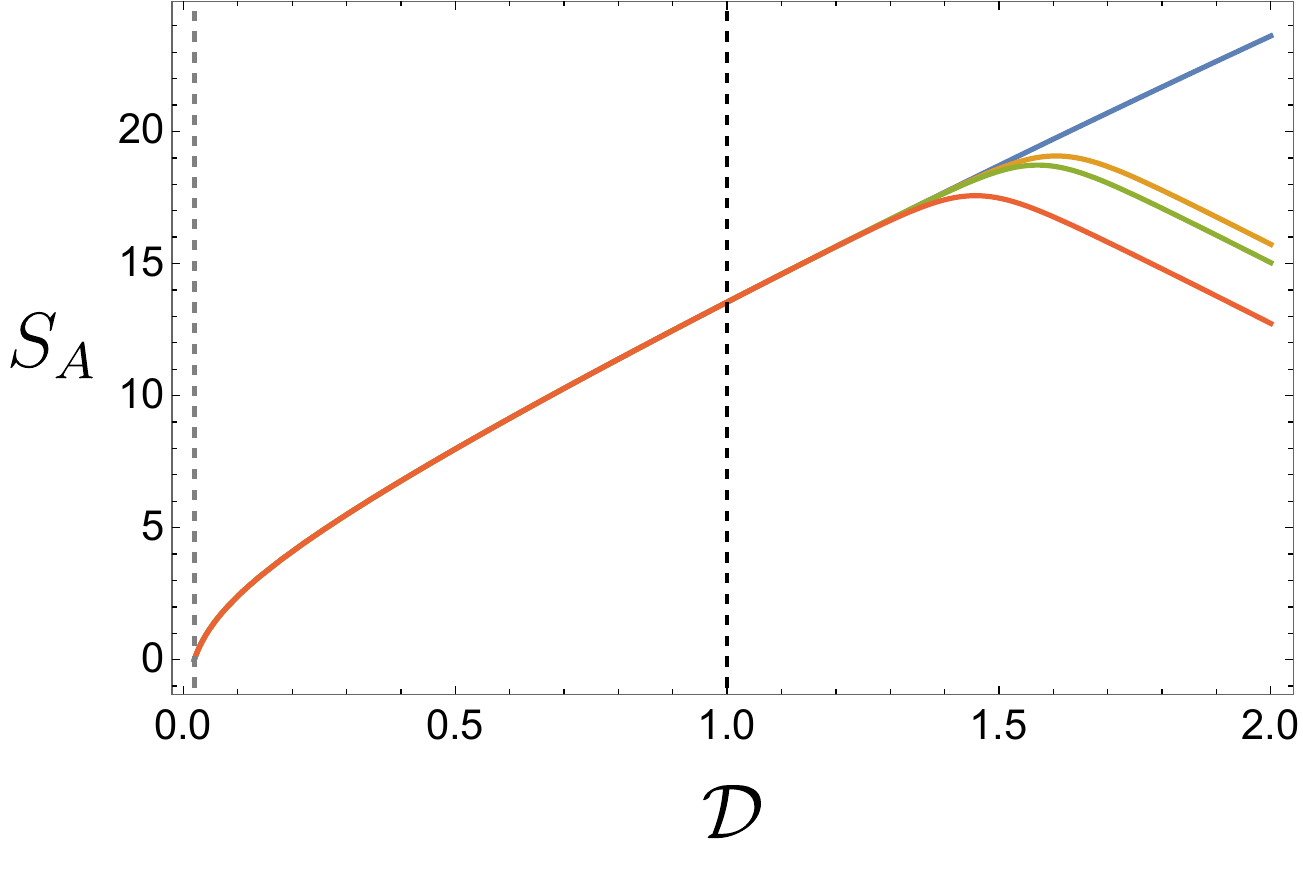}
	\includegraphics[width=3.35in]{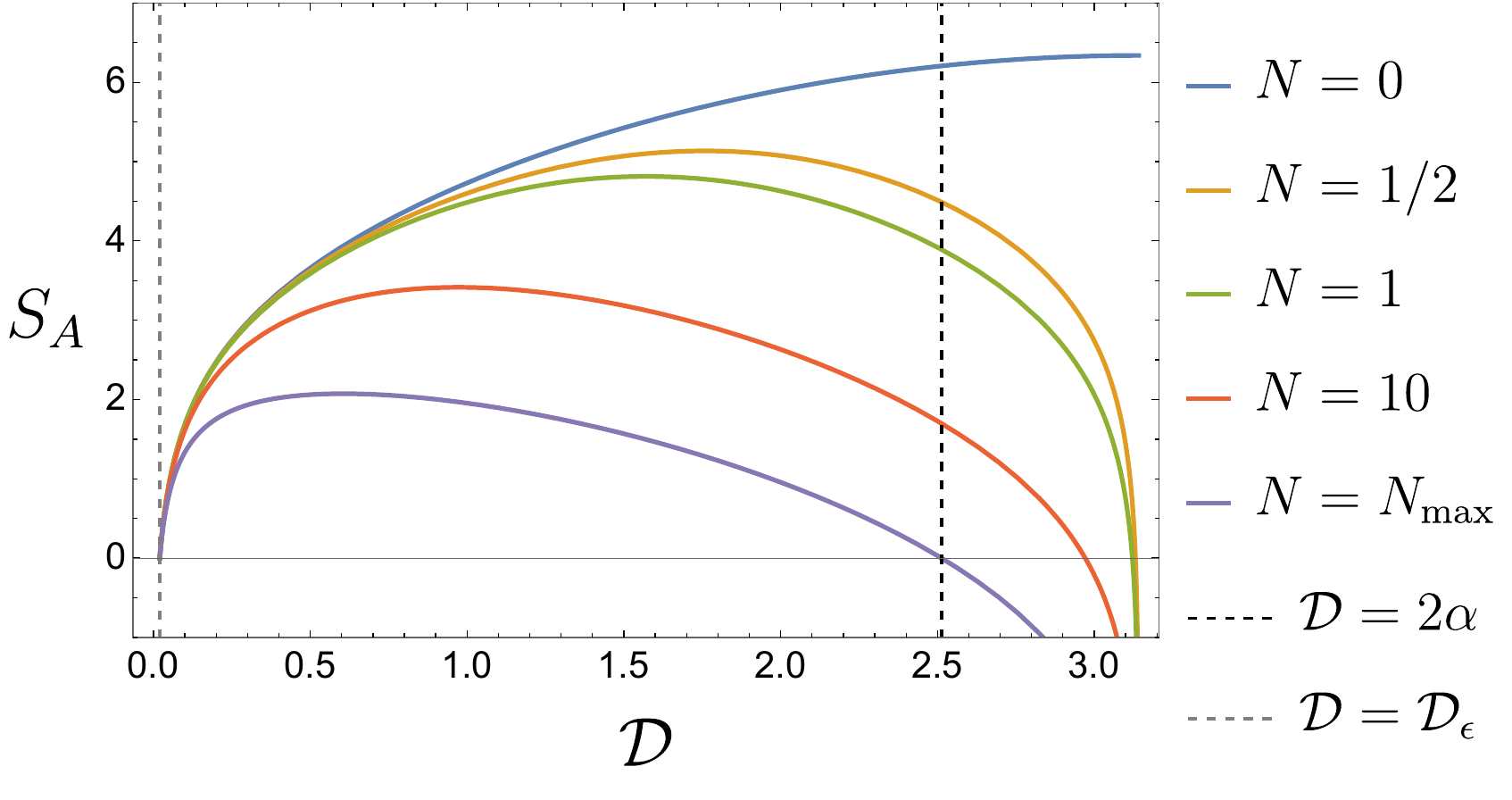}
	\caption{The entropy function $S_A$ (with taking $C=1$) as a function of the geodesic length in dS$_3$. Note that the length of spacelike geodesic anchored on the stretched horizon located at $r_\ast/L=\sin \alpha$ only takes the value with $\mD \in (D_\epsilon, 2\alpha)$. The numerical values of parameters are chosen as $\alpha=1/2, D_\epsilon=1/50$. The left plot corresponds to the large dimension limit with $\mu=10$ and the right one is generated by taking $\mu=1$.}
	\label{fig:SAasD}
\end{figure}

Taking the lessons from the AdS bulk spacetime, it is natural to expect that there is a similar cutoff scale $D_\epsilon$ appearing in de Sitter space. Although it is unclear yet what the geometric meaning of this bulk cut-off scale $D_{\epsilon}$ is, we conclude that it is necessary due to universal short distance divergence of bulk correlation functions and the non-negativity of entropy. Equipped with a cut-off scale $D_\epsilon$, we can require that the zero entropy is given by 
\begin{equation}
  S_{\mt{EE}} \big|_{\mD=\mD_\epsilon} = 0 \,,
\end{equation}
It indicates that the subsystem at a cut-off scale or the entire boundary on the stretched horizon corresponds to a pure state as one generally expects. Or equivalently, this corresponds to fixing the normalization of the dS Green's function in terms of 
\begin{equation}
\begin{split}
  N_{\mt{G}} &= \frac{ 4\pi \sinh ( \mu \pi) \sin \mD_\epsilon}{\sinh ( \mu (\pi -D_\epsilon)) + N \sinh (\mu D_{\epsilon})}  \approx 4\pi D_{\epsilon}   \,, \\
N_{\mt{A}} &= N \times \frac{4\pi \sinh ( \mu \pi)\sin \mD_\epsilon}{\sinh ( \mu (\pi -D_\epsilon)) + N \sinh (\mu D_{\epsilon})}  \,,\\
\end{split}
\end{equation}
keeping $N_{\mt{A}}/N_{\mt{E}} = N$. This relative ratio $N$ parametrizes the deviation of the dual state from the Bunch-Davies vacuum state. As a byproduct, it helps to reduce one free parameter in the proposed entropy formula \eqref{eq:defineentropy}. Correspondingly, the bulk cut-off scale $\mD_\ep$ also introduces a cut-off for the boundary. Recalling that he geodesic distance between two points on the stretched horizon is given by 
\begin{equation}
\cos\mD=\cos^2 \alpha\cosh\left(\frac{\Delta t}{L}\right)+\sin^2\alpha \cos\Delta\phi \,, 
\end{equation}
taking $\mD_\ep \sim  0$ is equivalent to $\Delta \sim 0$ for a interval on the constant time slice. Similar to the critical time given by the null geodesic \eqref{eq:bulknulltime}, the constraint $\mD \ge \mD_\epsilon$ imposes a stronger condition as follows: 
\begin{equation}\label{eq:zeroentropy}
\Delta \tilde{t} \le  \Delta \tilde{t}\big|_{\rm zero-entropy} =  \arccosh \left(  \frac{ \cos(D_\epsilon) - \sin^2 \alpha \cos (\Delta \phi)}{\cos^2\alpha }  \right)  \,.
\end{equation}
It gives rise to the minimal interval size: 
\begin{equation}\label{eq:zeroentropysize}
 \Delta \phi \ge \arccos\left(  \frac{(\cos \mD_\epsilon -1)}{\sin ^2\alpha}+1  \right) \approx  \frac{\mD_\epsilon}{\cos \alpha}  + \mathcal{O}(\mD_\epsilon^3) \,, 
\end{equation}
for a finite cut-off scale $\mD_\epsilon$. 

Finally, we note that while the short-distance divergence mirrors the AdS case, the antipodal divergence at $\mD=\pi$ is unique to de Sitter space. However, it is interesting to note that this type of antipodal divergence with $\mD=\pi$ arises only when
\begin{equation}
 \arcsin\frac{r_\ast}{L} \equiv \alpha  \to  \frac{\pi}{2}\, \quad \text{and}  \qquad \Delta \phi \to  \pi \,,
\end{equation}
as shown in eq.~\eqref{eq:defineDdS3}. This is realized by placing the holographic screen on the cosmological horizon and taking a half of the cosmological horizon. Consequently, placing the boundary theory on a stretched horizon ($r_\ast<L$) automatically avoids this antipodal divergence. Nonetheless, if the antipodal contribution dominates (\eg $N \to \infty$), the entropy could still become negative. As a result of the non-negativity of the entanglement entropy, the relative coefficient $N=N_{\mt{A}}/N_\mt{G}$ should be bounded from above. It is straightforward to find that the maximal value of $N$ is determined by $S_A(\mD=2\alpha)=0$, \ie          
\begin{equation}
\begin{split}
 N_{\rm max} &= \frac{\sin (2 \alpha ) \sinh ((\mD_\epsilon-\pi ) \mu )+\sin (\mD_\epsilon) \sinh ((\pi -2 \alpha ) \mu )}{\sin (2 \alpha ) \sinh (\mD_\epsilon \mu )-\sin (\mD_\epsilon) \sinh (2 \alpha  \mu )} \\
 &\approx \frac{\sin (2 \alpha ) \sinh (\pi  \mu )}{ \sinh (2 \alpha  \mu )-\mu  \sin (2 \alpha )} \frac{1}{\mD_\epsilon} +\mathcal{O}(\mD_\epsilon^0)\,,
\end{split}
\end{equation}

In summary, with the introduction of a cut-off scale $D_\epsilon$, the entropy function reduces to 
\begin{equation}\label{eq:defineSA01}
 \boxed{     S_{\mt{EE}}=- C \cdot\log\left(\frac{\sin \mD_\ep}{\sin\mD} \frac{\sinh\mu(\pi-\mD)+ N \sinh (\mu\mD)}{\sinh\mu(\pi-\mD_\ep)+ N \sinh (\mu\mD_\ep)} \right) } \,.
\end{equation}
As shown in figure~\ref{fig:SAasD}, one then can obtain the desired non-negativity of the entanglement entropy, namely 
\begin{equation}
  S_{\mt{EE}}  \ge  0  \,,
\end{equation}
provided the relative coefficient obeys
\begin{equation}
\boxed{ \text{Positivity condition:} \qquad N \le N_{\rm max}\approx \frac{\sin (2 \alpha ) \sinh (\pi  \mu )}{ \sinh (2 \alpha  \mu )-\mu  \sin (2 \alpha )} \frac{1}{\mD_\epsilon} } \,,
\end{equation}
This positivity bound is essential for the consistent definition of holographic entanglement entropy in the static patch holography, and further constraints arise when considering stronger entropic inequalities such as strong subadditivity, which we analyze next.

%%%%%%%%%%%%%%%%%%%%%%%%%%%%%%%%%%%%%%%%%%%%%
\subsection{Constraints from entropic inequalities}
%%%%%%%%%%%%%%%%%%%%%%%%%%%%%%%%%%%%%%%%%%%%%
Beyond non-negativity, entanglement entropy, \ie the von Neumann entropy of a density matrix $\rho_A$ satisfies a number of important inequalities that reflect fundamental aspects of quantum correlations. These entropic inequalities play a central role in quantum information theory and These entropic inequalities serve as consistency checks for any proposed notion of entropy, including its holographic counterpart.

One of the most basic is the {\it subadditivity} inequality. For a bipartite system composed of two subsystems $A$ and $B$, the entanglement entropy satisfies
\begin{equation}\label{eq:defineSA}
    S(A \cup B) \leq S(A) + S(B),
\end{equation}
which is saturated if and only if the state factorizes across the $A$-$B$ boundary. This inequality reflects the intuitive expectation that the total correlation cannot exceed the sum of individual correlations. A stronger and more subtle condition is the {\it strong subadditivity} (SSA), which is one of the most fundamental inequalities obeyed by the von Neumann entropy. For three adjacent subsystems $A$, $B$, and $C$, SSA states that
\begin{equation}\label{eq:defineSSA}
    S(A \cup B) + S(B \cup C) \geq S(B) + S(A \cup B \cup C).
\end{equation}
This inequality implies, for example, the monotonicity of mutual information and underlies the structure of quantum entanglement in multipartite systems.  It was rigorously proven Lieb and Ruskai \cite{Lieb:1973zz,Lieb:1973cp}. Notably, the strong subadditivity condition is strictly stronger than subadditivity: by setting $B = \varnothing$ in eq.~\eqref{eq:defineSSA}, one directly recovers the subadditivity inequality. Remarkably, SSA can be derived from an even more fundamental property of quantum systems, \ie the monotonicity of relative entropy. Given two density matrices $\rho$ and $\sigma$, the relative entropy $S(\rho || \sigma)$ is known to be non-increasing under partial trace (\ie data processing inequality), \ie tracing out degrees of freedom cannot increase distinguishability. This monotonicity leads directly to SSA, which in turn serves as a cornerstone for many important and interesting results in quantum information theory. 

In the context of quantum field theory, the implications of SSA have been actively investigated. Importantly, the strong subadditivity has been shown to hold for the Ryu-Takayanagi prescription \cite{Headrick:2007km} as well as the Hubeny-Rangamani-Takayanagi prescription \cite{Wall:2012uf} in the framework of the standard AdS/CFT correspondence. We therefore regard SSA as a critical benchmark for any candidate definition of holographic entanglement entropy in de Sitter space (see also its application to FLRW spacetimes in \cite{Noumi:2025lbb}). As we shall see, imposing SSA and related inequalities provides powerful constraints on any entropy functional built from de Sitter Green's functions.

Before turning to de Sitter holography, it is useful to recall that in AdS spacetimes with a finite radial cut-off the standard HRT prescriptions do {\it not} in general obey strong subadditivity. In particular, it was demonstrated in \cite{Lewkowycz:2019xse} for holographic theories with $T\bar{T}$ deformation that the {\it boosted SSA} associated with subsystems on different time slices can be violated, even though the purely static version remains valid (see \cite{Mori:2023swn,Franken:2024wmh} for related works). Because SSA follows from the monotonicity of relative entropy, such violations indicate that the geometric prescription in the bulk is no longer computing the von Neumann entropy of a density matrix, but rather encodes the non-local interactions induced by the finite cut-off. Similar breakdowns have been observed in half dS holography \cite{Kawamoto:2023nki,Chang:2024voo} and in the dS/dS correspondence \cite{Geng:2020kxh,Chang:2024voo}, where minimal (or extremal) surfaces anchored to a finite boundary again fail to respect (strong) subadditivity. Another possible source of violation is the time-dependence of the scalar-field vacuum in de Sitter space. Within the twist operator formalism, this suggests that the Green’s function of Euclidean (Bunch-Davies) vacuum is insufficient when approximating boundary correlators by bulk scalar propagators.

The lesson is clear: holographic area of extremal surfaces need not coincide with von Neumann entropy once locality is lost. Yet SSA is a fundamental property of von Neumann entropy, and can be violated only if the state prepared by the Euclidean path integral associated with a boundary subsystem $A$ fails to define a positive semi-definite operator (for instance, if negative modes are included). Our aim in this work is therefore to formulate a holographic dual that {\it does} reproduce the entropy of a generic density matrix for static patch observers in de Sitter space.

In particular, one generally expects that the de Sitter entropy $S_{\rm dS}$, given by the area of the cosmological horizon, represents the entanglement entropy of the maximal entropy state. This expectation has recently gained strong support: \cite{Chandrasekaran:2022cip} showed that including an observer in the description of the static patch of dS gravity converts the von Neumann algebra from a Type \Romannum{3} algebra to Type \Romannum{2}$_1$, thereby admitting a well-defined notion of entropy. Motivated by this, we will construct in the following an alternative definition of holographic entanglement entropy \eqref{eq:defineSA01}, based on bulk two-point functions, and subject it to the full suite of entropic inequalities, with strong subadditivity playing the central role.

%%%%%%%%%%%%%%%%%%%%%%%%%%%%%%%%%%%%
\subsubsection{Static SSA}
%%%%%%%%%%%%%%%%%%%%%%%%%%%%%%%%%%%%
\begin{figure}[t]
\centering
\includegraphics[width=5in]{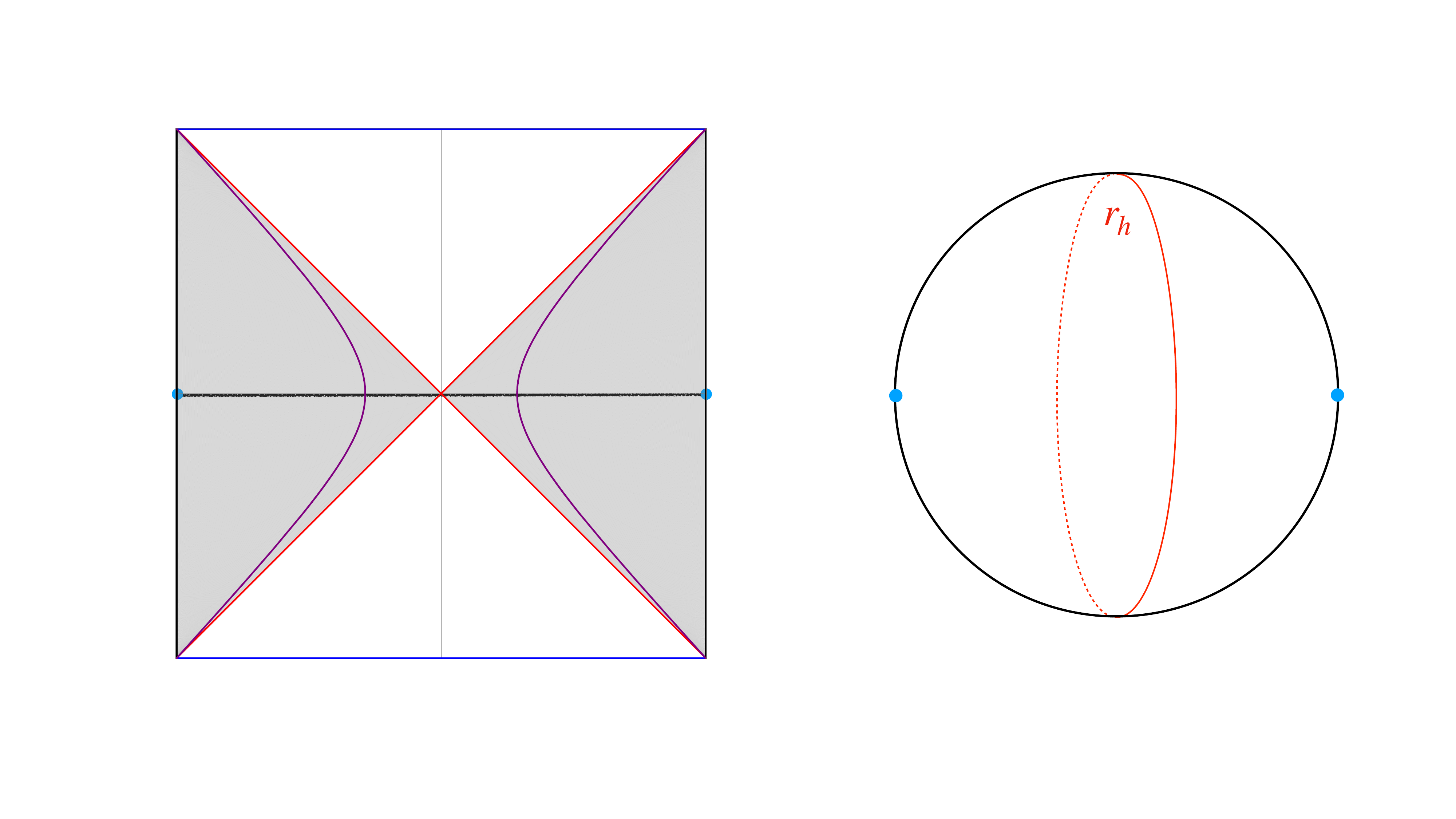}
\caption{Configuration for testing strong subadditivity with static intervals. We consider a constant-time slice of the stretched horizon and two overlapping subsystems $A$ and $B$.}\label{fig:staticSSA}
\end{figure}

We now demonstrate how entropy inequalities serve as diagnostics for candidate holographic entropy functionals in de Sitter holography. We begin with the simplest setup: boundary intervals defined on a constant-time slice $t = t_0$ of the stretched horizon. In this case, the entanglement entropy depends only on the interval length, which we denote by $S(A)= S(l_A)$.
To probe strong subadditivity in this static configuration, we consider two overlapping intervals $A$ and $B$ as shown in figure.~\ref{fig:staticSSA}. Their union $A \cup B$ has length $l+\delta l$, while their intersection $A \cap B$ has length $l$. Using the equivalent form of SSA \eqref{eq:defineSSA} as follows
\begin{equation}\label{eq:SSA}
    S(A) + S(B) \geq S(A \cup B) + S(A \cap B) \,,
\end{equation}
we obtain the inequality
\begin{equation}\label{eq:SSA_specific}
    2\, S\left(l+\frac{\delta l}{2}\right) \geq S(l+\delta l) + S(l) \,. 
\end{equation}
Assuming that $S(l)$ is twice differentiable, we expand each term in a Taylor series around $l$:
\begin{align}
    S\left(l+\frac{\delta l}{2}\right) &= S(l) + \frac{\delta l}{2}\, S'(l) + \frac{(\delta l)^2}{8}\, S''(l) + \mathcal{O}((\delta l)^3), \\
    S(l+\delta l) &= S(l) + \delta l\, S'(l) + \frac{(\delta l)^2}{2}\, S''(l) + \mathcal{O}((\delta l)^3) \,.
\end{align}
Substituting into eq.~\eqref{eq:SSA_specific} and simplifying the expression yields the infinitesimal version of static SSA \cite{Lewkowycz:2019xse}:
\begin{equation}\label{eq:infStaticSSA}
     \boxed{  S''(l) \leq 0 \,.}
\end{equation}

It is worth emphasizing that inequality \eqref{eq:SSA_specific} is equivalent to the statement that the entanglement entropy must be a concave function of the geometric parameter that characterizes the subsystem size \cite{Hirata:2006jx}. Concavity, in this context, encodes the requirement that the growth of entropy with increasing interval length becomes progressively slower, thereby enforcing subadditive behavior. Equivalently, concavity can be expressed as the following condition
\begin{equation}\label{eq:concavity}
    S\big(\lambda l_1 + (1-\lambda) l_2\big) \geq \lambda S(l_1) + (1-\lambda) S(l_2), \quad \text{for all } \lambda \in [0,1]\,,
\end{equation}
which must hold for any pair of interval lengths $l_1$ and $l_2$. A standard result from analysis asserts that for a twice-differentiable function, this condition is satisfied if and only if the second derivative is non-positive $  \frac{d^2 S}{dl^2} \leq 0$. Physically, this means that the rate of change of entanglement entropy decreases with increasing subsystem size. If $S''(l) > 0$ in any region, then $S(l)$ would grow faster than a linear function, leading to potential violations of subadditivity or strong subadditivity. Ensuring concavity of the entropy function is therefore essential for the consistency of any candidate entropic quantity. Particularly geometric functional in holographic constructions may not automatically enforce such constraints.

%%%%%%%%%%%%%%%%%%%%%%%%%%%%%%%%%%%%%%%%%%%%%%%%
Back to the dS bulk spacetime, it is straightforward to check that the geodesic length anchored on the stretched horizon satisfy 
\begin{equation}
\frac{d\mathcal{D}}{ d \Delta \phi} = \mD'= \frac{\sin \alpha  \cos \left(\frac{\Delta \phi }{2}\right)}{\sqrt{1-\sin^2\alpha  \sin ^2\left(\frac{\Delta \phi }{2}\right)}} \,,
\end{equation}
and
\begin{equation}
 \frac{d^2\mathcal{D}}{ (d \Delta \phi)^2}  = \mD''= -\frac{\sin \alpha \cos^2\alpha \sin \left(\frac{\Delta \phi }{2}\right)}{2 \left(1-\sin^2 \alpha  \sin ^2\left(\frac{\Delta \phi }{2}\right)\right)^{3/2}}  \le 0 \,,
\end{equation}
As a result, we have demonstrated that the geodesic length corresponding precisely to the RT formula in de Sitter space indeed satisfies the infinitesimal version of the spatial strong subadditivity. This outcome suggests that the entropy component arising exclusively from the global vacuum part consistently respects the fundamental concavity required by the static SSA condition. However, when including the antipodal contributions to the entropy, we find a subtle but significant competition emerges, which has the potential to violate the static SSA.

To explicitly illustrate this interplay between the global vacuum term $G_{\mt{G}}$ and the antipodal term $G_{\mt{A}}$, we separately define their corresponding entropy contributions:
\begin{equation}
S_{\mt{G}} = - \log(G_{\mt{G}}) \,, \quad S_{\mt{A}} = - \log(G_{\mt{A}}) \,.
\end{equation}
First, we can rigorously prove that the function $S_{\mt{G}}$ is indeed concave by verifying the following condition:
\begin{equation}
 \frac{d^2 S_{\mt{G}}}{d \Delta \phi^2}  =  \underbrace{\mD''}_{<0} \times \underbrace{(\mu  \coth ((\pi -\mD) \mu )+\cot \mD)}_{>0}+ \underbrace{\mD'^2 \left( \frac{\mu^2}{ \sinh^2((\pi -\mD) \mu ) }-  \frac{1}{\sin^2\mD }\right)}_{<0}  < 0 \,, 
\end{equation}
where each individual term has been analyzed to demonstrate the negative overall contribution explicitly. In contrast, the second derivative of the antipodal entropy function $S_{\mt{A}}$ reveals a more subtle behavior
\begin{equation}
 \frac{d^2 S_{\mt{A}}}{d \Delta \phi^2}  = \mD'' (\cot \mD -\mu  \coth (\mD \mu ))+\left(\mD'\right)^2 \left(\frac{\mu ^2}{\sinh ^2( \mD \mu )}-\frac{1}{\sin ^2\mD}\right)  \,. 
\end{equation}
which, depending on the values of $\mu$ and the geodesic length $\mD$, can yield both positive and negative values.
Specifically, at the boundary points, the geodesic distance and its derivatives have the following properties:
\begin{equation}
\begin{split}
\mD (\Delta\phi = 0 ) &=0\,, \qquad \mD'' (\Delta\phi = 0 ) =0 \,, \\
\mD(\Delta\phi = \pi) &= 2\alpha \,, 
\qquad 
\mD' (\Delta\phi = \pi ) =0 \,. 
\end{split}
\end{equation} 
Consequently, the behavior of $S_{\mt{A}}$ at these boundary points is explicitly derived as
\begin{equation}
\begin{split}
 \frac{d^2 S_{\mt{A}}}{d \Delta \phi^2}   &=
 \begin{cases}
     \mD'' (\cot \mD -\mu  \coth (\mD \mu ))  >0  \,, \qquad \Delta \phi = \pi \, \\
     \,\\
    -\frac{1}{3} \left(\mu ^2+1\right)  \mD'^2  < 0 \,\,, \quad \quad \quad \qquad \Delta \phi =  0\\
 \end{cases}  
\end{split}
\end{equation}
This result clearly highlights the critical violation of concavity, particularly around $\Delta \phi \sim \pi$.

\begin{figure}[t]
\centering
\includegraphics[width=4in]{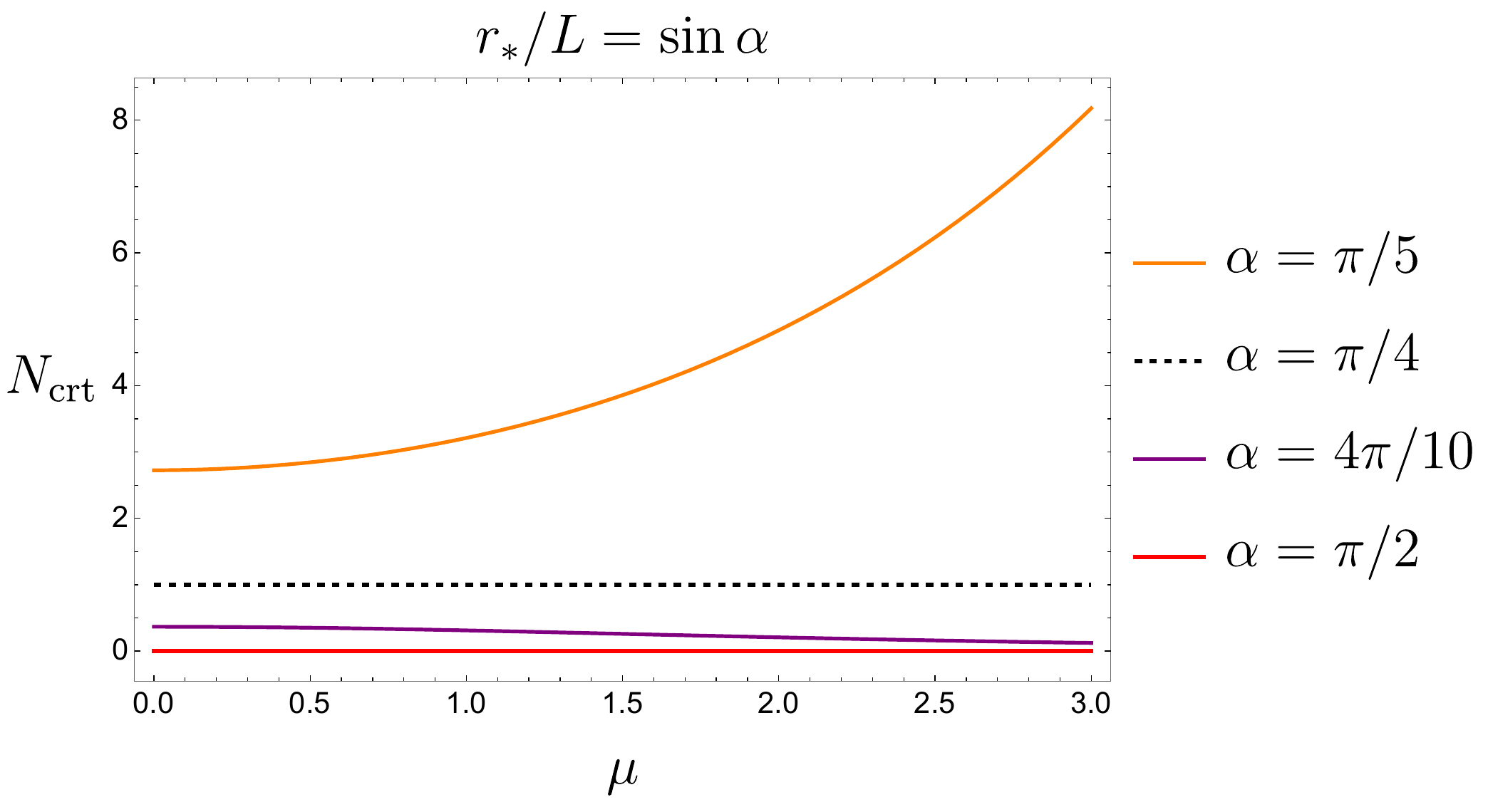}
\caption{The upper bound of $N$ as derived in eq.~\eqref{eq:staticSSAcondiiton}. It presents different behaviors for $\alpha > \frac{\pi}{4}$ and $\alpha < \frac{\pi}{4} $.}\label{fig:Ncrt}
\end{figure}

Given this potential violation of concavity, a careful analysis of the proposed holographic entropy function $S_{\mt{EE}}$ is required. To maintain consistency with the static SSA, it is essential to impose appropriate constraints on the relative coefficient $N = N_{\mt{A}} / N_{\mt{G}}$. To precisely derive this constraint, we explicitly compute the second derivative: 
\begin{equation}
\frac{d^2   S_{\mt{EE}} }{(d\Delta \phi)^2}  = 
\frac{N^2 e^{2  S_{\mt{G}} } S_{\mt{A}}''+e^{2 S_{\mt{A}}} S_{\mt{G}}'' + N e^{ S_{\mt{A}}+S_{\mt{G}}} \left(S_{\mt{A}}''-\left(S_{\mt{A}}'-S_{\mt{G}}'\right)^2+S_{\mt{G}}''\right)}{\left(N e^{S_{\mt{G}}}+e^{S_{\mt{A}}}\right)^2} \,. 
\end{equation}
To establish the most restrictive condition ensuring static SSA, we examine the case $\Delta \phi = \pi$ explicitly, where the derivative of the geodesic length vanishes, $\mD'(\Delta \phi = \pi) = 0$. Under this specific configuration, the second derivative simplifies significantly as follows
\begin{equation}
 \frac{d^2 S_{\mt{EE}}}{(d\Delta \phi)^2} \bigg|_{\Delta \phi= \pi}  = 
\mD'' \left(\frac{\mu  (\cosh ((\pi -\mD) \mu )-N  \cosh (\mD \mu ))}{\sinh ((\pi -\mD ) \mu )+ N \sinh (\mD \mu )}+\cot (\mD)\right)  \,.
\end{equation}
This leads to the following necessary condition:
\begin{equation*}
\cot (2 \alpha )+\frac{\mu  (\cosh ((\pi -2 \alpha ) \mu )-N  \cosh (2 \alpha  \mu ))}{\sinh ((\pi -2 \alpha ) \mu )+N  \sinh (2 \alpha  \mu )} \ge 0 \,.
\end{equation*}
Consequently, the static SSA imposes explicit bounds on the relative coefficient $N$, \ie 
\begin{equation}\label{eq:staticSSAcondiiton}
\boxed{\text{Static SSA:} \quad  -\frac{\sinh ((\pi -2 \alpha ) \mu )}{\sinh (2 \alpha  \mu )}   \le  N   \le \frac{\sinh ((\pi -2 \alpha ) \mu )+\mu  \tan (2 \alpha ) \cosh ((\pi -2 \alpha ) \mu )}{\mu  \tan (2 \alpha ) \cosh (2 \alpha  \mu )-\sinh (2 \alpha  \mu )}\,. } 
\end{equation}  
One can also numerically confirm that the strongest condition for the static SSA is given by $\Delta \phi= \pi$. This condition therefore provides the both necessary and sufficient condition for maintaining concavity and static SSA of the proposed holographic entanglement entropy formula. For later use, we denote the upper bound as $N_{\rm crt}$ as numerically shown in  in figure~\ref{fig:Ncrt}. It is interesting to notice that this critical value as a function of the location of the stretched horizon satisfies  
\begin{equation}
N_{\rm crt} (\alpha, \mu) \rightarrow
\begin{cases}
    >1 \,, \qquad \alpha <  \frac{\pi}{4}\,,\\
    =1 \,, \qquad \alpha = \frac{\pi}{4}\,,\\
    < 1 \,, \qquad \alpha > \frac{\pi}{4}\,.\\
\end{cases}
\end{equation}
As illustrated in figure~\ref{fig:Ncrt}, it indicates that the physical region satisfying the static SSA inequality shrinks as the stretched horizon approaches the cosmological horizon.

%%%%%%%%%%%%

\subsubsection{Boosted SSA}

\begin{figure}[t]
\centering
\includegraphics[width=4in]{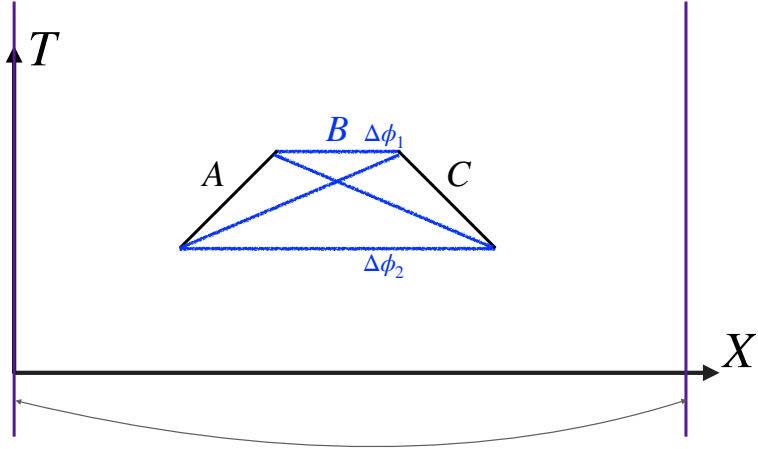}
\caption{Setup for boosted strong subadditivity. The subsystems $B$ and $A \cup B \cup C$ are placed on different time slices of the stretched horizon (represented as a cylinder).}
\label{fig:BSSA}
\end{figure}

We now investigate strong subadditivity in a Lorentz-boosted configuration, where subsystems are anchored on different time slices of the stretched horizon. In this case, the entanglement entropy is determined via the bulk geodesic length $\mD$, and the SSA inequality takes the form
\begin{equation}
\text{SSA} = 2 S(\mD_{AB}) - S(\mD_B) - S(\mD_{ABC}) \ge 0 \,.
\end{equation}
To study its infinitesimal version, we consider small separations between intervals, parameterized by $\delta = \Delta \phi_1 - \Delta \phi_2 \ll 1$.\footnote{The size of the interval must still respect the non-negativity constraint derived in eq.~\eqref{eq:zeroentropysize}, \ie $\delta \gtrapprox \mD_\epsilon / \cos \alpha$.} Expanding geodesic lengths perturbatively in $\delta$ gives
\begin{equation}
 \begin{split}
    \mD_{AB} \approx \mD_{B}  + \mD_{AB}' \, \delta  + \frac{\delta^2}{2}\mD_{AB}'' + \mathcal{O}(\delta^3)\,,\\
    \mD_{ABC} \approx \mD_{B}  + \mD_{ABC}' \, \delta  + \frac{\delta^2}{2}\mD_{ABC}'' + \mathcal{O}(\delta^3)\,.\\
 \end{split}
\end{equation}
Substituting into the SSA inequality yields
\begin{equation}
\begin{split}
\text{SSA} &= \left( 2 \mD'_{AB} - \mD'_{ABC}\right) S' \delta  \\
&+  \left(  (\mD''_{AB} - \frac{1}{2}\mD''_{ABC} ) S' + S'' \left( (\mD'_{AB})^2 -\frac{1}{2}(\mD'_{ABC})^2 \right) \right)   \delta^2 + \mathcal{O}(\delta^3) \,. 
\end{split}
\end{equation}
Here, the entropy derivatives are defined with respect to $\mD$:
\begin{equation}
S'= \frac{dS(\mD)}{d \mD} \,, \qquad S''= \frac{d^2S(\mD)}{d\mD^2} \,. 
\end{equation}
From the geometric relations, we observe
\begin{equation}
\mD'_{ABC} = 2 \mD'_{AB} \,,
\end{equation}
which simplifies the SSA expression to
\begin{equation}
\text{SSA} = \left( \left(\mD''_{AB} - \tfrac{1}{2} \mD''_{ABC}\right) S' - (\mD'_{AB})^2 S'' \right) \delta^2 + \mathcal{O}(\delta^3) \,.
\end{equation}
As a result, the condition for infinitesimal boosted SSA is recast as 
\begin{equation}\label{eq:infSSA}
\boxed{\text{Infinitesimal SSA Condition:} \qquad   \left(\mD''_{AB} - \frac{1}{2}\mD''_{ABC} \right) S' - (\mD'_{AB})^2 S''    \ge 0  \,.} 
\end{equation}
To check consistency, we now recover the static case shown in figure.~\ref{fig:staticSSA}, where $\delta_{ABC} = 2\delta_{AB}$ implies $\mD'_{ABC} = 2 \mD'_{AB}$ and $\mD''_{ABC} = 4 \mD''_{AB}$. In this case, the SSA reduces to
\begin{equation}
\text{Static SSA} = -\left(\mD'' \frac{dS}{d\mD} + (\mD')^2 \frac{d^2S}{d\mD^2}\right) \delta^2 + \mathcal{O}(\delta^3) ,.
\end{equation}
Noting the relation between $\Delta \phi$ and $\mD$:
\begin{equation}
\frac{d}{d\Delta \phi} = \mD' \frac{d}{d\mD} \,,
\end{equation}
we recover the concavity condition, \ie 
\begin{equation}
\boxed{\text{Static SSA} = \frac{d^2 S}{d(\Delta \phi)^2} = \mD'' S' + (\mD')^2 S'' \le 0 \,,}
\end{equation}
which agrees with the infinitesimal static SSA derived in eq.~\eqref{eq:infStaticSSA}. In term of the bulk geodesic length, the infinitesimal version of the static SSA is rewritten as 
\begin{equation}
\text{Static SSA} = \mD'' S' + (\mD')^2 S''  \propto S'' \sin (\mD) + \left( \frac{\cos (2 \alpha ) \cos (\mD)-1}{\cos (\mD)-\cos (2 \alpha )} \right) S'   \le 0 \,. 
\end{equation}
Noting the coefficient of $S'$ term is constrained by
\begin{equation}
\left( \frac{\cos (2 \alpha ) \cos (\mD)-1}{\cos (\mD)-\cos (2 \alpha )} \right)  \in (-\infty, -1] \,.
\end{equation}
It approaches negative infinity at $\mD \to 2\alpha$. In order to satisfy the static SSA, it is necessary to require
\begin{equation}
 S' \big|_{\mD = 2\alpha} \ge 0 \,,
\end{equation}
which is equivalent to 
\begin{equation}
S' \big|_{\mD = 2\alpha} = \cot (2 \alpha )+\frac{\mu  (\cosh ((\pi -2 \alpha ) \mu )-N  \cosh (2 \alpha  \mu ))}{\sinh ((\pi -2 \alpha ) \mu )+N  \sinh (2 \alpha  \mu )} \ge 0 \,. 
\end{equation}
This condition reproduces the result obtained earlier in eq.~\eqref{eq:staticSSAcondiiton}.

Moreover, we can further show that this necessary condition is also sufficient. While the analytic details are cumbersome, the logic is as follows. We begin by rewriting the static SSA inequality as a function of the geodesic length $\mD$ and reducing all terms to a common positive denominator. The numerator then takes the form of a quadratic polynomial in $N$, namely 
\begin{equation}
 \text{Static SSA} =   \frac{a N^2 + b N + c }{\text{Positive Denominator}}\le 0 \,, 
\end{equation}
with $a>0$ for $\mD \in (0, 2\alpha)$. For a fixed $\mD$, the inequality holds within the range $ \frac{-b - \sqrt{b^2-4ac}}{2a} \le N \le \frac{-b + \sqrt{b^2 -4ac}}{2a}$. Correspondingly, the sufficient condition is given by  
\begin{equation}
\max_{\mD} \left(\frac{-b - \sqrt{b^2 -4ac}}{2a} \right) \le N \le  \min_{\mD} \left(\frac{-b + \sqrt{b^2 -4ac}}{2a} \right) \,.
\end{equation}
A detailed analysis shows that the two critical values are, respectively, monotonically increasing and monotonically decreasing functions of $\mD$. Since the geodesic length is bounded by $\mD \le 2\alpha$, both the lower and upper bounds above are attained at $\mD = 2\alpha$. They coincide precisely with those in eq.~\eqref{eq:staticSSAcondiiton}. This demonstrates that the sufficient and necessary conditions are equivalent, thereby completing the proof.

%%%%%%%%%%%%%%%%%%%%%%%%%%%%%%%%%
%%%%%%%%%%%%%%%%%%%%%%%%%%%%%%%%%
\subsubsection*{Boundary null interval limit}

\begin{figure}[t]
\centering
\includegraphics[width=4.5in]{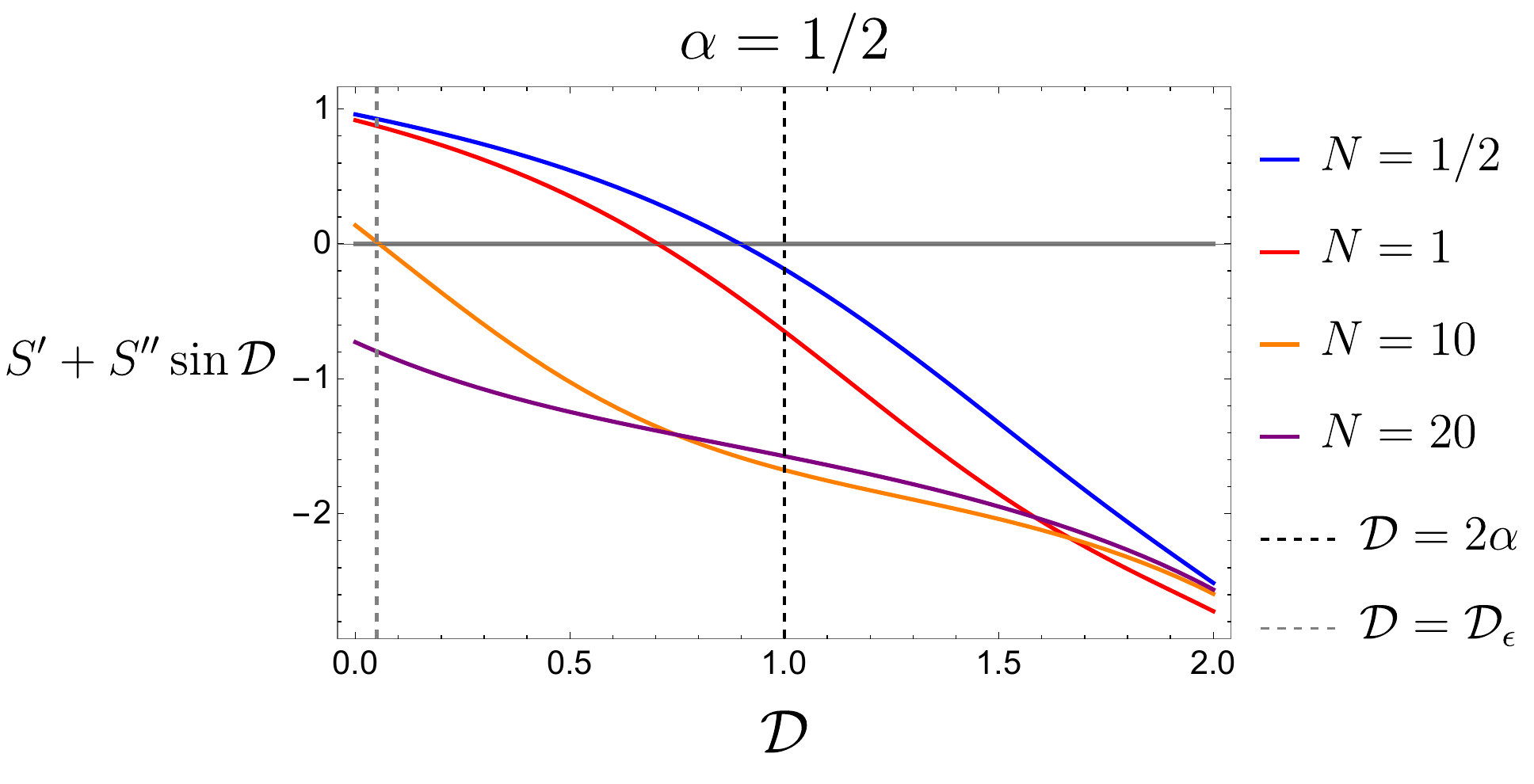}
\caption{Plot of the entropy function $S' + S'' \sin \mD$ versus geodesic length $\mD$ for different values of $N$, with $\alpha = 1/2$ and $\mD_\epsilon = 1/20$.}\label{fig:SSAcondition}
\end{figure}

We now turn to the special case in which the boundary intervals $A$ and $C$ approach null intervals, while $B$ and $A \cup B \cup C$ lie on different time slices. Since the null condition is defined with respect to the induced metric on the stretched horizon, we refer to this as the \emph{boundary null limit}. The induced metric is
\begin{equation}
ds^2 = -\cos^2\alpha \, dt^2 + L^2 \sin^2\alpha \, d\phi^2 \,. 
\end{equation}
The null condition in this metric fixes the sizes of the joint subsystems $A \cup B$ and $B \cup C$ as
\begin{equation}
\Delta \phi_{AB} = \frac{\Delta \phi_1 + \Delta \phi_2}{2} \,, \qquad \Delta t_{AB} = \frac{\Delta \phi_2 - \Delta \phi_1}{2} L\tan \alpha \,.
\end{equation}
The relevant derivatives of geodesic lengths in this limit are given by
\begin{equation}
    \begin{split}
     \mD_{AB}' &=  \frac{\sin \alpha  \cos \left(\frac{\Delta \phi_1 }{2}\right)}{2 \sqrt{1-\sin ^2\alpha  \sin ^2\left(\frac{\Delta \phi_1 }{2}\right)}} \,, \\
     \mD_{AB}'' &= -\frac{\sin \alpha \csc \left(\frac{\Delta \phi_1 }{2}\right) \left(\cos (2 \alpha ) \sin^2\left(\frac{\Delta \phi_1 }{2}\right)+1\right)}{8 \left(1-\sin^2\alpha  \sin ^2\left(\frac{\Delta \phi_1 }{2}\right)\right)^{3/2}} \,, \\
     \mD_{ABC}'' &=-\frac{\sin \alpha  \cos ^2\alpha  \sin \left(\frac{\Delta \phi_1 }{2}\right)}{2 \left(1-\sin ^2\alpha \sin ^2\left(\frac{\Delta \phi_1 }{2}\right)\right)^{3/2}} \,,.
    \end{split}
\end{equation}
Substituting these expressions into the infinitesimal boosted SSA condition \eqref{eq:infSSA}, and assuming $\Delta \phi < \pi$, the inequality simplifies to \footnote{The full variation defined in eq.~\eqref{eq:infSSA} is derived as 
\begin{equation}
\text{boosted SSA}= - \delta^2 \left(  \frac{\sin \alpha \cos^2 (\Delta \phi/2)}{8 \sin (\Delta \phi/2) \left(1- \sin^2\alpha \sin^2 (\Delta \phi/2)\right)^{3/2}} \right) \times  \left(    S' +  S'' \sin (\mD) \right) \,. 
\end{equation} 
where we ignore the prefactor because it is non-negative. This factor explains that we have a divergence in terms of $\text{SSA}  \sim  \frac{1}{\Delta \phi} \,, \mD \sim 0$. On the second hand, it leads to the vanishing case at 
\begin{equation}
\text{boosted SSA} \big|_{\Delta \phi= \pi}  = 0 \,.
\end{equation} 
}
\begin{equation}
 \boxed{
\begin{split}
 &\quad S' + 2 S'' \sin (\alpha ) \sin \left(\frac{\Delta \phi_1 }{2}\right) \sqrt{1-\sin^2 \alpha  \sin ^2\left(\frac{\Delta \phi_1 }{2}\right)}  \le 0\,,    \label{cond} 
\end{split}
}
\end{equation}
with 
\begin{equation}
 \begin{split}
    S' &= \frac{dS(\mD_B)}{d \mD_B} = \cot (\mD_B) + \frac{\mu  (\cosh ((\pi -\mD_B) \mu )- N \cosh (\mD_B \mu ))}{\sinh ((\pi -\mD_B) \mu )+N \sinh (\mD_B \mu )} \\
      S'' &=\frac{d^2S(\mD_B)}{d \mD_B^2} = - \frac{1}{\sin^2(\mD_B)}  +  \frac{\mu ^2 \left(N^2-2 N \cosh (\pi  \mu )+1\right)}{(\sinh ((\pi-\mD_B ) \mu )+N \sinh (\mD_B \mu ))^2} \,. \\
      \mD_B&=  \arccos\left(\sin^2\alpha \cos \Delta \phi_1 +\cos^2\alpha  \right).
 \end{split}
\end{equation}
For brevity, we denote $\mD_B \equiv \mD$ below. Using the bulk geodesic length presented above, we obtain a concise form of the boosted SSA condition:
\begin{equation}\label{eq:boostedSSAcondition}
 \boxed{   S' +  S'' \sin (\mD) \le 0   \,.} 
\end{equation}
In terms of the variable $\mD$, the left-hand side is depicted in figure~\ref{fig:SSAcondition}. Substituting the proposed entropy \eqref{eq:defineentropy}, the inequality becomes
\begin{equation}
\frac{\mu ^2 \sin \mD \left(N^2-2 N \cosh (\pi  \mu )+1\right)}{(\sinh (( \pi -\mD ) \mu ) + N \sinh (\mD \mu ))^2}+\frac{\mu  (\cosh ((\pi -\mD) \mu )-N \cosh (\mD \mu ))}{\sinh ((\pi -\mD) \mu )+N \sinh (\mD \mu )} - \tan \left( \frac{\mD}{2} \right) \le 0 \,. 
\end{equation}
Near the cutoff scale $\mD \sim \mD_\epsilon$, one finds
\begin{equation}
\frac{\mu \left( N - \cosh (\pi \mu)   \right) }{\sinh (\pi \mu)} + \mathcal{O}(D_\epsilon) \ge  0 \,,
\end{equation}
implying the leading-order condition
\begin{equation}\label{eq:boostedSSAresult}
\boxed{ \text{boosted SSA condition:} \qquad N \ge \cosh (\pi \mu)  \,. }
\end{equation}

\begin{figure}[t]
	\centering
	\includegraphics[width=3.2in]{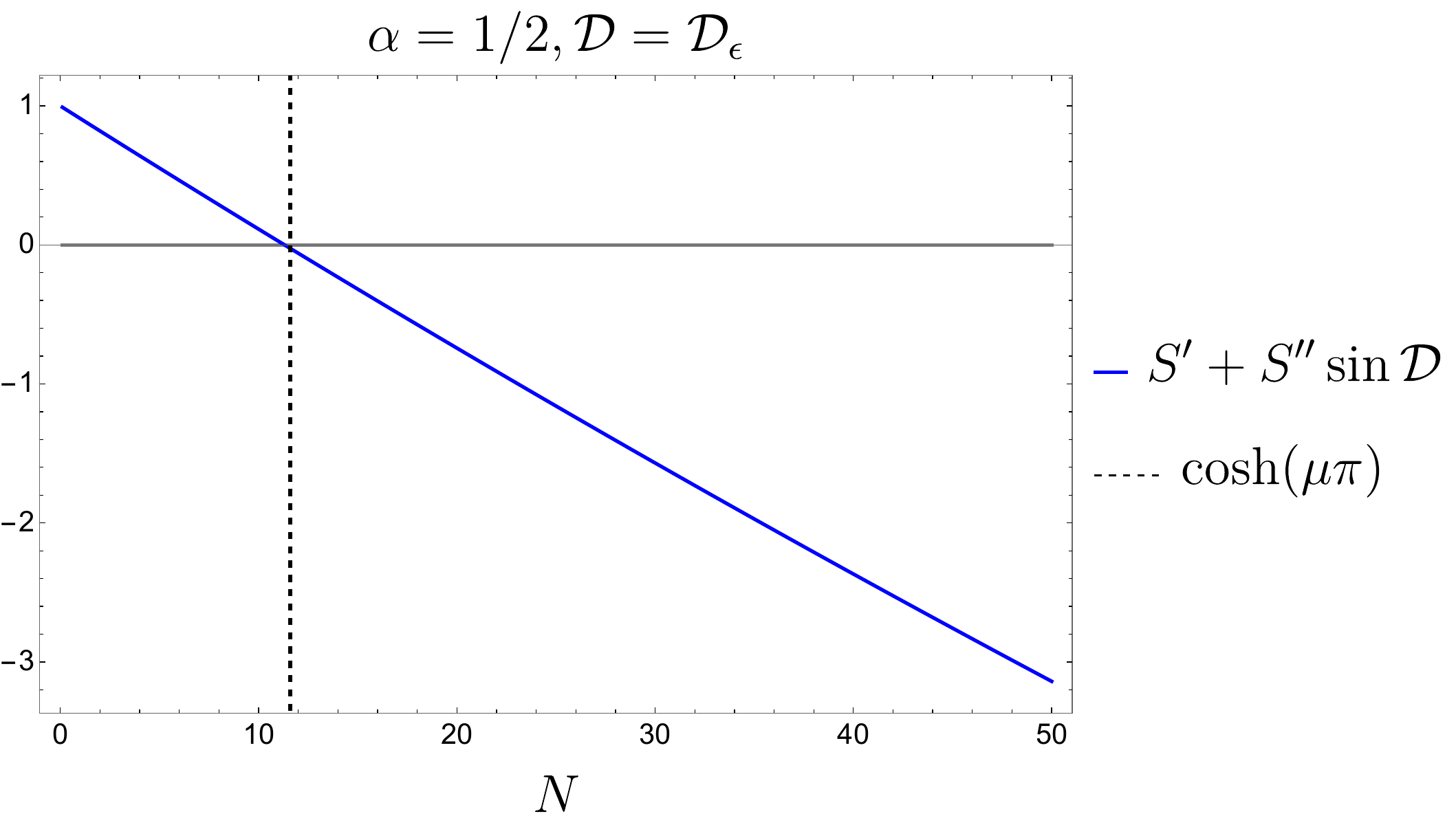}
    \includegraphics[width=2.8in]{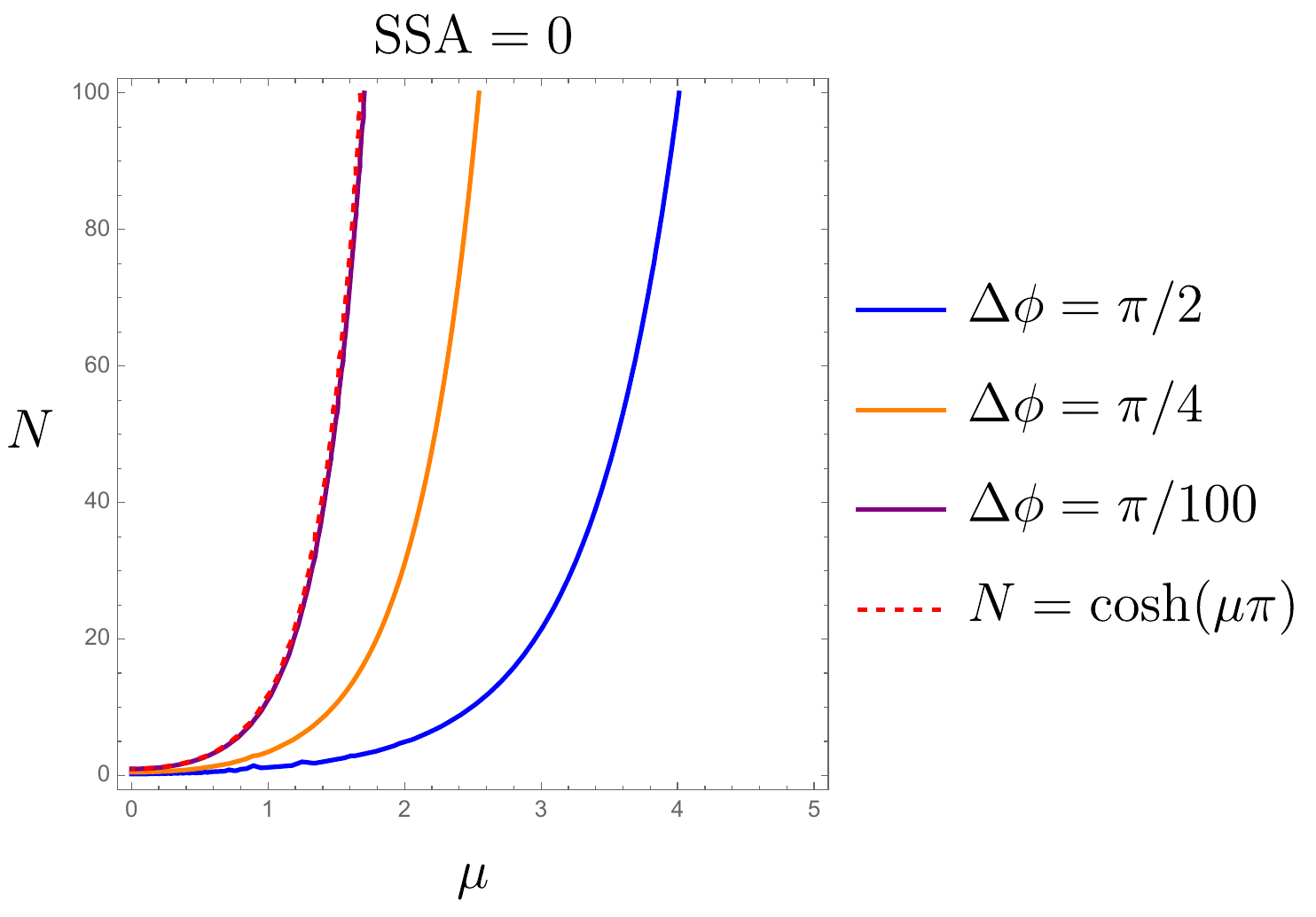}
\caption{Left: Entropy function $S' + S'' \sin \mD$ as a function of the relative coefficient $N$, evaluated at $\mD = \mD_\epsilon = 1/100$ with $\alpha = 1/2$. Right: The curves $N(\mu)$ that parametrizes the case with $\text{SSA}=0$ in eq.~\eqref{eq:infSSA}. The tightest condition corresponds to the critical configuration with $\mD=\Delta \phi=0$.}
	\label{fig:SSAcondition02}
\end{figure}

While this constraint ensures the validity of boosted SSA near $\Delta \phi \sim 0$, it serves only as a \emph{necessary} condition. To promote it to a \emph{sufficient} condition, we can further verify that the inequality in eq.~\eqref{eq:boostedSSAcondition} holds over the entire domain of $\mD$. For any fixed $\mD$, the boosted SSA requires 
\begin{equation}
\frac{\mu ^2 \sin \mD \left(N^2-2 N \cosh (\pi  \mu )+1\right)}{(\sinh (( \pi -\mD ) \mu ) + N \sinh (\mD \mu ))^2}+\frac{\mu  (\cosh ((\pi -\mD) \mu )-N \cosh (\mD \mu ))}{\sinh ((\pi -\mD) \mu )+N \sinh (\mD \mu )} - \tan \left( \frac{\mD}{2} \right) \le 0 \,. 
\end{equation}
After a bit more analysis\footnote{After the reduction of fractions to a common denominator, the numerator on the left hand side can be recast as a quadratic form: $a N^2 + b N  + C < 0$ with $a=2\mu \sin \mD -\sinh(2\mD \mu) \le 0 $.}, one finds that the strongest constraint on $N$ arises from maximizing the right-hand side over $\mD$:
\begin{equation}
\begin{split}
N &\ge \max_{\mD} \left( \frac{\sinh(\pi \mu) \sqrt{-2 \mu^2 \cos(2 \mD) + 2 \mu^2 + 1} + \sinh((2 \mD - \pi) \mu) - 2\mu \sin \mD \cosh(\pi \mu)}{\sinh(2\mD \mu) - 2\mu \sin \mD} \right) \\
& \qquad\quad  =\cosh(\mu \pi) \,,\\
\end{split}
\end{equation}
which is consistent with the necessary condition derived previously. We conclude that the necessary and sufficient condition for boosted SSA is precisely given by eq.~\eqref{eq:boostedSSAresult}. Figure~\ref{fig:SSAcondition02} provides a numerical check. 

To summarize, the boundary null interval limit highlights how boosted SSA imposes distinct constraints than the static case. By expanding the SSA expression around small variations, we derived an explicit inequality \eqref{eq:boostedSSAcondition} involving the first and second derivatives of the entropy. This not only reproduces the known static case as a special limit but also identifies a clear bound on the parameter $N$ controlling the relative weight of antipodal contributions. This condition \eqref{eq:boostedSSAresult} ensures consistency with entropic inequalities in time-dependent setups and provides a sharp diagnostic for the viability of the proposed holographic entanglement entropy \eqref{eq:defineSA01} in de Sitter holography.\footnote{Numerical checks suggest that the null case indeed provides the tightest bound on $N$, though more general boost angles can also be analyzed.}

%%%%%%%%%%%%%%%%%%%%%%%%%%%%%%%%
%%%%%%%%%%%%%%%%%%%%%%%%%%%%%%%%
\subsubsection{Constraints on holographic entanglement entropy}

We briefly summarize our result on the constraints on holographic entanglement entropy in this section.

Our construction is built from the generic two-point function of scalar operators in dS spacetime. Using the geodesic length $\mD$ in dS$_3$, the proposed holographic entanglement entropy in static patch holography is
\begin{equation}\label{eq:summarySEE}
S_{\mt{EE}}=- C \cdot\log\left(\frac{\sin \mD_\ep}{\sin\mD} \frac{\sinh\mu(\pi-\mD)+ N \sinh (\mu\mD)}{\sinh\mu(\pi-\mD_\ep)+ N \sinh (\mu\mD_\ep)} \right) \,,
\end{equation}
Here, the parameter $N$ quantifies the relative weight between the global dS vacuum and its antipodal counterpart, thereby encoding the choice of boundary conditions and the associated quantum states in static patch holography. To regulate the short-distance divergence of the geodesic length and ensure non-negativity of the entropy, we introduced a cutoff scale $\mD_\epsilon$.

Having established this candidate holographic entropy functional, we tested its consistency with the fundamental entropic inequalities. In the static setting, the necessary and sufficient condition for SSA imposes the following constraint on $N$:
\begin{equation}\label{eq:summarystaticSSA}
\text{Static SSA:} -\frac{\sinh ((\pi -2 \alpha ) \mu )}{\sinh (2 \alpha \mu )} \le N \le \frac{\sinh ((\pi -2 \alpha ) \mu )+\mu \tan (2 \alpha ) \cosh ((\pi -2 \alpha ) \mu )}{\mu \tan (2 \alpha ) \cosh (2 \alpha \mu )-\sinh (2 \alpha \mu )}\,.
\end{equation}
To probe more dynamical configurations, we analyzed the boosted SSA, in which the intervals are defined on different time slices of the boundary. This setup is sensitive to the causal structure of the static patch, and yields the following necessary and sufficient condition 
\begin{equation}\label{eq:boostedSSAcondition2}
\text{Boosted SSA condition:} \qquad N \ge \cosh (\pi \mu) \ge 1 \,.
\end{equation}
This bound is independent of $\alpha$ and imposes a stronger lower limit on $N$ compared to the static case.

\begin{figure}[t]
\centering
\includegraphics[width=3in]{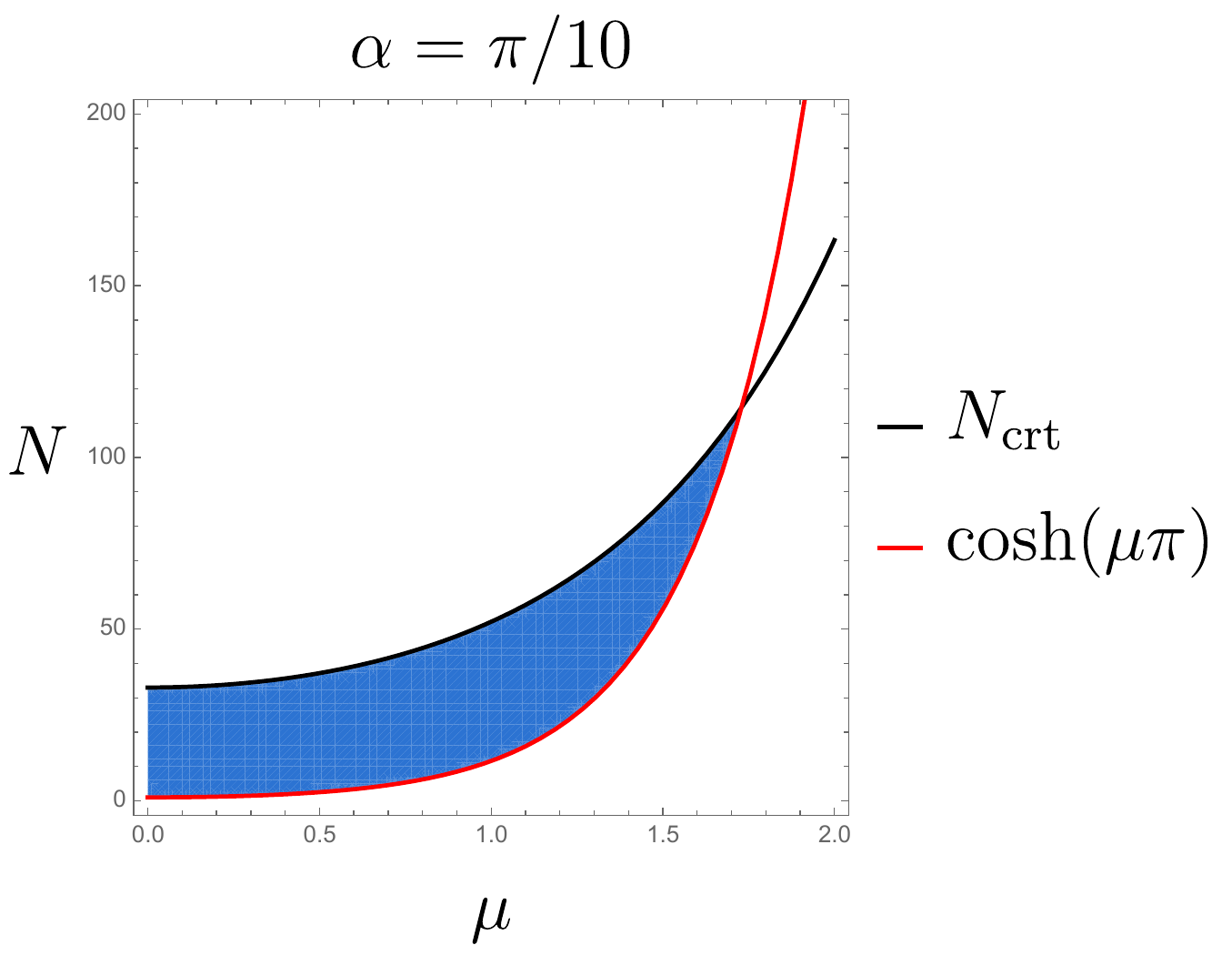}
\includegraphics[width=3in]{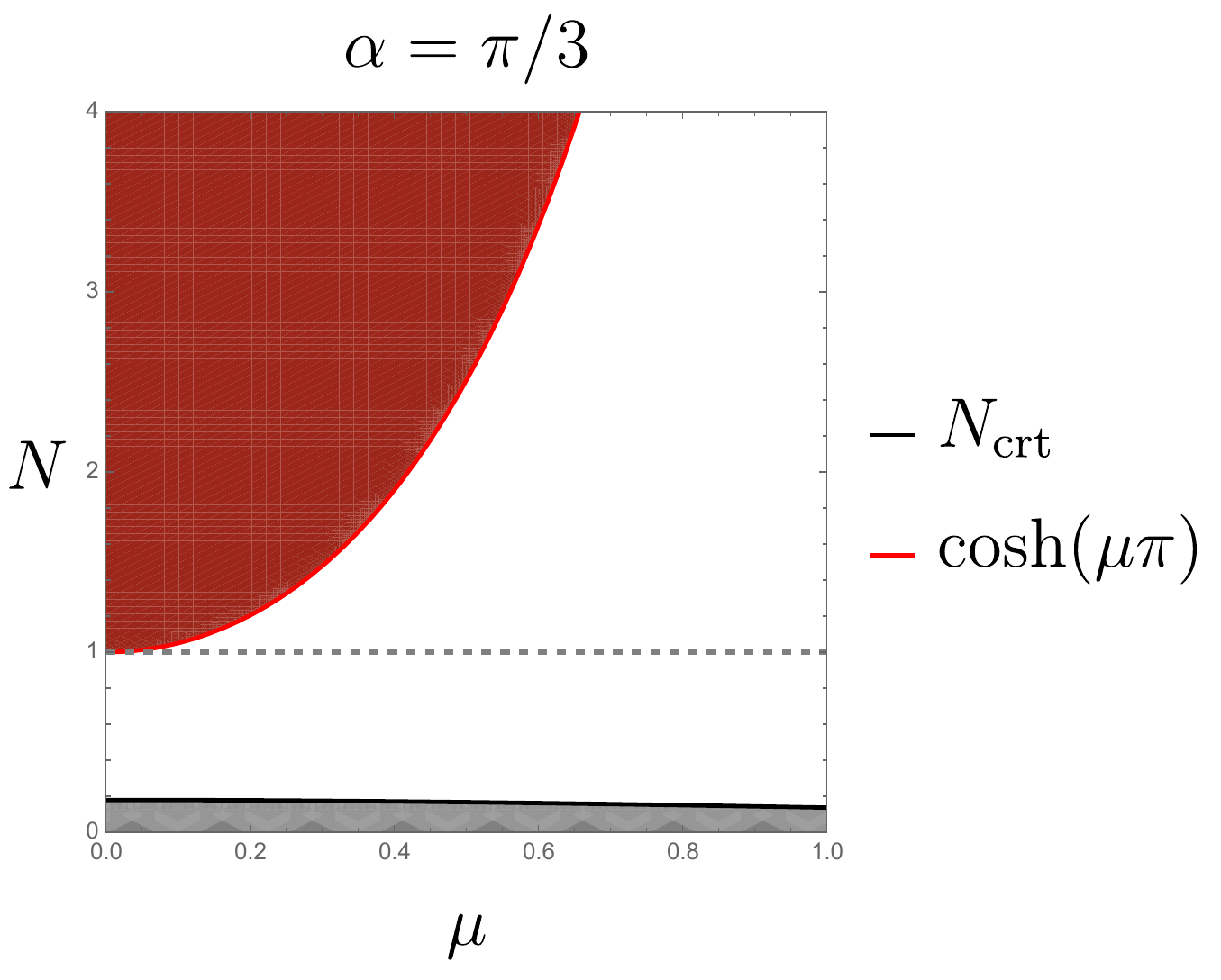}
\caption{Parameter regions consistent with both static and boosted SSA inequalities. Left: The blue shaded region represents the allowed regime when $\alpha \le \pi/4$. Right: No consistent region exists for $\alpha > \pi/4$.}
\label{fig:largesmallalpha}
\end{figure}

Together, these results place nontrivial and physically motivated constraints on the \emph{bulk parameter space} of static patch holography. In particular, they imply that only a restricted class of entropy function within the allowed range of $N$ can yield holographic entropies consistent with the basic definition of quantum entanglement entropy. More interesting, we should impose both the static and boosted SSA conditions simultaneously. However, the compatible parameter regime does not always exist, as illustrated in figure.~\ref{fig:largesmallalpha}. Noting the properties of the upper bound in eq.~\eqref{eq:summarystaticSSA}, \ie $N_{\rm{crt}}(\mu, \alpha) < 1$ when $\alpha > \frac{\pi}{4}$, one can find that the compatibility of two inequalities requires $\alpha \le  \frac{\pi}{4}$. That is, the stretched horizon cannot be chosen arbitrarily close to the cosmological horizon. The combined set of inequalities therefore yields the final constraint:
\begin{equation}
\boxed{\text{SSA conditions:}\quad \cosh (\mu \pi ) \le  N \le N_{\rm{crt}}(\mu, \alpha)   \quad \text{and} \quad \alpha \le  \tfrac{\pi}{4} } \,,
\end{equation}
with saturation occurring at $\alpha = \pi/4$ and $\mu=0$.

%%%%%%%%%%%%%%%%%%%%%%%%%%%%%%%%
\subsection{Twist operators with light mass}

\begin{figure}[t]
\centering
\includegraphics[width=3in]{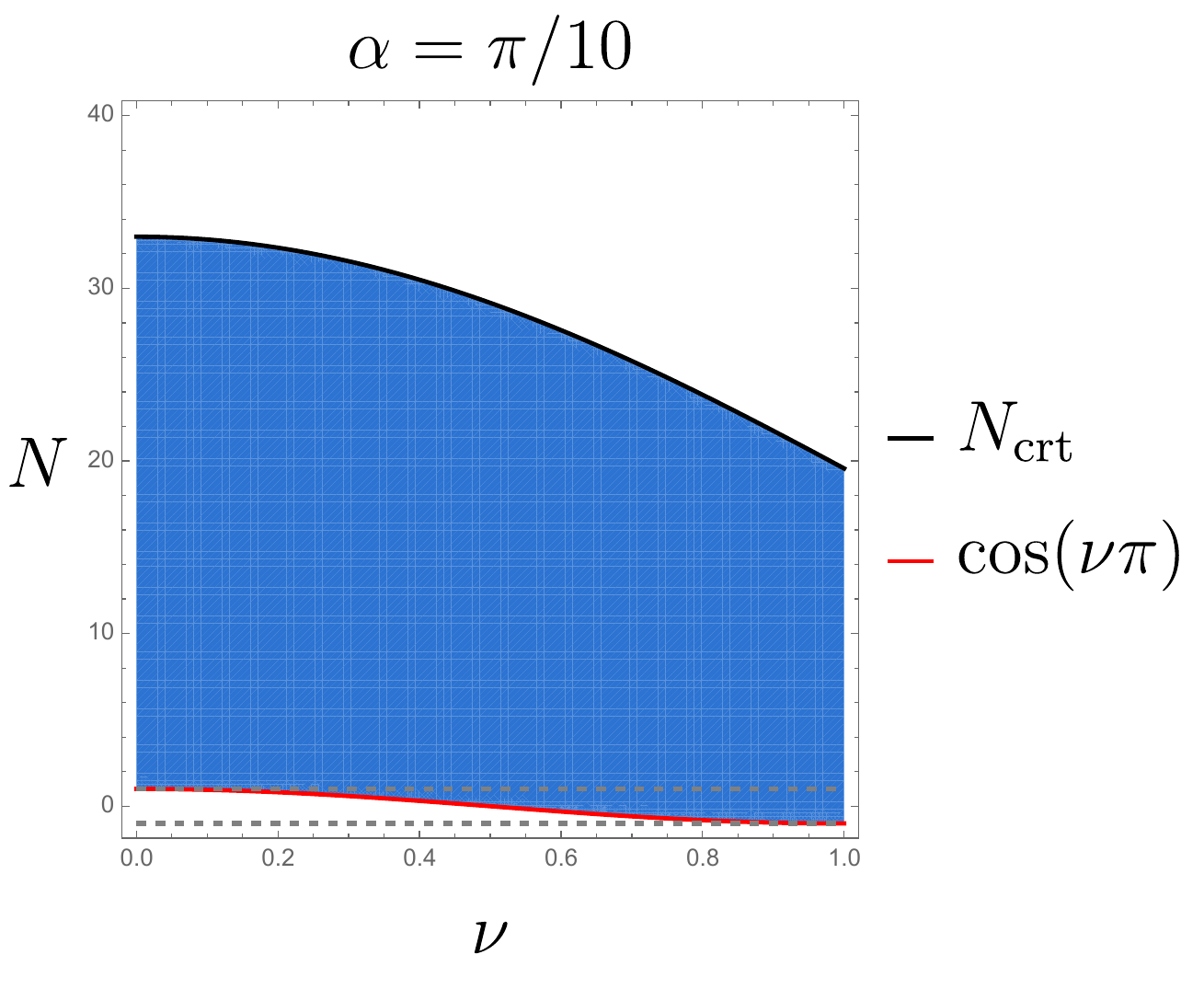}
\includegraphics[width=3in]{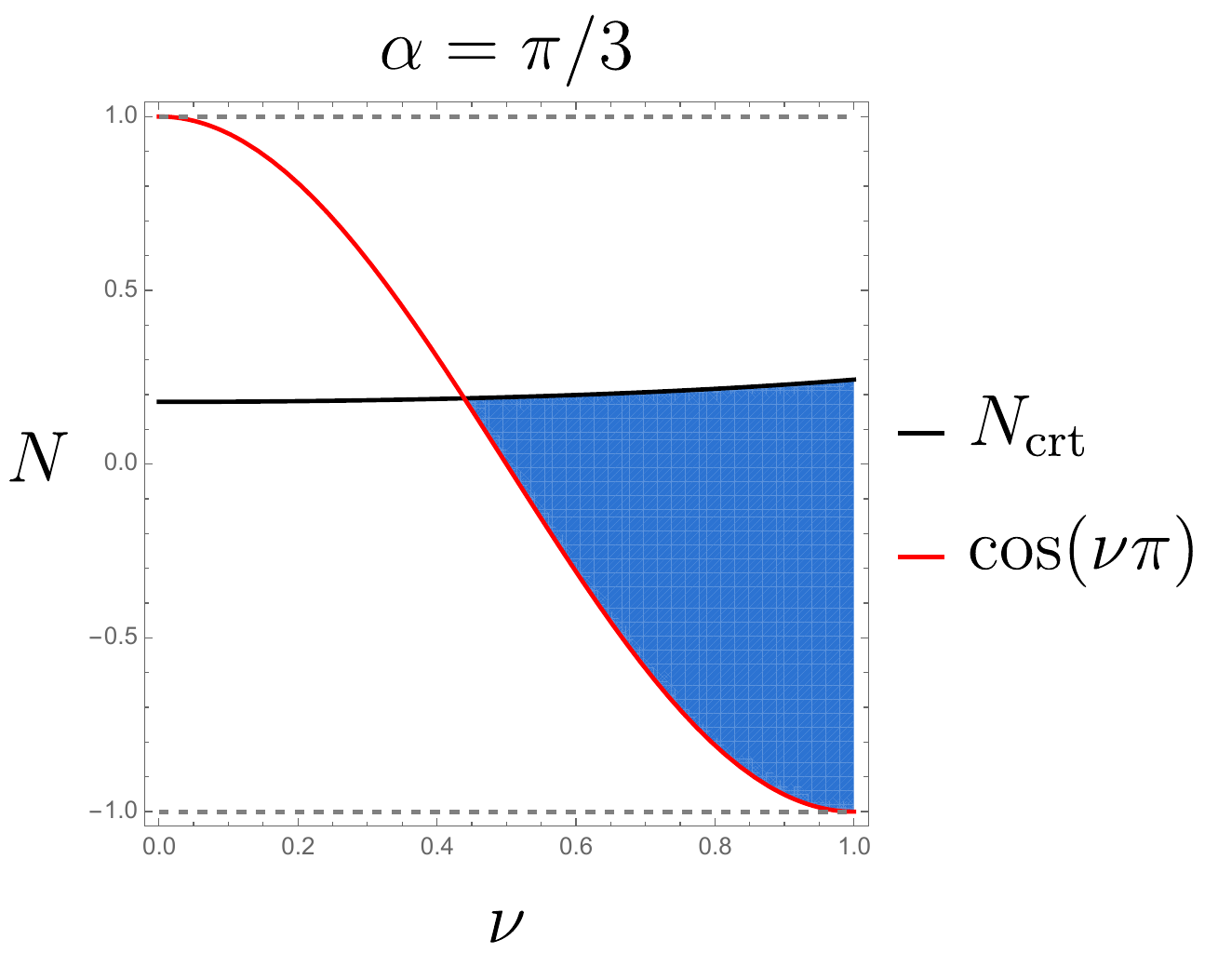}
\caption{For the entropy formula associated with a twist operator of light mass, the physically allowed region satisfying both the static and boosted SSA inequalities is indicated by the blue shaded area.}\label{fig:largesmallalpha02}
\end{figure}

In the previous analysis, we assumed that the conformal dimension of the twist operator is parametrized by a real parameter
\begin{equation}
\mu=\sqrt{m^2L^2-1} \,,
\end{equation}
which corresponds to a scalar operator with large mass. We now turn to the case of a light twist operator\footnote{Strictly speaking, the mass of the twist operator should be uniquely determined. However, since the precise QFT on the stretched horizon is not identified, we treat the mass as a free parameter and determine its allowed range from the entanglement inequalities.}. For convenience, we introduce a new real parameter $\nu$ defined by
\begin{equation}
0 \leq \nu = \sqrt{1 - m^2} \leq 1 \,,
\end{equation}
which is related to the mass through $\mu = i \nu$.

The corresponding entanglement entropy for the light mass case is obtained by analytically continuing eq.~\eqref{eq:summarySEE}, yielding
\begin{equation}\label{eq:SEEnu}
S_{\mt{EE}}=- C \cdot\log\left(\frac{\sin \mD_\ep}{\sin\mD} \frac{\sin \nu(\pi-\mD)+ N \sin(\nu\mD)}{\sin\nu(\pi-\mD_\ep)+ N \sin (\nu\mD_\ep)} \right) \,.
\end{equation}
In this case, one naturally expect that the static and boosted SSA inequalities can be derived by the same analytical continuation, \ie 
\begin{equation}
\boxed{\text{SSA Conditions:} \quad \cos(\pi\nu) \le N \le \frac{\sin ((\pi -2 \alpha ) \nu )+\nu \tan (2 \alpha ) \cos ((\pi -2 \alpha ) \nu )}{\nu \tan (2 \alpha ) \cos (2 \alpha \nu )-\sin (2 \alpha \nu )} \,,}
\end{equation}
Numerically, we also have verified that the analytic continuation $\mu \to i \nu $ correctly reproduces the constraints on $N$. However, unlike the heavy mass case discussed previously, combining the two inequalities always allows for an overlap between the static and boosted SSA constraints. This can be easily confirmed by noting that, at the edge of the parameter space $\nu = 1$, the upper and lower bounds satisfy
\begin{equation}
N_{\rm crt}(\nu=1, \alpha) = -1 + \frac{2 \pi}{4 \alpha - \sin (4 \alpha )} > -1 = \cos(\pi\nu)\big|_{\nu =1} \,.
\end{equation}
For comparison, we present the numerical results illustrating the physical regime of the bulk parameter space in figure~\ref{fig:largesmallalpha02}.

%%%%%%%%%%%%%%%%%%%%%%%%%%%%%%%%%%%%%%%%%%%%
\subsection{Finite entropy inequalities: bulk causality vs boundary causality}

In the previous subsections, our analysis of entropic inequalities was confined to the infinitesimal regime, where the separation between relevant intervals was taken to be parametrically small. In this limit, the boosted strong subadditivity conditions reduce to local differential constraints on the entropy function, allowing us to extract sharp bounds on the parameter $N=N_{\mt{A}}/N_{\mt{G}}$ that controls the relative weight of antipodal contributions. However, infinitesimal tests probe only the short-distance structure of the entropy functional and need not guarantee its validity for finite-size intervals, where nonlocal features of the bulk geometry become important. To address this, we now turn to the finite version of entropy inequalities, in which the relevant subsystems are separated by macroscopic angular and temporal intervals on the stretched horizon.

\begin{figure}[t]
	\centering
	\includegraphics[width=4in]{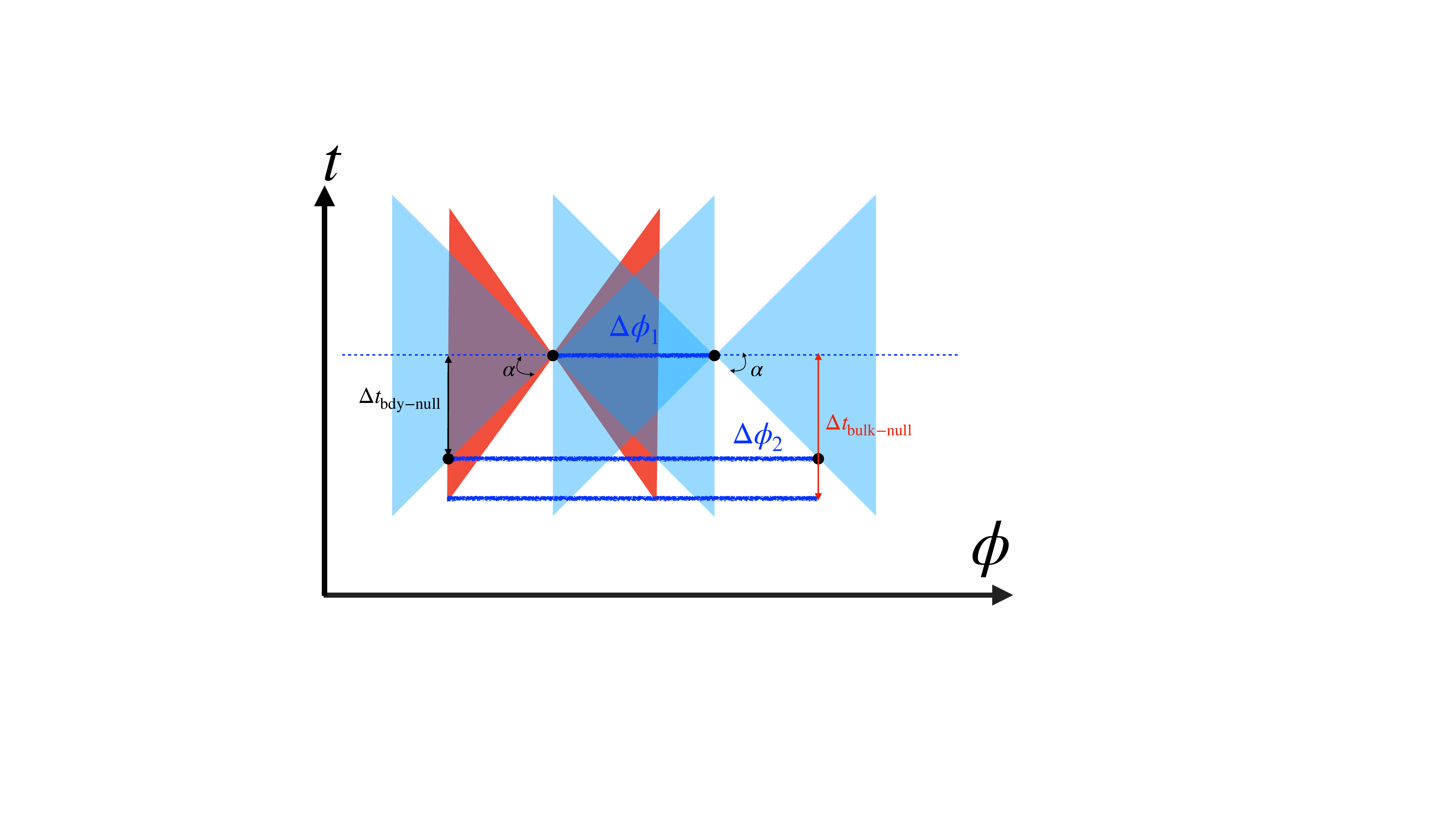}
	\caption{Boundary causality vs bulk causality. The shaded region denotes the region which is spacelike separated from the given point on the stretched horizon. The blue region presents the domain of dependence associated with the boundary Lorentzian symmetry. The red region is referred to as the domain induced from the bulk dS spacetime.}
	\label{fig:Causality}
\end{figure}

A key aspect associated with the finite-interval analysis is the mismatch between bulk and boundary causal structures in dS holography. This is also one crucial difference between dS holography and the standard AdS/CFT correspondence. In asymptotically AdS spacetimes that satisfy the averaged null energy, there are no ``bulk shortcuts'' that let signals arrive earlier than allowed by the boundary lightcone. This is the Gao-Wald gravitational time delay: bulk causal curves between boundary points never beat the boundary null geodesic, and boundary spacelike separation remains noncausal through the bulk.

As shown earlier, the induced metric on the stretched horizon corresponds to a flat boundary spacetime where a natural ``boundary lightcone'' for the dual QFT is given by respecting the Lorentzian symmetry. However, the geodesics in the bulk determine a bulk lightcone whose boundary is given by null geodesics with $\mD=0$. The intersection between the bulk lightcone and the stretched horizon thus defines a distinct causality structure on the boundary. Although it is tempting to expect that dual boundary theory in dS holography could respect the Lorentzian symmetry, we will show that the induced causality structure from dS bulk spacetime is preferred by the holographic entropy functional. As highlighted before, the mismatch of the bulk causality and boundary causality has been decoded in the proposed definition of the holographic entanglement entropy formula for dS bulk spacetime. More explicitly, let us consider any boundary interval parametrized by a spatial size $\Delta \phi$ and temporal distance $\Delta t$, the boost invariant distance with respect to the two endpoints of the interval read
\begin{equation}
  \text{Boundary length}: \qquad \Delta L = \sqrt{\Delta X^2 - \Delta T ^2} = L  \cos \alpha \sqrt{ \tan^2 \alpha \, (\Delta \phi )^2  - (\Delta \tilde{t})^2} \,. 
\end{equation}
Whereas the holographic entanglement entropy in dS bulk spacetime is defined by using the bulk geodesic distance associated with the same endpoints of the boundary interval, \ie 
\begin{equation}
 \text{Bulk distance}: \qquad \mD = \arccos\left(\sin^2\alpha \cos (\Delta \phi )+\cos^2\alpha \cosh (\Delta \tilde{t}) \right) \,.
\end{equation}
which does not respect the boundary boost invariance.
The different distance measures correspond to two distinct null separated intervals, namely 
\begin{align}
    \text{Boundary null interval}:& \qquad \Delta t_{\text{bdy null}}=L \cdot\Delta \phi  \, \tan\alpha \,,\\
    \text{Bulk null interval}:& \qquad \Delta t_{\text{bulk null}}=L\cdot\arccosh\left[\frac{ 1-\sin^2\alpha\cos\Delta \phi }{\cos^2\alpha}\right] \,. 
\end{align}
\footnote{It is of course interesting to apply the same method to investigate the bulk causality and boundary causality in the AdS/CFT correspondence. However, a subtle point is that the spacelike geodesic in AdS$_3$ not always approach the null geodesics, namely 
\begin{equation}
 \lim_{\mD \to 0}\left(  \text{spacelike geodesic} \right)  \ne \text{bulk null geodesic} \,,
\end{equation}
because of the timelike boundary in AdS$_3$. In fact, all null geodesics starting on the boundary in AdS$_3$ all terminate at the antipodal point on the boundary.}
As a result, the intrinsic boundary causality structure is distinct from the one induced from dS bulk spacetime, as illustrated in figure \ref{fig:Causality}. 
For any subsystem with a fixed size $\Delta \phi$, it is straightforward to find the following inequality\footnote{Because $ \partial_{\Delta \phi } \left( \Delta t_{\text{bdy null}} - \Delta t_{\text{bulk null}} \right)> 0$, its minium is given by $ \min \left( \Delta t_{\text{bdy null}} - \Delta t_{\text{bulk null}} \right) =0 $ at $\Delta \phi=0$.}: 
\begin{equation}
\Delta t_{\text{bulk null}} \le \Delta t_{\text{bdy null}} \,,
\end{equation}
which is saturated only when $\Delta\phi =0$. Consequently, we find that the causal domains associated with the boundary metric is always \emph{bigger} than that projected from the bulk. 
In other words, the dS bulk spacetime admits causal ``shortcuts'' that are not constrained by the boundary light-cone structure. Equivalently, bulk causality in dS is stronger than boundary causality: bulk causal curves can arrive earlier than any curve confined to the boundary and can even connect boundary points that are spacelike separated with respect to the boundary flat metric.

\begin{figure}[t]
	\centering
	\includegraphics[width=4in]{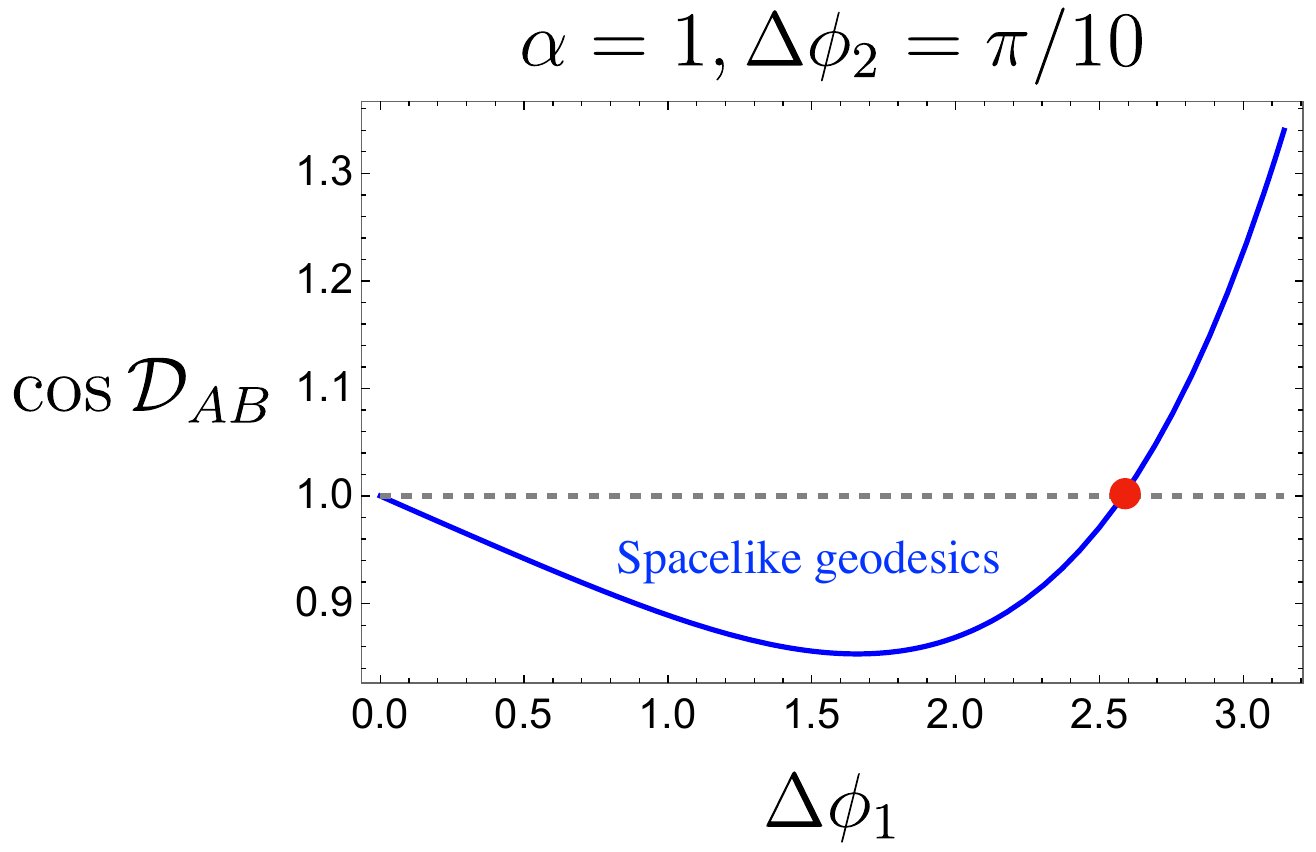}
	\caption{The geodesic length connecting the endpoints of interval $A \cup B$. Here the interval A is chosen to be null interval along the boundary lightcone. The spacelike geodesic $\mD_{AB}$ may not exist.}
	\label{fig:CosDAB}
\end{figure} 

For infinitesimal intervals with $\Delta\phi \sim 0$, the null time scales defined by the bulk geodesic condition and by the induced boundary metric coincide,
\begin{equation}
\Delta t_{\text{bulk null}}  \approx   L \Delta\phi  \, \tan \alpha + \mathcal{O}((\Delta \phi )^2)  \approx \Delta t_{\text{bdy null}} \,.
\end{equation}
which explains why no distinction was required in our earlier derivation of the infinitesimal boosted SSA constraints. On the contrary, for finite-size subsystems, the bulk and boundary notions of null separation differ appreciably, and this discrepancy has direct implications for the validity of SSA. Considering the same configuration as shown in figure.~\ref{fig:BSSA}, we can boost the auxiliary intervals $A$ and $C$ to be the extremal cases, \ie either null intervals along the boundary lightcone or the intervals connected by bulk null geodesics. Considering the null intervals $A, C$ along the boundary lightcone, the geodesic distance associated with the joint system $A, B$ reads 
\begin{equation}
\cos\mD_{AB}=\cos^2\alpha\cosh\left[\frac{\Delta\phi_2-\Delta\phi_1}{2}\tan\alpha\right]+\sin^2\alpha\cos\left( \frac{\Delta\phi_2+\Delta\phi_1}{2} \right) \,,
\end{equation}
with identifying $\Delta t_{AB} = \Delta t_{A} = \Delta t_{C} = \Delta t_{\text{bdy null}}(\Delta \phi_A)$. There are two reasons that this extremal configuration is not physical. First of all, the spacelike geodesic $\mD_{AB}$ does not exist if 
\begin{equation}
\Delta t_{\text{bdy null}}(\Delta \phi_A) > \Delta t_{\text{bulk null}}  (\Delta \phi_{AB}) \,, 
\end{equation}
as illustrated in figure \ref{fig:CosDAB}. Secondly, we find that the finite boosted SSA in this finite configuration can be violated even if the corresponding infinitesimal condition is satisfied. By contrast, defining extremal intervals with respect to the bulk null geodesics preserves the boosted SSA inequality in all cases we have tested. It would be interesting to examine the complete condition of the finite SSA using the proposed holographic entanglement entropy in static patch holography.  

Regardless of the detailed functional form of the holographic entanglement entropy in dS holography, we expect that, in the spirit of the replica trick, it should be associated with the length of the bulk geodesic connecting the boundary endpoints. The causality analysis developed above remains valid in this broader context. The key lesson is that the causal structure inherited from the de Sitter bulk is the one physically compatible with holographic entanglement entropy in static patch holography, whereas the intrinsic boundary lightcone extends beyond the actual causal domain of the dual field theory.
%%%%%%%%%%%%%%%%%%%%%%%%%%%%%%%%
%%%%%%%%%%%%%%%%%%%%%%%%%%%%%%%%
\section{Conclusion and Discussion}\label{sec:dis}
In this work, we have initiated a systematic study of holographic entanglement entropy in the context of static patch holography for de Sitter space. Building on the replica trick formulation and the conjectured holographic dictionary relating boundary correlation functions to bulk Green’s functions in de Sitter spacetime, we proposed a new prescription for the entanglement entropy of boundary subregions residing on the stretched horizon of the static patch, namely eq.~\eqref{eq:summarySEE}. We then analyzed the fundamental inequalities that any consistent entanglement entropy must satisfy, \ie subadditivity and strong subadditivity. We found that these inequalities impose nontrivial constraints on the bulk parameter space, specifically on the relative coefficient $N$ and the stretched horizon location $\alpha$ in static patch holography. These constraints ensure the compatibility of our proposed holographic entanglement entropy with the axioms of quantum information theory. Our results demonstrate that entanglement inequalities serve as a powerful diagnostic tool in de Sitter holography, comparable to their role in AdS/CFT, and provide new quantitative insights into the viability of static patch holography. Before concluding the paper, we turn to discuss several subtle points in our analysis and open questions for future studies. 

\paragraph*{The existence of holographic entanglement entropy in dS.}\par\vspace{0.2\baselineskip}
Earlier studies that attempted to understand the holographic entanglement entropy in de Sitter bulk spacetime by a similar formula of RT formula, \ie defining the area of minimal surface as the ``holographic entanglement entropy''. However, it was shown that such a definition generically violates (strong) subadditivity \cite{Kawamoto:2023nki,Chang:2024voo}. This violation was taken as an indication that any putative dual field theory of de Sitter space must be intrinsically nonlocal, lacking a proper notion of entanglement entropy in the standard sense of quantum information theory. From this perspective, the failure of SSA was viewed as a fundamental obstruction to defining a consistent entanglement structure in de Sitter holography.

Nevertheless, in this work, we still try to construct a proper definition of entanglement entropy for the static patch holography. An important motivation for our current analysis arises from recent progress in understanding von Neumann algebras associated with de Sitter space. In particular, \cite{Chandrasekaran:2022cip}  constructed an algebra of observables appropriate to an observer confined to a static patch, defined by gravitationally dressed operators anchored to the observer’s worldline. The resulting von Neumann algebra is of Type II$_1$, which admits a natural definition of entropy. Within this framework, empty de Sitter space corresponds to a maximum-entropy state, while semiclassical states yield entropies that reproduce the generalized entropy $S_{\rm gen} = \frac{A}{4 \GN} + S_{\rm out}$ up to an additive constant determined by the renormalization scheme. Crucially, both entanglement entropy and relative entropy are well-defined in this setting, providing a  algebraic interpretation of de Sitter entropy. A central feature of this construction is the role of the observer: the entropy is not a global property of the spacetime but reflects the information accessible to a single observer restricted to their static patch. This observer-dependence gives a natural entanglement interpretation of the dS entropy and highlights the importance of formulating holographic entanglement entropy directly in the static patch. Our work in this paper thus aim to construct a consistent definition of holographic entanglement entropy in the static patch holography and to reconcile the entropy inequalities from quantum information theory with the proposed holographic formula.

\paragraph*{Undetermined boundary parameters.}\par\vspace{0.2\baselineskip}
The first subtlety in the proposed entropy formula relates to the appearance of undetermined boundary parameters in the expression for holographic entanglement entropy in the static patch. More specifically, the construction depends on the dimension of the twist operator, parametrized by $\mu$ (or $\nu$ in the light-mass regime), as well as the overall normalization coefficient $C$.

In the well-established AdS/CFT correspondence, these parameters are unambiguously fixed by the structure of the holographic boundary field theory. By taking the replica limit $n \to 1$, the entanglement entropy of CFT is derived from the two-point function of twist operators located at the endpoints, namely
\begin{equation}\label{EErep}
S_{\mt{EE}}(\hat{\rho}_A) \equiv \lim_{n \to 1} \frac{1}{1-n} \log \langle \sigma_n(X_1) \bar{\sigma}_n( X_2) \rangle \,,
\end{equation}
which is the AdS counterpart of the propose entropy formula \eqref{eq:defineEEindS} in dS space. The conformal dimension of the twist operator is determined by the replica trick, taking the universal form $\Delta_n = \frac{c}{12}\left(n-\frac{1}{n}\right)$, which allows one to recover the Ryu–Takayanagi prescription. The overall normalization is likewise determined by the central charge, ensuring consistency with the Brown–Henneaux relation $c=\frac{3L}{2G}$. In the semiclassical limit of a holographic field theory, this approach is equivalent to taking the large central charge limit $\Delta_n \sim c=\frac{3L}{2G} \to \infty$. By applying the saddle-point approximation, the holographic propagator or two-point function can be described by the geodesic length, \ie 
\begin{equation}
\begin{split}
\left\langle\mathcal{O}\left(X_1\right) \mathcal{O}\left(X_2\right)\right\rangle &= \int \mathcal{D} \mathcal{P}  \, e^{-\Delta  \cdot \mathcal{L}(\mathcal{P})} \approx \sum_{\text {geodesics }} e^{-\Delta \cdot \mathcal{L}} \,, 
\end{split}
\end{equation}
where $\mathcal{L}$ denote the length of geodesic connecting two points. The {\it geodesic approximation} in AdS$_3$ thus yields the RT formula: 
\begin{equation}
\begin{split}
\text{AdS$_3$/CFT$_2$:} \qquad  S_{\mt{EE}}(\hat{\rho}_A) &\approx   \lim_{n \to 1} \frac{1}{1-n} \log  \left( e^{-\Delta_n \mathcal{L}}  \right)  = \frac{c}{6} \mathcal{L}\left(X_1, X_2 \right) \,. 
\end{split}
\end{equation}
in the AdS$_3$/CFT$_2$ correspondence. In contrast, the situation in de Sitter holography is more subtle. The absence of a well-defined boundary field theory on the stretched horizon prevents us from fixing either $\mu$ (or $\nu$) or $C$ from first principles. At present, they must be regarded as free parameters in the proposed entropy formula, to be constrained only indirectly by consistency conditions such as entropy inequalities. 

More importantly, the essential ingredient of the replica trick is the analytic continuation of the R\'enyi entropies \eqref{eq:Renyi} to the replica limit $n \to 1$, which yields the von Neumann entropy. The universal prefactor $C_n=\tfrac{1}{1-n}$ is uniquely fixed by this requirement: it is the only choice for which the $n\to 1$ continuation reproduces the correct entanglement entropy. In the AdS/CFT correspondence, this structure is guaranteed by the explicit dependence of the twist operator dimensions on $n$, \ie  $\Delta_n=\tfrac{c}{12}(n-\tfrac{1}{1-n})$, which makes the $n\to 1$ limit well-defined. By contrast, in the present de Sitter construction we lack such an explicit $n$-dependence, since the dimension of the twist operator on the replica manifold is not determined yet. As a result, we are unable to perform the replica limit in a controlled manner, and our prescription necessarily leaves an undetermined boundary parameter $C$ as the prefactor of the entropy formula in the $n\to 1$ limit.

While the replica trick and the geodesic approximation motivate the form of the holographic entanglement entropy, a complete holographic dictionary for de Sitter space would be required to determine the precise operator dimensions and normalization coefficients. Establishing such a dictionary remains an important open direction for clarifying the microscopic underpinnings of static patch holography.

\paragraph*{The cut-off scale $\mD_{\epsilon}$}\par\vspace{0.2\baselineskip}
Another subtlety of our construction arises from the introduction of the cutoff scale $D_\epsilon$. This regulator was necessary both to remove the divergence of the proposed entanglement entropy and to ensure its non-negativity. In the AdS/CFT correspondence, the role of such a regulator is geometrically transparent: the cutoff surface is naturally placed near the conformal boundary, and $\epsilon$ parametrizes the distance scale to that asymptotic boundary. This interpretation directly relates the divergence structure of holographic entanglement entropy to the short-distance divergences of the boundary QFT.

By contrast, the situation in de Sitter space is far less clear. Our cutoff $D_\epsilon$ is defined intrinsically from the bulk perspective, but its geometric and physical meaning remains obscure. Unlike AdS, de Sitter space lacks a timelike conformal boundary, and the stretched horizon that plays the role of the holographic screen does not come equipped with a canonical notion of UV regularization. It is therefore not obvious whether $D_\epsilon$ should be thought of as parametrizing a physical distance to the cosmological horizon\footnote{The distance between the stretched horizon and cosmological scale has been determined by the parameter $\sin \alpha = \tfrac{r_\ast}{L}$.}, as a renormalization scheme for the Green’s functions of twist operators, or as an effective encoding of short-distance divergence in the putative dual field theory. Understanding whether $D_\epsilon$ has a geometric realization, or whether it should be regarded as an emergent scale associated with the static patch holography, is an open question that deserves further investigation.

\paragraph*{Future problems.}\par\vspace{0.2\baselineskip}
In this work, we derived conditions relating to the mixing coefficient of the Green's functions of the dS global vacuum and its antipodal counterpart, as well as the location of the stretched horizon. These conditions are converted into the corresponding boundary conditions for scalar operators on the stretched horizon. A deeper exploration of such boundary conditions would provide a more precise formulation of the holographic dictionary in static patch holography and thus represents a promising direction for future study.

Another important issue concerns the interpretation of our proposed entropy from the perspective of quantum information theory. At the semiclassical level, the Ryu–Takayanagi formula is known to satisfy the monogamy of mutual information \cite{Hayden:2011ag}, which serves as a signature of holographic theories admitting a semiclassical gravity dual.  It would be compelling to investigate whether the entropy proposed here also satisfies similar information-theoretic inequalities, and to what extent such constraints can further refine or restrict the structure of static patch holography.
%%%%%%%%%%%%%%%%%%%%%%%%%%%%%%
%%%%%%%%%%%%%%%%%%%%%%%%%%%%%%
\acknowledgments
We would like to thank T. Noumi, T. Mori and T. Takayanagi for useful comments and discussion. This work is supported by MEXT KAKENHI Grant-in-Aid for Transformative Research Areas (A) through the ``Extreme Universe'' collaboration: Grant Number 21H05187. SMR is supported by Peking University under startup Grant No.~7101303985. YS is supported by Grant-in-Aid for JSPS Fellows No.\ 23KJ1337. Y.S. thanks the attendees of the Kyushu IAS-iTHEMS conference: Non-perturbative methods in QFT for fruitful discussions, especially N. Benjamin, S. Komatsu, Y. Kusuki, Y. Nakayama, T. Tada and  Y. Wang.

%%%%%%%%%%%%%%%
\appendix

\section{Laplacian Operator in dS Space}\label{app:Laplacian}
We regard $\mathrm{dS}_{d+1}$, with coordinates $x^\mu$ and covariant derivative $\nabla_\mu$, as a hypersurface embedded in a $(d+2)$-dimensional flat spacetime with coordinates $X^A$ and covariant derivative $D_A$. The hypersurface is defined by the constraint
\begin{equation}
F(X^A)= \eta_{AB}X^A X^B -  L^2 =-X_0^2 + X_1^2 + \cdots + X_{d+1}^2 -  L^2  =0 \,. 
\end{equation}
The normal vector to the hypersurface is then
\begin{equation}
 n_A \propto \partial_A F(X^A)  \quad \longrightarrow \quad n_A = \frac{\eta_{AB} X^B}{L} = \frac{X_A}{L} \,.
\end{equation}
The induced metric on $\mathrm{dS}_{d+1}$ is obtained as
\begin{equation}
  g_{\mu\nu} := \frac{\partial X^A}{\partial x^\mu}\frac{\partial X^B}{\partial x^\nu} \eta_{AB} = e^A_\mu e^B_\nu \eta_{AB}  \,,
\end{equation}
where the tangent vectors are defined by $e^A_\mu = \frac{\partial X^A}{\partial x^\mu}$ and satisfy $e^A_\mu n_A =0$. In terms of the normal vector, the induced metric (first fundamental form) can be expressed as
\begin{equation}
 g^{AB}= g^{\mu\nu}e^A_\mu e^B_\nu   = \eta^{AB} - \epsilon \, n^A n^B \,.  
\end{equation}
The second fundamental form (extrinsic curvature) is given by
\begin{equation}
K_{\mu\nu}:= e^A_\mu e^B_\nu  \( D_{A} n_B   \) \,, \quad \longrightarrow \quad K_{\mu\nu}= \frac{\eta_{AB}}{L}e^A_\mu e^B_\nu  = \frac{1}{L}g_{\mu\nu} \,. 
\end{equation}

\subsection{Laplacian operator $\nabla^2 G$}

We now compute the Laplacian acting on a Green’s function $G(\mP)$, which depends only on the de Sitter invariant length $\mP(x,x')$:
\begin{equation}
\begin{split}
\nabla^\mu \nabla_\mu G(\mP(x,x')) &=  \nabla^\mu \(  \frac{d G}{ d\mP} \nabla_\mu \mP(x,x') \) = \frac{d G}{ d\mP} \nabla^2 \mP + \frac{d^2 G}{ d \mP^2} \nabla^\mu \mP \nabla_\mu \mP \,,
\end{split}
\end{equation}
where
\begin{equation}
\mP(x,x') = \mP(X(x),X'(x')) = \frac{\eta_{AB} X^A X'^B}{L^2} \,.
\end{equation}
First, one finds
\begin{equation}
 \nabla_\mu \mP(x,x') =\nabla_\mu \mP(X(x),X'(x')) =  \frac{\partial X^A}{\partial x^\mu} \frac{\partial \mP}{\partial X^A} =\frac{\partial X^A}{\partial x^\mu}  \frac{X_A'}{L^2}  =e^A_\mu \frac{X_A'}{L^2}\,. 
\end{equation}
It then follows that
\begin{equation}
 \begin{split}
   \left(\nabla \mP(x,x') \right)^2 &= e^A_\mu \frac{X_A'}{L^2} g^{\mu\nu} e^B_\mu \frac{X_B'}{L^2} =  \left(  \eta^{AB} -  \, n^A n^B   \right) \times \frac{X_A'X_B'}{L^4} \\
   &=  \left(  L^2 - \frac{X^A X_A' X^B X_B'}{L^2}  \right)\frac{1}{L^4} \,,
 \end{split}
\end{equation}
which gives the first useful identity:
\begin{equation}
 \text{First identity:} \qquad   \left(\nabla \mP \right)^2  = \frac{1- \mP^2}{L^2} \,.
\end{equation}
For the Laplacian of the de Sitter invariant length $\mP$, we have
\begin{equation}
 \begin{split}
   \nabla^2 \mP &= \nabla^\mu  \nabla_\mu\left(   \frac{\eta_{AB} X^A X'^B}{L^2}     \right) = \nabla^\mu \left(  \frac{\eta_{AB} e^A_\mu X'^B}
   {L^2}   \right) =  \frac{\eta_{AB} X'^B} {L^2}    \nabla^\mu  e^A_\mu \,.
 \end{split}
\end{equation}
Using the Gauss–Weingarten relation (see \eg \cite{Poisson:2009pwt}),
\begin{equation}
 \nabla_\mu e^A_\nu  = \Gamma^\gamma_{\mu\nu} e^A_\gamma - \epsilon K_{\mu\nu} n^A \,, 
\end{equation}
we can obtain
\begin{equation}
 \nabla^\mu e^A_\mu = - \epsilon K n^A \,. 
\end{equation}
For the $\mathrm{dS}_{d+1}$ hypersurface we have
\begin{equation}
 n^A = \frac{X^A}{L} \,, \qquad  K = \frac{d+1}{L} \,, 
\end{equation}
which leads to
\begin{equation}
\begin{split}
 \nabla^2 \mP   &=  \frac{\eta_{AB} X'^B} {L^2}    \nabla^\mu  e^A_\mu  = \frac{\eta_{AB} X'^B} {L^2}  \(  - \frac{d+1}{L} \frac{X^A}{L} \) =- \frac{d+1}{L^2} \mP \,,
\end{split}
\end{equation}
This yields the second useful identity:
\begin{equation}
 \text{Second identity:} \qquad  \nabla^2 \mP  = - \frac{d+1}{L^2} \mP \,.
\end{equation}
Combining the two equalities, we arrive at
\begin{equation}
\boxed{ \nabla^2 G( \mP(x,x')) = \frac{d G}{ d \mP} \nabla^2 \mP + \frac{d^2 G}{ d \mP^2} (\nabla \mP )^2 =  - \frac{d+1}{L^2} \mP\frac{d G}{ d \mP}  +\frac{1- \mP^2}{L^2} \frac{d^2 G}{ d \mP^2} \,.}
\end{equation}
This relation gives the explicit form of the Laplacian acting on de Sitter Green’s functions. The general solutions of the resulting differential equation determine the two-point functions of the twist operator utilized in the main text.

%%%%%%%%%%%%%%%%%%%%%%%%%%%%%%%%%%%%%%%%%%%%%%%%%%%%%%%
\bibliographystyle{JHEP}
\bibliography{paper.bib}

\providecommand{\href}[2]{#2}\begingroup\raggedright\begin{thebibliography}{10}

\bibitem{Callan:1994py}
C.~G. Callan, Jr. and F.~Wilczek, {\it {On geometric entropy}},  {\em Phys.
  Lett. B} {\bf 333} (1994) 55--61,
  [\href{http://arxiv.org/abs/hep-th/9401072}{{\tt hep-th/9401072}}].

\bibitem{Calabrese:2004eu}
P.~Calabrese and J.~L. Cardy, {\it {Entanglement entropy and quantum field
  theory}},  {\em J. Stat. Mech.} {\bf 0406} (2004) P06002,
  [\href{http://arxiv.org/abs/hep-th/0405152}{{\tt hep-th/0405152}}].

\bibitem{Maldacena:1997re}
J.~M. Maldacena, {\it {The Large N limit of superconformal field theories and
  supergravity}},  {\em Adv. Theor. Math. Phys.} {\bf 2} (1998) 231--252,
  [\href{http://arxiv.org/abs/hep-th/9711200}{{\tt hep-th/9711200}}].

\bibitem{Ryu:2006bv}
S.~Ryu and T.~Takayanagi, {\it {Holographic derivation of entanglement entropy
  from AdS/CFT}},  {\em Phys. Rev. Lett.} {\bf 96} (2006) 181602,
  [\href{http://arxiv.org/abs/hep-th/0603001}{{\tt hep-th/0603001}}].

\bibitem{Ryu:2006ef}
S.~Ryu and T.~Takayanagi, {\it {Aspects of Holographic Entanglement Entropy}},
  {\em JHEP} {\bf 08} (2006) 045,
  [\href{http://arxiv.org/abs/hep-th/0605073}{{\tt hep-th/0605073}}].

\bibitem{Lewkowycz:2013nqa}
A.~Lewkowycz and J.~Maldacena, {\it {Generalized gravitational entropy}},  {\em
  JHEP} {\bf 08} (2013) 090, [\href{http://arxiv.org/abs/1304.4926}{{\tt
  arXiv:1304.4926}}].

\bibitem{Lieb:1973zz}
E.~H. Lieb and M.~B. Ruskai, {\it {A Fundamental Property of Quantum-Mechanical
  Entropy}},  {\em Phys. Rev. Lett.} {\bf 30} (1973) 434--436.

\bibitem{Lieb:1973cp}
E.~H. Lieb and M.~B. Ruskai, {\it {Proof of the strong subadditivity of
  quantum-mechanical entropy}},  {\em J. Math. Phys.} {\bf 14} (1973)
  1938--1941.

\bibitem{Headrick:2007km}
M.~Headrick and T.~Takayanagi, {\it {A Holographic proof of the strong
  subadditivity of entanglement entropy}},  {\em Phys. Rev. D} {\bf 76} (2007)
  106013, [\href{http://arxiv.org/abs/0704.3719}{{\tt arXiv:0704.3719}}].

\bibitem{Zamolodchikov:2004ce}
A.~B. Zamolodchikov, {\it {Expectation value of composite field T anti-T in
  two-dimensional quantum field theory}},
  \href{http://arxiv.org/abs/hep-th/0401146}{{\tt hep-th/0401146}}.

\bibitem{McGough:2016lol}
L.~McGough, M.~Mezei, and H.~Verlinde, {\it {Moving the CFT into the bulk with
  $ T\overline{T} $}},  {\em JHEP} {\bf 04} (2018) 010,
  [\href{http://arxiv.org/abs/1611.03470}{{\tt arXiv:1611.03470}}].

\bibitem{Guica:2019nzm}
M.~Guica and R.~Monten, {\it {$T\bar T$ and the mirage of a bulk cutoff}},
  {\em SciPost Phys.} {\bf 10} (2021), no.~2 024,
  [\href{http://arxiv.org/abs/1906.11251}{{\tt arXiv:1906.11251}}].

\bibitem{Sanches:2016sxy}
F.~Sanches and S.~J. Weinberg, {\it {Holographic entanglement entropy
  conjecture for general spacetimes}},  {\em Phys. Rev. D} {\bf 94} (2016),
  no.~8 084034, [\href{http://arxiv.org/abs/1603.05250}{{\tt
  arXiv:1603.05250}}].

\bibitem{Lewkowycz:2019xse}
A.~Lewkowycz, J.~Liu, E.~Silverstein, and G.~Torroba, {\it {$ T\overline{T} $
  and EE, with implications for (A)dS subregion encodings}},  {\em JHEP} {\bf
  04} (2020) 152, [\href{http://arxiv.org/abs/1909.13808}{{\tt
  arXiv:1909.13808}}].

\bibitem{Grado-White:2020wlb}
B.~Grado-White, D.~Marolf, and S.~J. Weinberg, {\it {Radial Cutoffs and
  Holographic Entanglement}},  {\em JHEP} {\bf 01} (2021) 009,
  [\href{http://arxiv.org/abs/2008.07022}{{\tt arXiv:2008.07022}}].

\bibitem{Strominger:2001pn}
A.~Strominger, {\it {The dS / CFT correspondence}},  {\em JHEP} {\bf 10} (2001)
  034, [\href{http://arxiv.org/abs/hep-th/0106113}{{\tt hep-th/0106113}}].

\bibitem{Doi:2022iyj}
K.~Doi, J.~Harper, A.~Mollabashi, T.~Takayanagi, and Y.~Taki, {\it
  {Pseudoentropy in dS/CFT and Timelike Entanglement Entropy}},  {\em Phys.
  Rev. Lett.} {\bf 130} (2023), no.~3 031601,
  [\href{http://arxiv.org/abs/2210.09457}{{\tt arXiv:2210.09457}}].

\bibitem{Doi:2023zaf}
K.~Doi, J.~Harper, A.~Mollabashi, T.~Takayanagi, and Y.~Taki, {\it {Timelike
  entanglement entropy}},  {\em JHEP} {\bf 05} (2023) 052,
  [\href{http://arxiv.org/abs/2302.11695}{{\tt arXiv:2302.11695}}].

\bibitem{Susskind:2021omt}
L.~Susskind, {\it {De Sitter Holography: Fluctuations, Anomalous Symmetry, and
  Wormholes}},  {\em Universe} {\bf 7} (2021), no.~12 464,
  [\href{http://arxiv.org/abs/2106.03964}{{\tt arXiv:2106.03964}}].

\bibitem{Kawamoto:2023nki}
T.~Kawamoto, S.-M. Ruan, Y.-k. Suzuki, and T.~Takayanagi, {\it {A half de
  Sitter holography}},  {\em JHEP} {\bf 10} (2023) 137,
  [\href{http://arxiv.org/abs/2306.07575}{{\tt arXiv:2306.07575}}].

\bibitem{Anninos:2024wpy}
D.~Anninos, D.~A. Galante, and C.~Maneerat, {\it {Cosmological observatories}},
   {\em Class. Quant. Grav.} {\bf 41} (2024), no.~16 165009,
  [\href{http://arxiv.org/abs/2402.04305}{{\tt arXiv:2402.04305}}].

\bibitem{Silverstein:2022dfj}
E.~Silverstein, {\it {Black hole to cosmic horizon microstates in string/M
  theory: timelike boundaries and internal averaging}},  {\em JHEP} {\bf 05}
  (2023) 160, [\href{http://arxiv.org/abs/2212.00588}{{\tt arXiv:2212.00588}}].

\bibitem{Batra:2024kjl}
G.~Batra, G.~B. De~Luca, E.~Silverstein, G.~Torroba, and S.~Yang, {\it
  {Bulk-local dS$_{3}$ holography: the matter with $ T\overline{T} $ +
  \ensuremath{\Lambda}$_{2}$}},  {\em JHEP} {\bf 10} (2024) 072,
  [\href{http://arxiv.org/abs/2403.01040}{{\tt arXiv:2403.01040}}].

\bibitem{Arias:2019pzy}
C.~Arias, F.~Diaz, and P.~Sundell, {\it {De Sitter Space and Entanglement}},
  {\em Class. Quant. Grav.} {\bf 37} (2020), no.~1 015009,
  [\href{http://arxiv.org/abs/1901.04554}{{\tt arXiv:1901.04554}}].

\bibitem{Arias:2019zug}
C.~Arias, F.~Diaz, R.~Olea, and P.~Sundell, {\it {Liouville description of
  conical defects in dS$_4$, Gibbons-Hawking entropy as modular entropy, and
  dS$_3$ holography}},  {\em JHEP} {\bf 04} (2020) 124,
  [\href{http://arxiv.org/abs/1906.05310}{{\tt arXiv:1906.05310}}].

\bibitem{Narayan:2022afv}
K.~Narayan, {\it {de Sitter space, extremal surfaces, and time entanglement}},
  {\em Phys. Rev. D} {\bf 107} (2023), no.~12 126004,
  [\href{http://arxiv.org/abs/2210.12963}{{\tt arXiv:2210.12963}}].

\bibitem{Narayan:2023zen}
K.~Narayan, {\it {Further remarks on de Sitter space, extremal surfaces, and
  time entanglement}},  {\em Phys. Rev. D} {\bf 109} (2024), no.~8 086009,
  [\href{http://arxiv.org/abs/2310.00320}{{\tt arXiv:2310.00320}}].

\bibitem{Nomura:2017fyh}
Y.~Nomura, P.~Rath, and N.~Salzetta, {\it {Spacetime from Unentanglement}},
  {\em Phys. Rev. D} {\bf 97} (2018), no.~10 106010,
  [\href{http://arxiv.org/abs/1711.05263}{{\tt arXiv:1711.05263}}].

\bibitem{Murdia:2022giv}
C.~Murdia, Y.~Nomura, and K.~Ritchie, {\it {Black hole and de Sitter
  microstructures from a semiclassical perspective}},  {\em Phys. Rev. D} {\bf
  107} (2023), no.~2 026016, [\href{http://arxiv.org/abs/2207.01625}{{\tt
  arXiv:2207.01625}}].

\bibitem{Franken:2023pni}
V.~Franken, H.~Partouche, F.~Rondeau, and N.~Toumbas, {\it {Bridging the static
  patches: de Sitter holography and entanglement}},  {\em JHEP} {\bf 08} (2023)
  074, [\href{http://arxiv.org/abs/2305.12861}{{\tt arXiv:2305.12861}}].

\bibitem{Franken:2023jas}
V.~Franken, H.~Partouche, F.~Rondeau, and N.~Toumbas, {\it {Closed FRW
  holography: a time-dependent ER=EPR realization}},  {\em JHEP} {\bf 05}
  (2024) 219, [\href{http://arxiv.org/abs/2310.20652}{{\tt arXiv:2310.20652}}].

\bibitem{Chang:2024voo}
J.-C. Chang, S.~He, Y.-X. Liu, and L.~Zhao, {\it {Holographic
  TT{\textasciimacron} deformation of the entanglement entropy in
  (A)dS3/CFT2}},  {\em Phys. Rev. D} {\bf 112} (2025), no.~2 026013,
  [\href{http://arxiv.org/abs/2409.08198}{{\tt arXiv:2409.08198}}].

\bibitem{Calabrese:2009qy}
P.~Calabrese and J.~Cardy, {\it {Entanglement entropy and conformal field
  theory}},  {\em J. Phys. A} {\bf 42} (2009) 504005,
  [\href{http://arxiv.org/abs/0905.4013}{{\tt arXiv:0905.4013}}].

\bibitem{Nomura:2016ikr}
Y.~Nomura, N.~Salzetta, F.~Sanches, and S.~J. Weinberg, {\it {Toward a
  Holographic Theory for General Spacetimes}},  {\em Phys. Rev. D} {\bf 95}
  (2017), no.~8 086002, [\href{http://arxiv.org/abs/1611.02702}{{\tt
  arXiv:1611.02702}}].

\bibitem{Cotler:2019nbi}
J.~Cotler, K.~Jensen, and A.~Maloney, {\it {Low-dimensional de Sitter quantum
  gravity}},  {\em JHEP} {\bf 06} (2020) 048,
  [\href{http://arxiv.org/abs/1905.03780}{{\tt arXiv:1905.03780}}].

\bibitem{Arenas-Henriquez:2022pyh}
G.~Arenas-Henriquez, F.~Diaz, and P.~Sundell, {\it {Logarithmic corrections,
  entanglement entropy, and UV cutoffs in de Sitter spacetime}},  {\em JHEP}
  {\bf 08} (2022) 261, [\href{http://arxiv.org/abs/2206.10427}{{\tt
  arXiv:2206.10427}}].

\bibitem{Hikida:2021ese}
Y.~Hikida, T.~Nishioka, T.~Takayanagi, and Y.~Taki, {\it {Holography in de
  Sitter Space via Chern-Simons Gauge Theory}},  {\em Phys. Rev. Lett.} {\bf
  129} (2022), no.~4 041601, [\href{http://arxiv.org/abs/2110.03197}{{\tt
  arXiv:2110.03197}}].

\bibitem{Narovlansky:2023lfz}
V.~Narovlansky and H.~Verlinde, {\it {Double-scaled SYK and de Sitter
  Holography}},  \href{http://arxiv.org/abs/2310.16994}{{\tt
  arXiv:2310.16994}}.

\bibitem{Gubser:1998bc}
S.~S. Gubser, I.~R. Klebanov, and A.~M. Polyakov, {\it {Gauge theory
  correlators from noncritical string theory}},  {\em Phys. Lett. B} {\bf 428}
  (1998) 105--114, [\href{http://arxiv.org/abs/hep-th/9802109}{{\tt
  hep-th/9802109}}].

\bibitem{Witten:1998qj}
E.~Witten, {\it {Anti-de Sitter space and holography}},  {\em Adv. Theor. Math.
  Phys.} {\bf 2} (1998) 253--291,
  [\href{http://arxiv.org/abs/hep-th/9802150}{{\tt hep-th/9802150}}].

\bibitem{Maldacena:2002vr}
J.~M. Maldacena, {\it {Non-Gaussian features of primordial fluctuations in
  single field inflationary models}},  {\em JHEP} {\bf 05} (2003) 013,
  [\href{http://arxiv.org/abs/astro-ph/0210603}{{\tt astro-ph/0210603}}].

\bibitem{Anninos:2011ui}
D.~Anninos, T.~Hartman, and A.~Strominger, {\it {Higher Spin Realization of the
  dS/CFT Correspondence}},  {\em Class. Quant. Grav.} {\bf 34} (2017), no.~1
  015009, [\href{http://arxiv.org/abs/1108.5735}{{\tt arXiv:1108.5735}}].

\bibitem{Hikida:2022ltr}
Y.~Hikida, T.~Nishioka, T.~Takayanagi, and Y.~Taki, {\it {CFT duals of
  three-dimensional de Sitter gravity}},  {\em JHEP} {\bf 05} (2022) 129,
  [\href{http://arxiv.org/abs/2203.02852}{{\tt arXiv:2203.02852}}].

\bibitem{Susskind:2021dfc}
L.~Susskind, {\it {Black Holes Hint towards De Sitter Matrix Theory}},  {\em
  Universe} {\bf 9} (2023), no.~8 368,
  [\href{http://arxiv.org/abs/2109.01322}{{\tt arXiv:2109.01322}}].

\bibitem{Shaghoulian:2021cef}
E.~Shaghoulian, {\it {The central dogma and cosmological horizons}},  {\em
  JHEP} {\bf 01} (2022) 132, [\href{http://arxiv.org/abs/2110.13210}{{\tt
  arXiv:2110.13210}}].

\bibitem{Shaghoulian:2022fop}
E.~Shaghoulian and L.~Susskind, {\it {Entanglement in De Sitter space}},  {\em
  JHEP} {\bf 08} (2022) 198, [\href{http://arxiv.org/abs/2201.03603}{{\tt
  arXiv:2201.03603}}].

\bibitem{Bousso:1999xy}
R.~Bousso, {\it {A Covariant entropy conjecture}},  {\em JHEP} {\bf 07} (1999)
  004, [\href{http://arxiv.org/abs/hep-th/9905177}{{\tt hep-th/9905177}}].

\bibitem{Bousso:2014sda}
R.~Bousso, H.~Casini, Z.~Fisher, and J.~Maldacena, {\it {Proof of a Quantum
  Bousso Bound}},  {\em Phys. Rev. D} {\bf 90} (2014), no.~4 044002,
  [\href{http://arxiv.org/abs/1404.5635}{{\tt arXiv:1404.5635}}].

\bibitem{Bousso:2014uxa}
R.~Bousso, H.~Casini, Z.~Fisher, and J.~Maldacena, {\it {Entropy on a null
  surface for interacting quantum field theories and the Bousso bound}},  {\em
  Phys. Rev. D} {\bf 91} (2015), no.~8 084030,
  [\href{http://arxiv.org/abs/1406.4545}{{\tt arXiv:1406.4545}}].

\bibitem{Bousso:2002ju}
R.~Bousso, {\it {The Holographic principle}},  {\em Rev. Mod. Phys.} {\bf 74}
  (2002) 825--874, [\href{http://arxiv.org/abs/hep-th/0203101}{{\tt
  hep-th/0203101}}].

\bibitem{tHooft:1993dmi}
G.~'t~Hooft, {\it {Dimensional reduction in quantum gravity}},  {\em Conf.
  Proc. C} {\bf 930308} (1993) 284--296,
  [\href{http://arxiv.org/abs/gr-qc/9310026}{{\tt gr-qc/9310026}}].

\bibitem{Susskind:1994vu}
L.~Susskind, {\it {The World as a hologram}},  {\em J. Math. Phys.} {\bf 36}
  (1995) 6377--6396, [\href{http://arxiv.org/abs/hep-th/9409089}{{\tt
  hep-th/9409089}}].

\bibitem{Hartle:1983ai}
J.~B. Hartle and S.~W. Hawking, {\it {Wave Function of the Universe}},  {\em
  Phys. Rev. D} {\bf 28} (1983) 2960--2975.

\bibitem{Wall:2012uf}
A.~C. Wall, {\it {Maximin Surfaces, and the Strong Subadditivity of the
  Covariant Holographic Entanglement Entropy}},  {\em Class. Quant. Grav.} {\bf
  31} (2014), no.~22 225007, [\href{http://arxiv.org/abs/1211.3494}{{\tt
  arXiv:1211.3494}}].

\bibitem{Hubeny:2007xt}
V.~E. Hubeny, M.~Rangamani, and T.~Takayanagi, {\it {A Covariant holographic
  entanglement entropy proposal}},  {\em JHEP} {\bf 07} (2007) 062,
  [\href{http://arxiv.org/abs/0705.0016}{{\tt arXiv:0705.0016}}].

\bibitem{Balasubramanian:2002zh}
V.~Balasubramanian, J.~de~Boer, and D.~Minic, {\it {Notes on de Sitter space
  and holography}},  {\em Class. Quant. Grav.} {\bf 19} (2002) 5655--5700,
  [\href{http://arxiv.org/abs/hep-th/0207245}{{\tt hep-th/0207245}}].

\bibitem{Cotler:2023xku}
J.~Cotler and A.~Strominger, {\it {Cosmic ER=EPR in dS/CFT}},
  \href{http://arxiv.org/abs/2302.00632}{{\tt arXiv:2302.00632}}.

\bibitem{Banks:1998dd}
T.~Banks, M.~R. Douglas, G.~T. Horowitz, and E.~J. Martinec, {\it {AdS dynamics
  from conformal field theory}},
  \href{http://arxiv.org/abs/hep-th/9808016}{{\tt hep-th/9808016}}.

\bibitem{Balasubramanian:1998sn}
V.~Balasubramanian, P.~Kraus, and A.~E. Lawrence, {\it {Bulk versus boundary
  dynamics in anti-de Sitter space-time}},  {\em Phys. Rev. D} {\bf 59} (1999)
  046003, [\href{http://arxiv.org/abs/hep-th/9805171}{{\tt hep-th/9805171}}].

\bibitem{Balasubramanian:1998de}
V.~Balasubramanian, P.~Kraus, A.~E. Lawrence, and S.~P. Trivedi, {\it
  {Holographic probes of anti-de Sitter space-times}},  {\em Phys. Rev. D} {\bf
  59} (1999) 104021, [\href{http://arxiv.org/abs/hep-th/9808017}{{\tt
  hep-th/9808017}}].

\bibitem{Harlow:2011ke}
D.~Harlow and D.~Stanford, {\it {Operator Dictionaries and Wave Functions in
  AdS/CFT and dS/CFT}},  \href{http://arxiv.org/abs/1104.2621}{{\tt
  arXiv:1104.2621}}.

\bibitem{Bousso:2001mw}
R.~Bousso, A.~Maloney, and A.~Strominger, {\it {Conformal vacua and entropy in
  de Sitter space}},  {\em Phys. Rev. D} {\bf 65} (2002) 104039,
  [\href{http://arxiv.org/abs/hep-th/0112218}{{\tt hep-th/0112218}}].

\bibitem{Spradlin:2001pw}
M.~Spradlin, A.~Strominger, and A.~Volovich, {\it {Les Houches lectures on de
  Sitter space}},  in {\em {Les Houches Summer School: Session 76: Euro Summer
  School on Unity of Fundamental Physics: Gravity, Gauge Theory and Strings}},
  pp.~423--453, 10, 2001.
\newblock \href{http://arxiv.org/abs/hep-th/0110007}{{\tt hep-th/0110007}}.

\bibitem{Chapman:2022mqd}
S.~Chapman, D.~A. Galante, E.~Harris, S.~U. Sheorey, and D.~Vegh, {\it {Complex
  geodesics in de Sitter space}},  {\em JHEP} {\bf 03} (2023) 006,
  [\href{http://arxiv.org/abs/2212.01398}{{\tt arXiv:2212.01398}}].

\bibitem{Balasubramanian:1999zv}
V.~Balasubramanian and S.~F. Ross, {\it {Holographic particle detection}},
  {\em Phys. Rev. D} {\bf 61} (2000) 044007,
  [\href{http://arxiv.org/abs/hep-th/9906226}{{\tt hep-th/9906226}}].

\bibitem{Louko:2000tp}
J.~Louko, D.~Marolf, and S.~F. Ross, {\it {On geodesic propagators and black
  hole holography}},  {\em Phys. Rev. D} {\bf 62} (2000) 044041,
  [\href{http://arxiv.org/abs/hep-th/0002111}{{\tt hep-th/0002111}}].

\bibitem{Aparicio:2011zy}
J.~Aparicio and E.~Lopez, {\it {Evolution of Two-Point Functions from
  Holography}},  {\em JHEP} {\bf 12} (2011) 082,
  [\href{http://arxiv.org/abs/1109.3571}{{\tt arXiv:1109.3571}}].

\bibitem{Balasubramanian:2012tu}
V.~Balasubramanian, A.~Bernamonti, B.~Craps, V.~Ker{\"a}nen, E.~Keski-Vakkuri,
  B.~M{\"u}ller, L.~Thorlacius, and J.~Vanhoof, {\it {Thermalization of the
  spectral function in strongly coupled two dimensional conformal field
  theories}},  {\em JHEP} {\bf 04} (2013) 069,
  [\href{http://arxiv.org/abs/1212.6066}{{\tt arXiv:1212.6066}}].

\bibitem{Noumi:2025lbb}
T.~Noumi, F.~Sano, and Y.-k. Suzuki, {\it {Holographic Entanglement Entropy in
  the FLRW Universe}},  \href{http://arxiv.org/abs/2504.10457}{{\tt
  arXiv:2504.10457}}.

\bibitem{Mori:2023swn}
T.~Mori and B.~Yoshida, {\it {Exploring causality in braneworld/cutoff
  holography via holographic scattering}},  {\em JHEP} {\bf 10} (2023) 104,
  [\href{http://arxiv.org/abs/2308.00739}{{\tt arXiv:2308.00739}}].

\bibitem{Franken:2024wmh}
V.~Franken and T.~Mori, {\it {Horizon causality from holographic scattering in
  asymptotically dS$_{3}$}},  {\em JHEP} {\bf 12} (2024) 199,
  [\href{http://arxiv.org/abs/2410.09050}{{\tt arXiv:2410.09050}}].

\bibitem{Geng:2020kxh}
H.~Geng, {\it {Non-local entanglement and fast scrambling in de-Sitter
  holography}},  {\em Annals Phys.} {\bf 426} (2021) 168402,
  [\href{http://arxiv.org/abs/2005.00021}{{\tt arXiv:2005.00021}}].

\bibitem{Chandrasekaran:2022cip}
V.~Chandrasekaran, R.~Longo, G.~Penington, and E.~Witten, {\it {An algebra of
  observables for de Sitter space}},  {\em JHEP} {\bf 02} (2023) 082,
  [\href{http://arxiv.org/abs/2206.10780}{{\tt arXiv:2206.10780}}].

\bibitem{Hirata:2006jx}
T.~Hirata and T.~Takayanagi, {\it {AdS/CFT and strong subadditivity of
  entanglement entropy}},  {\em JHEP} {\bf 02} (2007) 042,
  [\href{http://arxiv.org/abs/hep-th/0608213}{{\tt hep-th/0608213}}].

\bibitem{Hayden:2011ag}
P.~Hayden, M.~Headrick, and A.~Maloney, {\it {Holographic Mutual Information is
  Monogamous}},  {\em Phys. Rev. D} {\bf 87} (2013), no.~4 046003,
  [\href{http://arxiv.org/abs/1107.2940}{{\tt arXiv:1107.2940}}].

\bibitem{Poisson:2009pwt}
E.~Poisson, {\em {A Relativist's Toolkit: The Mathematics of Black-Hole
  Mechanics}}.
\newblock Cambridge University Press, 12, 2009.

\end{thebibliography}\endgroup
\end{document}